\documentclass[preprint,amsmath,amssymb,aps]{revtex4-1}

\selectfont

\usepackage[colorlinks]{hyperref}
\usepackage{graphicx}       
\usepackage{dcolumn}        
\usepackage{bm}             
\usepackage{amssymb}
\usepackage{amstext}
\usepackage{tensor}
\usepackage[utf8]{inputenc}
\usepackage[colorlinks]{hyperref}
\usepackage{color}
\usepackage[usenames,dvipsnames,svgnames,table]{xcolor}
\usepackage{subfigure}
\usepackage{multirow}

\newcommand{\eps}{\epsilon}

\newcommand{\bx}{\mathbf{x}}

\newcommand{\beq}{\begin{equation}}
\newcommand{\eeq}{\end{equation}}

\begin{document}

 \title{Binary black hole shadows, chaotic scattering and the Cantor set}

\author{Jake O.~Shipley}
\email{joshipley1@sheffield.ac.uk}
\affiliation{Consortium for Fundamental Physics, School of Mathematics and Statistics, University of Sheffield, Hicks Building, Hounsfield Road, Sheffield S3 7RH, United Kingdom.}

\author{Sam R.~Dolan}
\email{s.dolan@sheffield.ac.uk}
\affiliation{Consortium for Fundamental Physics, School of Mathematics and Statistics, University of Sheffield, Hicks Building, Hounsfield Road, Sheffield S3 7RH, United Kingdom.}

\date{\today}

\begin{abstract}
We investigate the qualitative features of binary black hole shadows using the model of two extremally charged black holes in static equilibrium (a Majumdar--Papapetrou solution). Our perspective is that binary spacetimes are natural exemplars of {\it chaotic scattering}, because they admit more than one fundamental null orbit, and thus an uncountably infinite set of perpetual null orbits which generate scattering singularities in initial data. Inspired by the three-disc model, we develop an appropriate symbolic dynamics to describe planar null geodesics on the double black hole spacetime. We show that a one-dimensional (1D) black hole shadow may constructed through an iterative procedure akin to the construction of the Cantor set; thus the 1D shadow is self-similar. Next, we study non-planar rays, to understand how angular momentum affects the existence and properties of the fundamental null orbits. Taking slices through 2D shadows, we observe three types of 1D shadow: regular, Cantor-like, and highly chaotic. The switch from Cantor-like to regular occurs where outer fundamental orbits are forbidden by angular momentum. The highly chaotic part is associated with an unexpected feature: stable and bounded null orbits, which exist around two black holes of equal mass $M$ separated by $a_1 < a < \sqrt{2} a_1$, where $a_1 = 4M/\sqrt{27}$. To show how this possibility arises, we define a certain potential function and classify its stationary points. We conjecture that the highly chaotic parts of the 2D shadow possess the Wada property. Finally, we consider the possibility of following null geodesics through event horizons, and chaos in the maximally extended spacetime.
\end{abstract}

\maketitle

\section{Introduction\label{sec:introduction}}
In the near future, astronomers aim to view the `shadow' of a black hole for the first time. The Event Horizon Telescope \cite{EHT, BlackHoleCam}, a global network of radio telescopes using long baseline interferometry \cite{Falcke:1999pj}, will form a high-resolution image of the environment of the centre of our galaxy. It is anticipated that such images will allow us to directly infer black hole mass, spin and inclination \cite{Psaltis:2014dea}; to test the no-hair conjecture \cite{Loeb:2013lfa}; and to probe general relativity in the strong-field regime \cite{Ricarte:2014nca, Johannsen:2015hib, Johannsen:2015}.

\begin{figure}
 \includegraphics[width=1\textwidth]{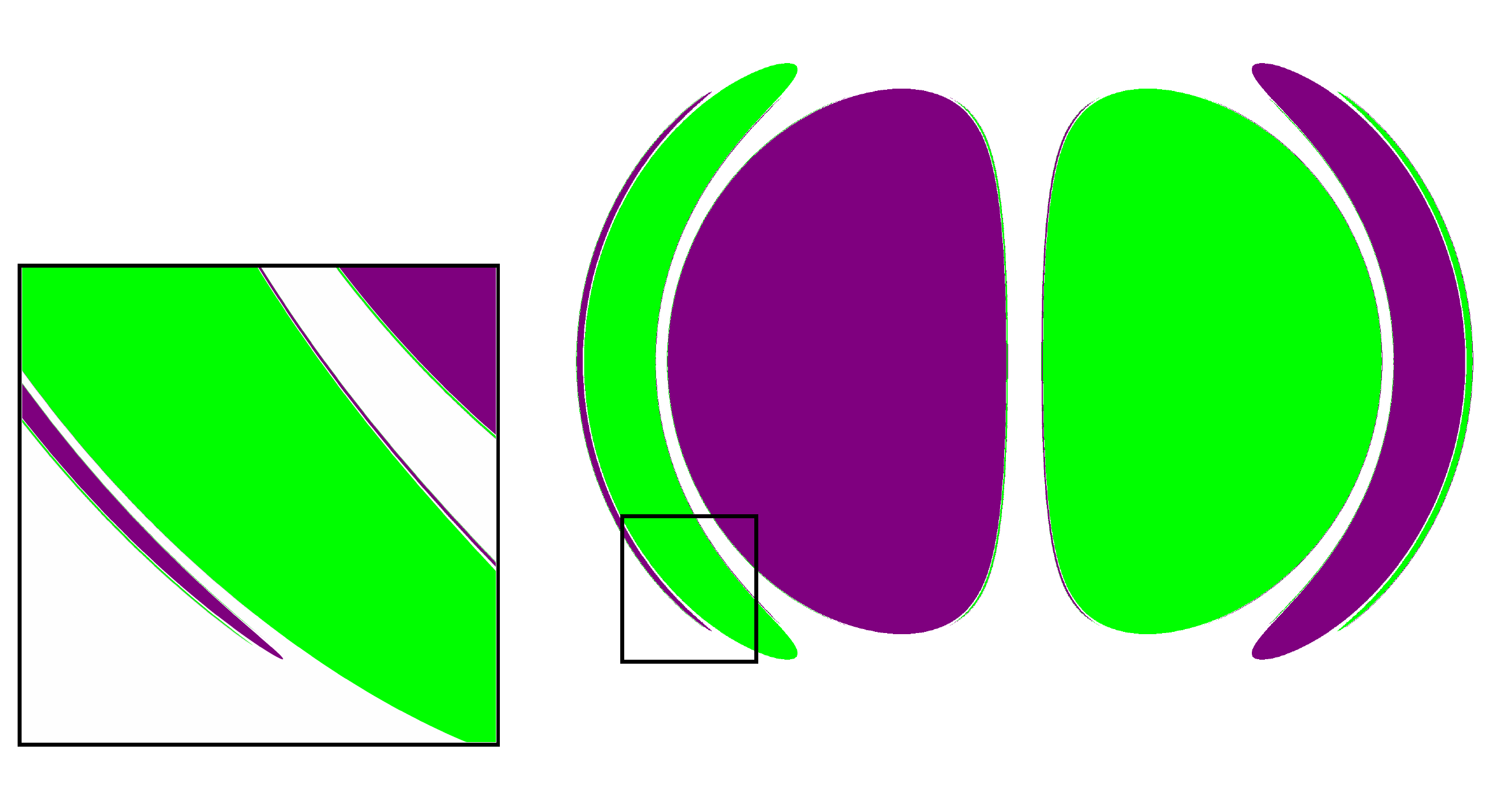}
 \caption{The shadow of a pair of black holes (equal in mass and extremally charged) viewed side on, hinting at a self-similar hierarchy of `eyebrow-like' structures: partial arcs surrounding the primary shadow.}
 \label{fig:binaryshadow}
\end{figure}

Black holes also occur in binary pairs, as confirmed in spectacular style by the first direct detection of gravitational waves \cite{Abbott:2016blz, TheLIGOScientific:2016wfe}. What kind of shadow is cast by a \emph{pair} of black holes in a close orbit? Recent work by Bohn {\it et al.}~\cite{Bohn:2014xxa} has confirmed that a binary shadow is qualitatively different to a superposition of two singleton shadows. It possesses `eyebrow-like' features, as anticipated by Refs.~\cite{Nitta:2011in, Yumoto:2012kz}: partial arcs surrounding the primary shadows. In fact, as we shall describe here, the (idealised) binary shadow is expected to have \emph{self-similar} properties (see e.g.~Fig.~\ref{fig:binaryshadow}) and thus it may possess, in principle, an infinite hierarchy of eyebrows. Or, in the words of De Morgan, echoing Swift, ``Great fleas have little fleas upon their backs to bite 'em, and little fleas have lesser fleas, and so \emph{ad infinitum}.''

Thanks to advances in numerical relativity, it is now possible for specialists to study realistic binary black hole spacetimes without resort to approximations or surrogate models \cite{Bohn:2014xxa}. Yet, as our principal aim here is to explore the \emph{qualitative} features of binary shadows, we will focus on a simpler closed-form model: a fixed pair of extremally charged black holes. This is an example of a Majumdar--Papapetrou (MP) geometry, a static solution of the Einstein--Maxwell equations in which gravitational attraction and electrostatic repulsion are in balance. The properties of MP geometries were studied in detail by Hartle \& Hawking \cite{Hartle:1972ya} and Chandrasekhar \cite{Chandrasekhar:1989vk}. Contopoulos revealed that the binary MP spacetime exhibits self-similarity and chaotic dynamics \cite{Contopoulos:1990, Contopoulos:1991, Contopoulos:1993, Contopoulos:1999js, Contopoulos:Harsula:2004, Contopoulos:2005, Contopoulos:book}. Influential perspectives on the role of chaos in binary systems in relativity have followed from Yurtsever \cite{Yurtsever:1994yb}, Dettmann \cite{Dettmann:1994dj,Dettmann:1995ex}, Cornish \cite{Cornish:1996de, Cornish:1996yg}, Frankel \cite{Cornish:1997hs}, Levin \cite{Levin:1999zx,Levin:1999wk} and many others \cite{Hobill, Sota:1995ms, deMoura:1999zd, Alonso:2007ts, Hanan:2007}.

The MP class of solutions has two particularly nice properties. First, the line element ($ds^2 = g_{\mu \nu} dx^\mu dx^\nu$) may be written in a standard coordinate system $\{t,x,y,z\}$ in isotropic form,
\beq
ds^2 = -U^{-2}(\bx) dt^2 + U^2(\bx) d\bx \cdot d\bx.  \label{eq:MP}
\eeq
Here, $U(\bx)$ satisfies Laplace's equation $\nabla^2 U = 0$, and thus solutions may be generated by linear superposition. Second, the null geodesics of the geometry are determined by a Maupertuis principle, as they are the extremal paths of the action functional
\beq
S = \int U^2(\bx) dl , \quad \quad dl = \sqrt{d\bx \cdot d\bx} = \sqrt{dx^2+dy^2+dz^2} \; . \label{eq:S}
\eeq
In other words, $U^2(\bx)$ may be interpreted as a (variable) refractive index. Thus, not only is this system amenable to standard Lagrangian/Hamiltonian methods, it is also conceivable that a material system mimicking MP geometries could one day be created in the laboratory.

In the Newtonian context, the problem of two fixed centres (Euler's three-body problem) is \emph{integrable}: the equations governing particle motion are separable in spheroidal coordinates, and so there arises an additional constant of motion (Whittaker's constant). By contrast, the geodesic equations for two fixed black holes are not integrable \cite{Chandrasekhar:1989vk, Contopoulos:1990}, and we should anticipate richer phenomena.

\begin{figure}
  \includegraphics[width=7cm]{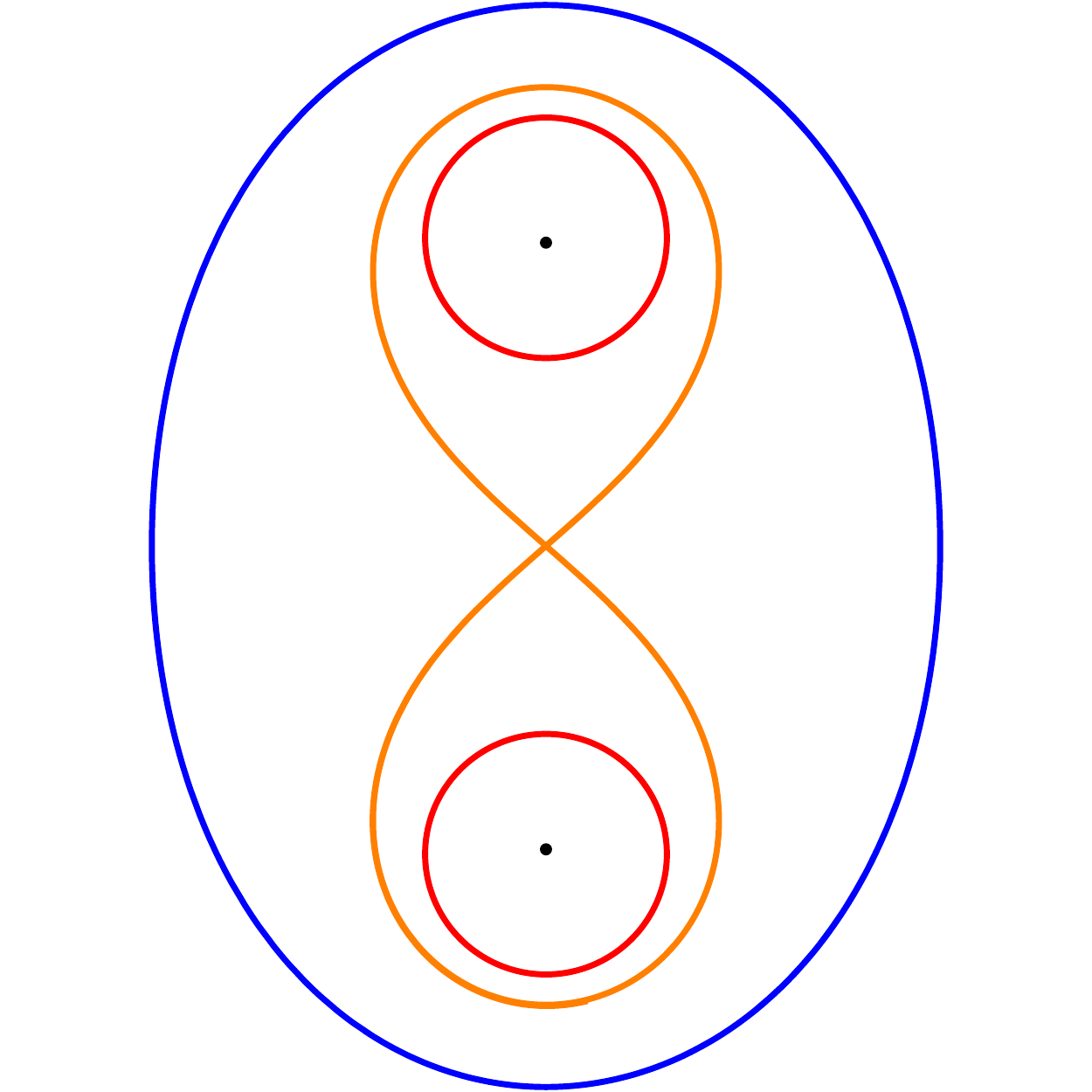}
 \caption{Fundamental planar null orbits [coloured lines] of two fixed black holes (the double black hole Majumdar--Papapetrou geometry). N.B.~In this coordinate system, the black hole event horizons appear as points.}
 \label{fig:fundamental}
\end{figure}

In this article we will advance the view that binary black hole shadows are a fascinating example of \emph{chaotic scattering} in nature. Chaotic scattering occurs in a wide range of contexts, such as chemical reactions, the Newtonian three-body problem, the motion of point vortices, particle motion in electric and magnetic fields, and geometric optics \cite{Ott:Tel:1993}. In Eckhardt's definition \cite{Eckhardt:1988}, scattering in a Hamiltonian system is \emph{irregular} (or chaotic) if there exists, on some manifold of initial data, an infinity of distinct `scattering singularities' of measure zero, typically arranged into a fractal set. A `scattering singularity' is an initial value for which the scattering process is not defined, and some physical quantity such as deflection angle or time delay becomes singular. In regions of irregular scattering, a small variation in initial conditions leads to a completely different outcome in the scattering process. In the case of two fixed black holes, the `singularities' correspond to those light rays (null geodesics) that asymptote towards {\it perpetual orbits}: unstable null orbits which are neither scattered, nor absorbed by either black hole. An important subset of the perpetual orbits are the {\it periodic orbits} \cite{Contopoulos:1990, Contopoulos:Harsula:2004, Contopoulos:book}. The `fundamental' periodic orbits in the plane for two fixed black holes of equal mass are shown in Fig.~\ref{fig:fundamental}.

\begin{figure}
 \includegraphics[width=7cm]{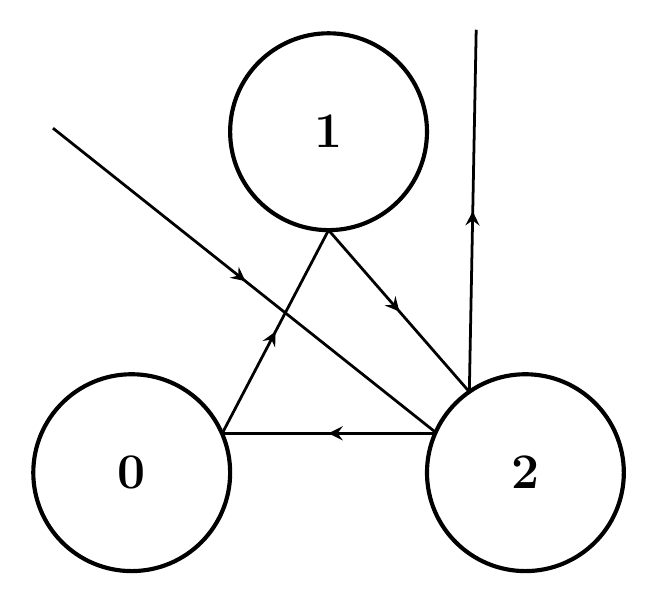}
 \caption{An example of a trajectory in Eckhardt's three-disc model \cite{Eckhardt:1988}. In symbolic dynamics a trajectory is encoded by a sequence of digits; there are several ways to do this. In `collision dynamics' (our term), this trajectory is $2012$, as it collides with the discs in this order. In `decision dynamics' (our term), the trajectory is $0001$ (or left-left-left-out), referring to the `decision' made after each collision. In this article we apply `decision dynamics' to light rays around a binary black hole.}
 \label{fig:3disc}
\end{figure}

The classic exemplar of chaotic scattering is Eckhardt's three-disc model \cite{Eckhardt:1987}, studied in detail by Gaspard \& Rice~\cite{Gaspard:1989}, in which an incoming particle undergoes perfectly elastic collisions with three fixed discs until it escapes towards infinity (see Fig.~\ref{fig:3disc}). The qualitative features of this model may be understood via symbolic dynamics \cite{ChaosBook}. Each disc may be assigned a label: 0, 1, or 2. A trajectory is labelled by a sequence of digits, corresponding to the discs with which it collides. For example, 021 would label a trajectory hitting first disc 0, then disc 2, then disc 1, before escaping from the system. In principle, there are sequences of arbitrary length. The `perpetual' trajectories, which do not escape the system, are labelled with an infinite sequence of digits. Among these are `asymptotically periodic' trajectories with recurring sequences, which are embedded in the former as the rational numbers are embedded in the reals.

There is an alternative way of labelling trajectories in the three-disc problem \cite{Eckhardt:1987}. After each collision, a trajectory can be continued in three ways: we may decide to continue to the disc on the left (0), to the disc on the right (2), or out of the system (1). We note that a `perpetual' trajectory corresponds to an infinite ternary sequence which does not feature the digit 1. Thus, the symbolic representation of perpetual orbits maps directly on to the usual ternary representation of the Cantor set. To draw a convenient distinction, we shall call the latter approach {\it decision dynamics}, and the approach of the previous paragraph {\it collision dynamics}. Note that in decision dynamics, neighbouring digits are permitted to be the same, whereas this is forbidden in collision dynamics.

\begin{figure}
 \includegraphics[width=8cm]{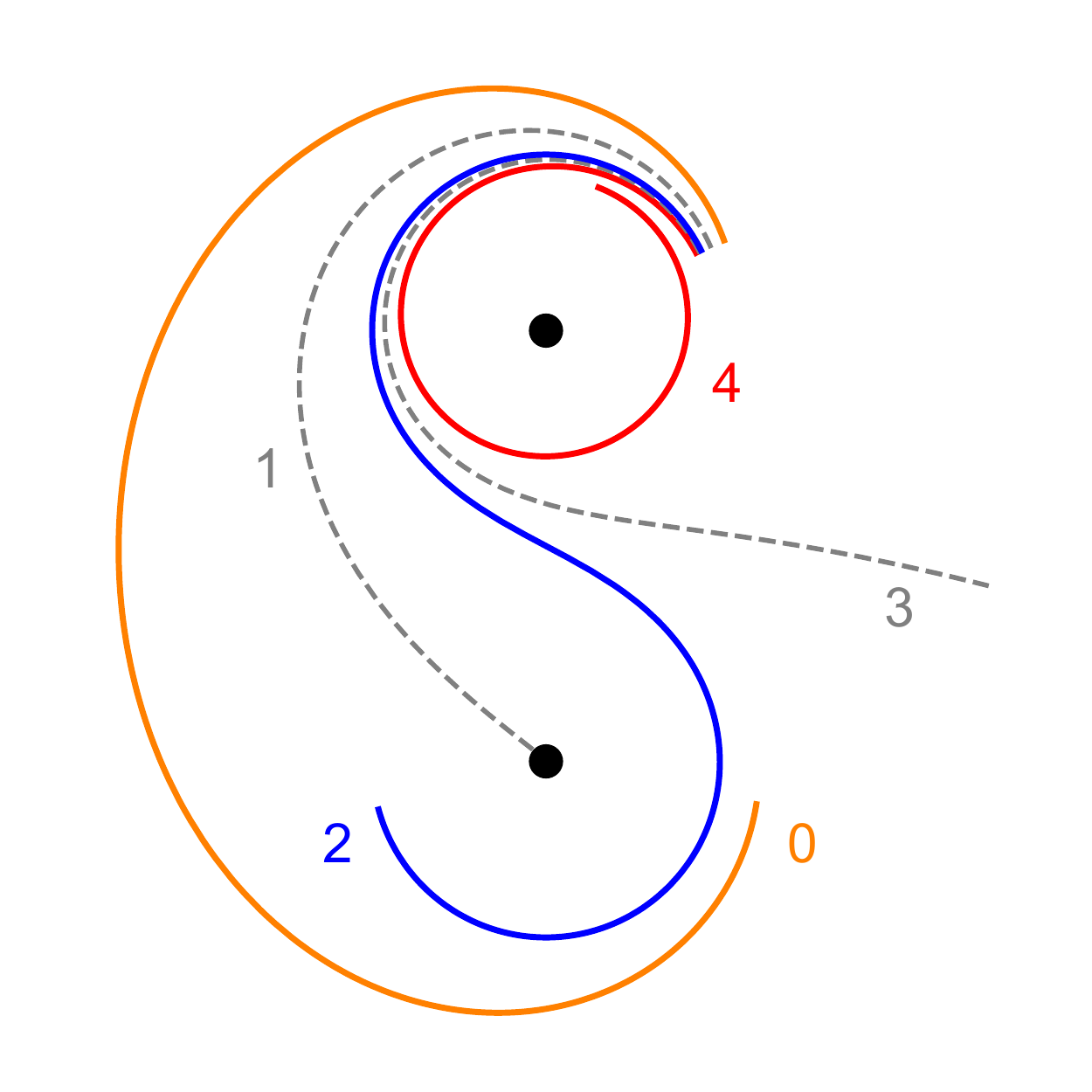}
 \caption{Null rays around a pair of black holes (planar case). In `decision dynamics', a null geodesic in a (initially narrow) congruence passing around the upper black hole faces a choice: it may (0) pass around the lower black hole in the same sense;  or (2) in the opposite sense; or (4) around the upper black hole in the same sense. Between these cases, it may plunge into the black hole (1), or escape to infinity (3).}
 \label{fig:decision}
\end{figure}

In this article, we shall develop decision dynamics to better understand the binary shadow problem. 
The key idea is illustrated in Fig.~\ref{fig:decision}. A geodesic passing around the upper black hole may continue by passing around the lower black hole in the same sense (0); the lower black hole in the opposite sense (2); or the upper black hole in the same sense (4). In between these possibilities, it may fall into the black hole (1), or escape to infinity (3).  

The article is organised as follows. In brief, in Sec.~\ref{sec:geodesics} we consider rays in a plane and one-dimensional (1D) shadows; in Sec.~\ref{sec:2Dshadows} we progress to non-planar rays and two-dimensional (2D) shadows; and in Sec.~\ref{sec:extensions} we extend to consider rays through event horizons. We conclude in Sec.~\ref{sec:conclusions} with a discussion. In more detail, in Sec.~\ref{sec:geodesics} we introduce the Majumdar--Papapetrou spacetime; derive its geodesic equations in the Hamiltonian formalism (\ref{subsec:geodesics}); define the black hole shadow (\ref{subsec:shadows}); explore a 1D shadow associated with the centre of mass (\ref{subsec:1dshadow}); introduce symbolic dynamics (\ref{subsec:symbolicdyn}); construct the 1D shadow with a Cantor-like iterative procedure (\ref{subsec:1dfractal}); and demonstrate chaotic scattering (\ref{subsec:chaotic}). In Sec.~\ref{sec:2Dshadows} we study non-planar null geodesics (\ref{subsec:nonplanar}) for the binary MP spacetime; highlight the existence of stable bounded null orbits (\ref{subsec:bounded}); and present a gallery of 2D shadows (\ref{subsec:gallery}) and their 1D slices (\ref{subsec:slices}). In Sec.~\ref{sec:extensions} we show how to track geodesics through event horizons (\ref{subsec:through}); and briefly explore shadows in alternative spacetimes (\ref{subsec:alternative}). 

\emph{Conventions:} We use units in which $G=c=1$, and the spacetime signature $+2$. Partial (covariant) derivatives are denoted with commas (semi-colons). Lowercase Greek letters ($\mu, \nu, \ldots$) denote spacetime indices $0\ldots3$, and lowercase Roman letters ($i, j, \ldots$) denote spatial indices $1\ldots3$. The terms `null geodesic' and `light ray' are used interchangeably.

\section{Planar geodesics, 1D shadows and chaotic scattering\label{sec:geodesics}}
In this section we consider the null geodesics of a double black hole MP spacetime \cite{Majumdar:1947eu} in standard coordinates $\{t,x,y,z\}$, with the black holes placed on the $z$ axis and the centre of mass at the origin. The line element is given by (\ref{eq:MP}) and electromagnetic vector potential by $A_\mu = [U^{-1}(\bx), 0, 0, 0]$, where
\beq
U(\bx) = 1 + \frac{M_-}{\sqrt{x^2 + y^2 + (z - z_-)^2}} +  \frac{M_+}{\sqrt{x^2 + y^2 + (z - z_+)^2}} .
\eeq
Here $M_+$ and $M_-$ are the black hole masses and $x=y=0$, $z_\pm = \pm a M_{\mp} / (M_+ + M_-)$ their positions, with $a$ their (coordinate) separation. The `points' at $x=y=0$, $z=z_\pm$ are actually null surfaces of finite area, corresponding to the black hole horizons \cite{Hartle:1972ya}. We shall consider in particular the equal mass case, with $M_+ = M_- \equiv M$ and $z_{\pm} = \pm a / 2$. Henceforth, we will  use units in which $M=1$.

 \subsection{Null geodesics\label{subsec:geodesics}}
A geodesic is a path in spacetime which extremizes the interval between two events. In the standard approach \cite{MTW}, one starts from the action functional $S[q^\mu(\lambda)] = \int L(q^\mu, \dot{q}^\mu) d \lambda$ with the Lagrangian
\beq
L(q^\mu, \dot{q}^\mu) \equiv \frac{1}{2} g_{\mu \nu} \dot{q}^{\mu} \dot{q}^{\nu}, 
\eeq
where $q^\mu(\lambda)$ is a spacetime path, and $\dot{q}^\mu = \frac{dq^\mu}{d\lambda}$ denotes the derivative with respect to an affine parameter $\lambda$. The canonical momentum is $p_\mu \equiv \frac{\partial L}{\partial \dot{q}^\mu} = g_{\mu \nu} \dot{q}^\mu$, and thus the corresponding Hamiltonian is
\beq
H(q^\mu, p_\mu) = \frac{1}{2} g^{\mu \nu} p_{\mu} p_{\nu} \label{eq:hamiltonian}
\eeq
where $g^{\mu \nu}$ is the inverse metric. The geodesics are the solutions to Hamilton's equations: $\dot{q}^\mu = \partial H / \partial p_\mu$ and $\dot{p}_\mu = -\partial H / \partial q^\mu$. Along geodesics, the Lagrangian and Hamiltonian functions are constant, with $L = 0 = H$ in the null case. On the MP spacetime, $H = \frac{1}{2} \left(-U^2 p_t^2 + U^{-2} [p_x^2 + p_y^2 + p_z^2] \right)$ and the null geodesic equations are
\beq
\dot{p}_t = 0, \quad \dot{t} = - U^2 p_t, \quad \dot{p}_x = \frac{\partial (U^2)}{\partial x}  \, p_t^2 , \quad \dot{x} = U^{-2} p_x, \label{eq:hamilton}
\eeq
with similar expressions for the $y$ and $z$ components. Rescaling the constant of motion $p_t$ is equivalent to rescaling the affine parameter $\lambda$; thus, we may set $p_t = -1$ in the following without loss of generality.

Our geometry is symmetric around the axis connecting the two black holes. Thus, we should anticipate a conserved azimuthal angular momentum $p_\phi$. This is seen most easily in cylindrical polar coordinates $\{t,\rho= \sqrt{x^2 + y^2},\phi,z\}$ in which $U$, and thus $H$,  
are independent of $\phi$. After setting $p_t = -1$, we have $\dot{t} = U^{2}$,
\beq
\dot{p}_\rho = \frac{\partial (U^{2})}{\partial \rho} + \frac{p_\phi^2}{\rho^3 U^2}, \quad \dot{\rho} = U^{-2} p_\rho, \quad
\dot{\phi} = \frac{p_\phi}{\rho^2 U^2}, \quad \dot{p}_z = \frac{\partial (U^2)}{\partial z}, \quad \dot{z} = U^{-2} p_z,
\label{eq:nonplanar}
\eeq
and $\dot{p}_t = 0 = \dot{p}_\phi$.

An advantage of using an affine parameter $\lambda$, rather than coordinate time $t$, to parametrize geodesics is that we can maximally extend geodesics through event horizons (after a change of coordinate system; see Sec.~\ref{subsec:through}). A disadvantage is that we may not so easily follow \emph{congruences} of geodesics, as each geodesic in the congruence has its own affine parameter, whereas $t$ is defined globally in the exterior spacetime. In a static spacetime, this is easily remedied. A straightforward approach is to apply the chain rule $\frac{d}{d t} = \frac{1}{\dot{t}} \frac{d}{d \lambda} = U^{-2} \frac{d}{d \lambda}$ to the equations above. Alternatively, we may restrict to a six-dimensional phase space $\{p_i, q^j\}$ (where $i,j=1,2,3$) with independent variable $t$, and introduce the alternative Hamiltonian for null geodesics,
\beq
H' = \left[ \frac{g^{ij} p_i p_j}{-g^{00}} \right]^{1/2} = U^{-2} \left( p_x^2 + p_y^2 + p_z^2 \right)^{1/2} .
\eeq 
We note that, along null geodesics, $H' = -p_t =1$. Now we may use the Legendre transformation to obtain the corresponding Lagrangian, $L' = X (X - 1)$ where $X = U^2 \sqrt{(dx/dt)^2 + (dy/dt)^2 + (dz/dt)^2}$. It is straightforward to show that $X=1$ along null geodesics (as $X=H$), and so the same geodesic equations may be obtained from the Lagrangian $L'' = X$; that is, from the action
\beq
S = \int U^2(x) \sqrt{\left(\frac{dx}{dt}\right)^2 + \left(\frac{dy}{dt}\right)^2 + \left(\frac{dz}{dt}\right)^2} \, dt = \int U^2(x) dl .
\eeq
Thus we have reached Eq.~(\ref{eq:S}), the Maupertuis principle.

One may reach the same conclusion in a more direct fashion by noting that the MP spacetime is conformally related to an ultrastatic spacetime $ds^2 = -dt^2 + U^4 d\bx \cdot d\bx$.

\subsection{Shadows\label{subsec:shadows}}
Here we briefly consider the definition of a `black hole shadow'. Bohn \emph{et al.} \cite{Bohn:2014xxa} employ a ray-casting approach, and conceive that ``a shadow is a region of the image where geodesics are traced backwards in time from the camera to a black hole''. The camera device provides a natural isomorphism between a point on a 2D image, and a null geodesic. A `pixel' on the image is part of the black hole shadow if and only if the corresponding null geodesic asymptotically approaches a black hole horizon. In other language, the shadow is the basin of attraction \cite{Dettmann:1994dj, Dettmann:1995ex} for the black holes in the initial data.

With a slight change of emphasis, we may draw our shadow on a two-sphere instead. Let $E$ be a spacetime event (e.g.~the opening of a pinhole camera). Around $E$, construct a Riemann normal coordinate system \cite{Poisson:2004relativist} with $E$ at the origin. Now consider the two-surface formed by the intersection of the past light cone of $E$ and the hypersurface $t = -\eps$. In the limit $\eps \rightarrow 0$, this is a two-sphere of radius $\eps$ (N.B.~spacetime is locally flat, and $g_{\mu \nu} = \eta_{\mu \nu} + O(x^2)$ in this coordinate system). A point on the sphere is associated with a null geodesic that passes through the point and is normal to the two-sphere, and outward-pointing. The point on the sphere is in the shadow if and only if the associated null geodesic asymptotically approaches a black hole horizon.

Of course, a shadow may be defined with respect to some other initial data surface. For example, one might consider rays normal to a collimator. We could define a shadow on any $n$-dimensional hypersurface if $n < d$ (where $d$ is the number of spatial dimensions), by associating points on the hypersurface with null geodesics that are normal to it (and passing through it the same sense).

Two more general possibilities are not considered here. First, one may wish to define a shadow with respect to some null congruence which is \emph{not} hypersurface-orthogonal. Second, in the cases above, the initial data surface has an intrinsic geometry; however, it may also be possible to consider initial data on a manifold without a metric structure.

\subsection{A one-dimensional shadow\label{subsec:1dshadow}}

Let us now examine an example of a 1D shadow, in a highly symmetric scenario. We shall consider light rays starting at the centre of mass between two fixed black holes of equal mass $M_\pm =1$, separated by a coordinate distance $a$, and confined to the $(x,z)$-plane. As initial data, we use
\beq
x=y=z=0, \quad p_y=0, \quad p_x = U_0^2 \cos \alpha, \quad p_z = U_0^2 \sin \alpha, \quad U_0 = 1 + \frac{4M}{a}, \label{eq:ic}
\eeq
with $p_t = -1$, where $\alpha$ is the initial angle. In other words, we seek the black hole shadow defined for rays normal to an (infinitessimal) ring surrounding the centre of mass.

\begin{figure}[h]
 \centering
 \begin{tabular}{c}
 \includegraphics[width=8cm]{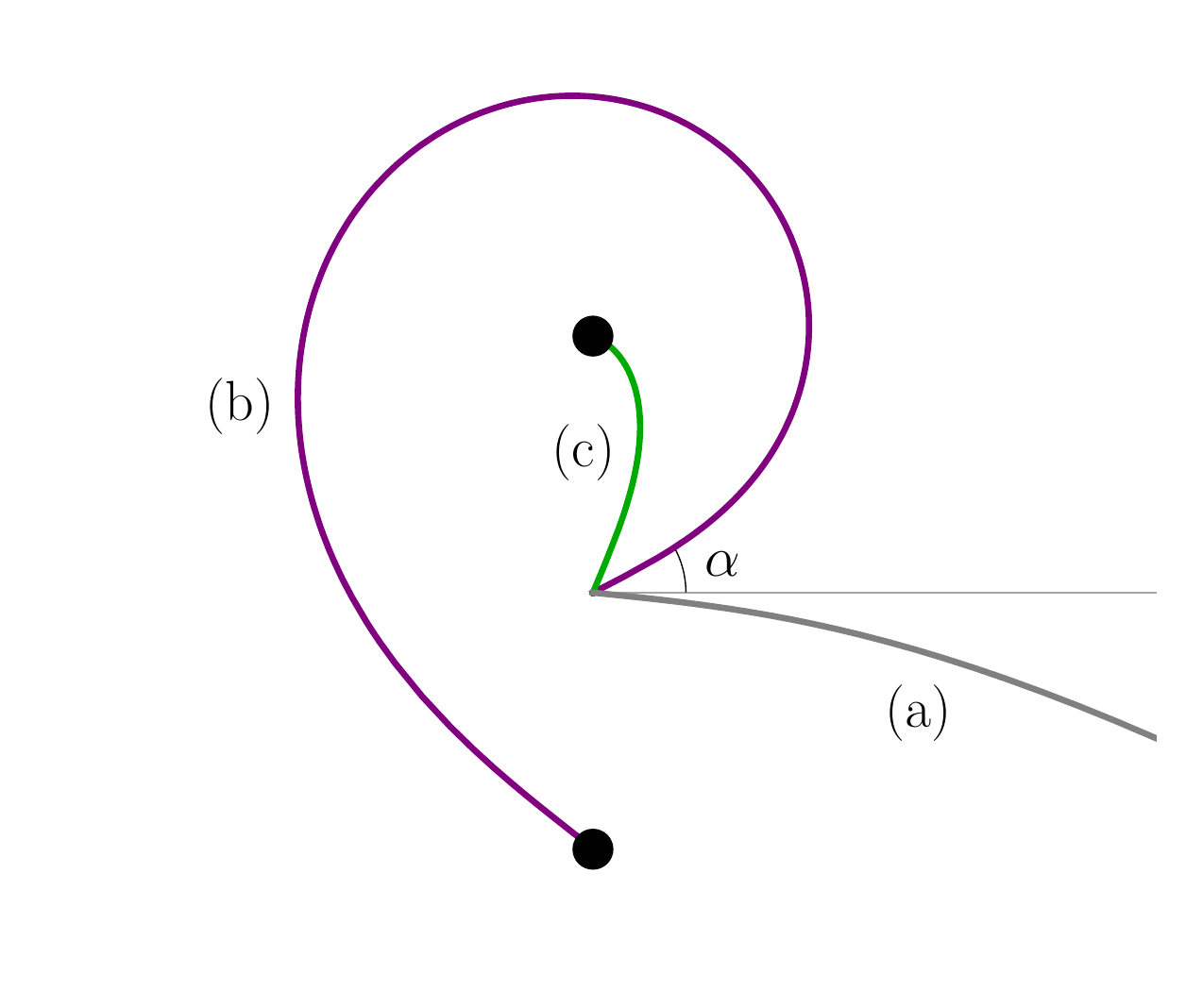}
 \end{tabular}
 \begin{tabular}{c}
 \includegraphics[width=8cm]{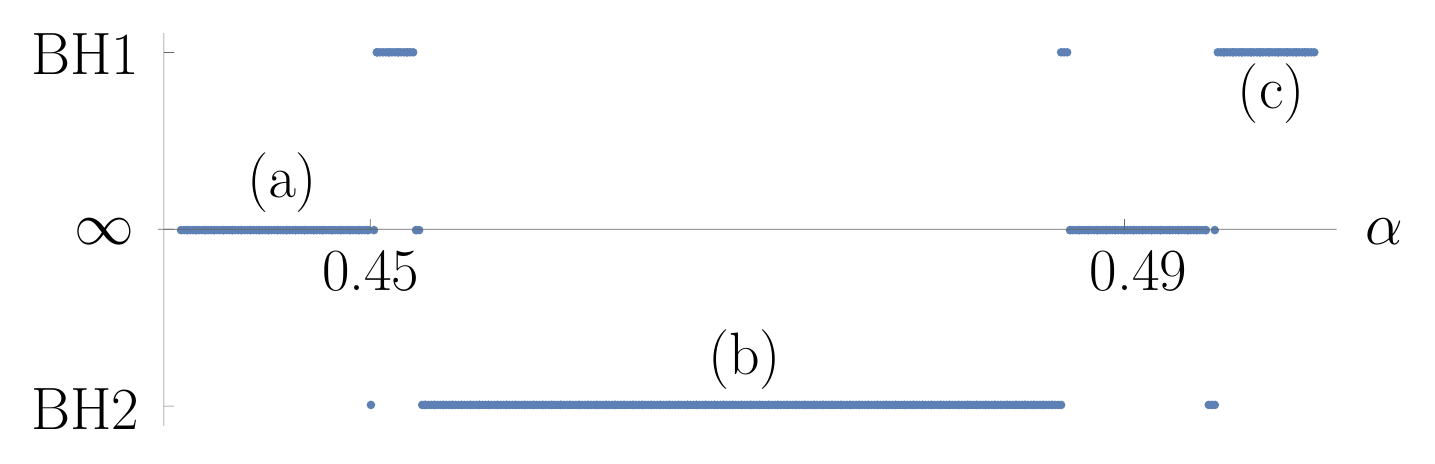}
 \end{tabular}
 \caption{Examples of null rays emanating from the centre of mass of two equal-mass extremally charged black holes. Here (a), (b) and (c) denote rays escaping to infinity, falling into the lower BH, and into the upper BH, respectively. The right plot shows that there are further possibilities between (a)/(b) and (b)/(c). For example, between (b) and (c) are rays that escape to infinity by passing between the BHs.}
 \label{fig:trajectorytypes}
\end{figure}

As shown in Fig. \ref{fig:trajectorytypes}, the fate of the light ray depends on the initial angle $\alpha$. If $\alpha \sim 0$ (or $\alpha \sim \pi$), the ray will escape to infinity; if $\alpha \sim \pi/2$, the ray will plunge directly into the upper black hole. With some intermediate value of $\alpha$, a ray may orbit around the upper hole and then plunge into the lower hole; or it may pass between the holes and escape to infinity.

We used Mathematica's function {\tt NDSolve} to obtain numerical solutions of Eq.~(\ref{eq:hamilton}) with initial data (\ref{eq:ic}), taking  $M=1$, $a=2$ as default values.

Figure \ref{fig:trajectorytypes} shows the fate of the ray as a function of $\alpha$. In pioneering work \cite{Contopoulos:1990}, Contopoulos labelled rays as Type I \& II (falling into the upper and lower BHs) and Type III (escaping to infinity). Here, we use the three similar labels $+1$ (into upper black hole), $-1$ (into lower black hole) and $0$ (escape to infinity).

\subsection{Symbolic dynamics\label{subsec:symbolicdyn}}

One way of demonstrating that a system is chaotic is through the use of symbolic dynamics, which describes the topology of trajectories in phase space \cite{Cornish:1996de}. Symbolic dynamics provides a coordinate-invariant method of characterizing chaos, which is particularly important when considering scattering problems in general relativity \cite{Dettmann:1994dj, Dettmann:1995ex}. Furthermore, a symbolic code allows us to study the dynamics \emph{analytically}, despite the fact that the geodesic equations themselves are non-integrable. In the binary black hole system, we are particularly interested in the role played by the (unstable) perpetual orbits, which correspond to infinitely long symbolic sequences. As we shall see, the initial conditions asymptoting to perpetual orbits form a Cantor-like set.

We develop the symbolic code by considering a null geodesic in a congruence which has reached a `decision', as shown in Fig.~\ref{fig:decision}. Faced with a decision, the geodesic may follow a path around the other black hole in the same sense $(0)$; around the other black hole in the opposite sense $(2)$; or around the same black hole in the same sense $(4)$. In addition, the particle could fall into one of the black holes $(1)$, or escape to infinity $(3)$. 

(N.B.~To avoid double-counting, the possibility of plunging into the `other' black hole is not enumerated  (cf.~Fig.~\ref{fig:decision}). Instead, this possibility is accounted for at a previous or subsequent decision point. Rays that plunge \emph{directly} into a black hole, or escape directly, can either be excluded from consideration (as they do not generate interesting structure), or assigned an empty decision dynamics sequence and some auxiliary label.)

\begin{figure}[h]
 \includegraphics[width=3.8cm]{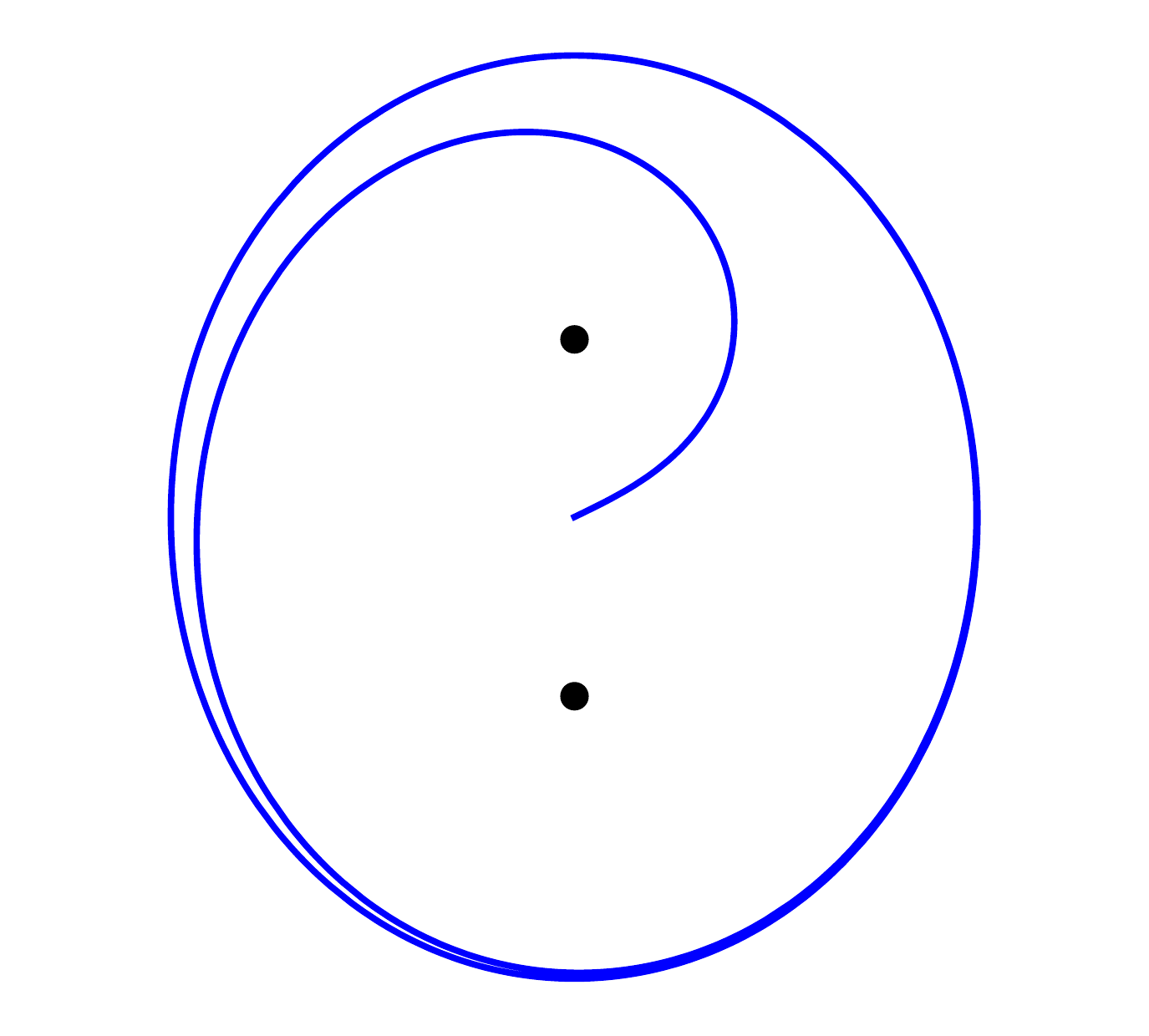}
 \includegraphics[width=3.8cm]{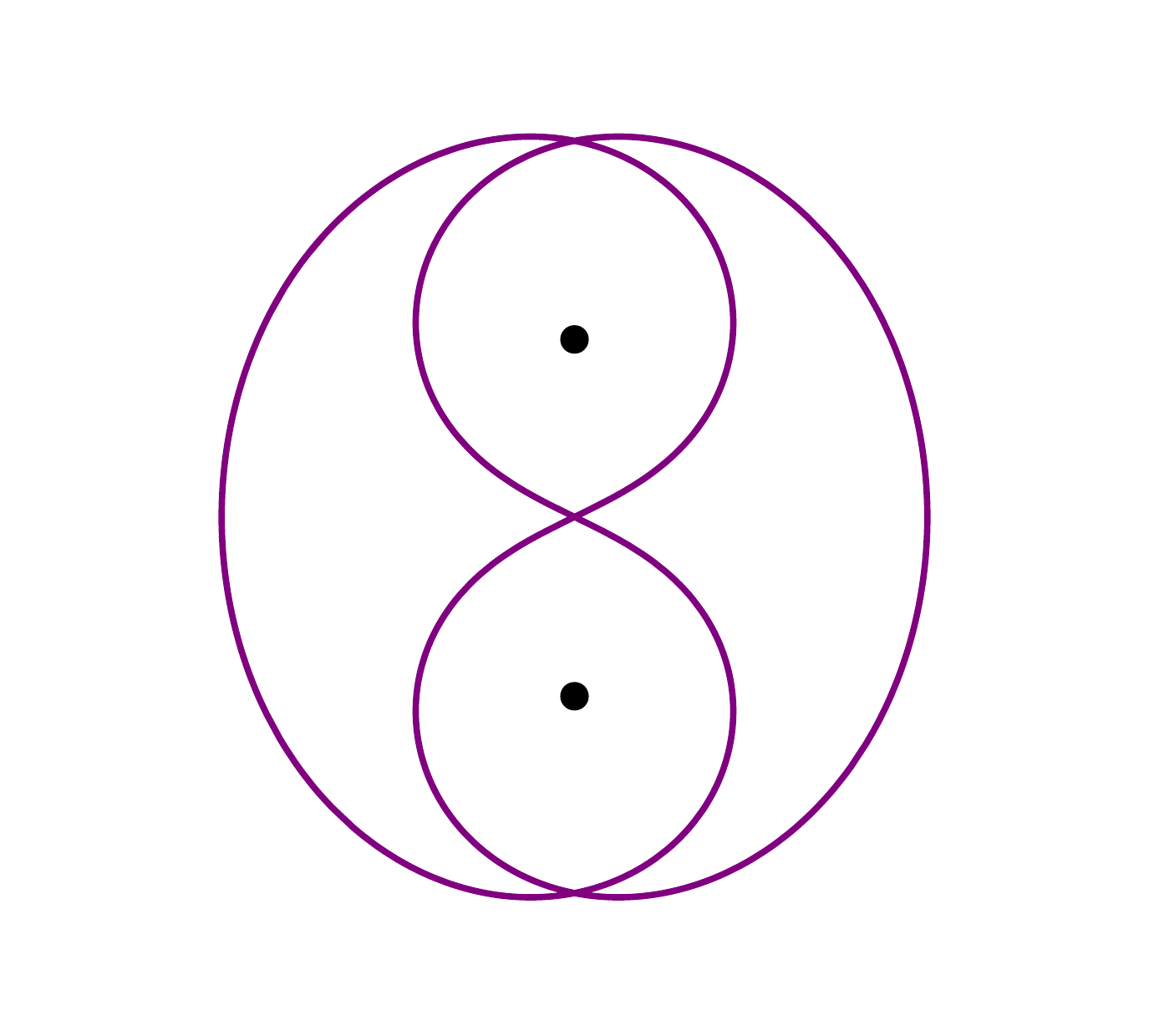}
 \includegraphics[width=3.8cm]{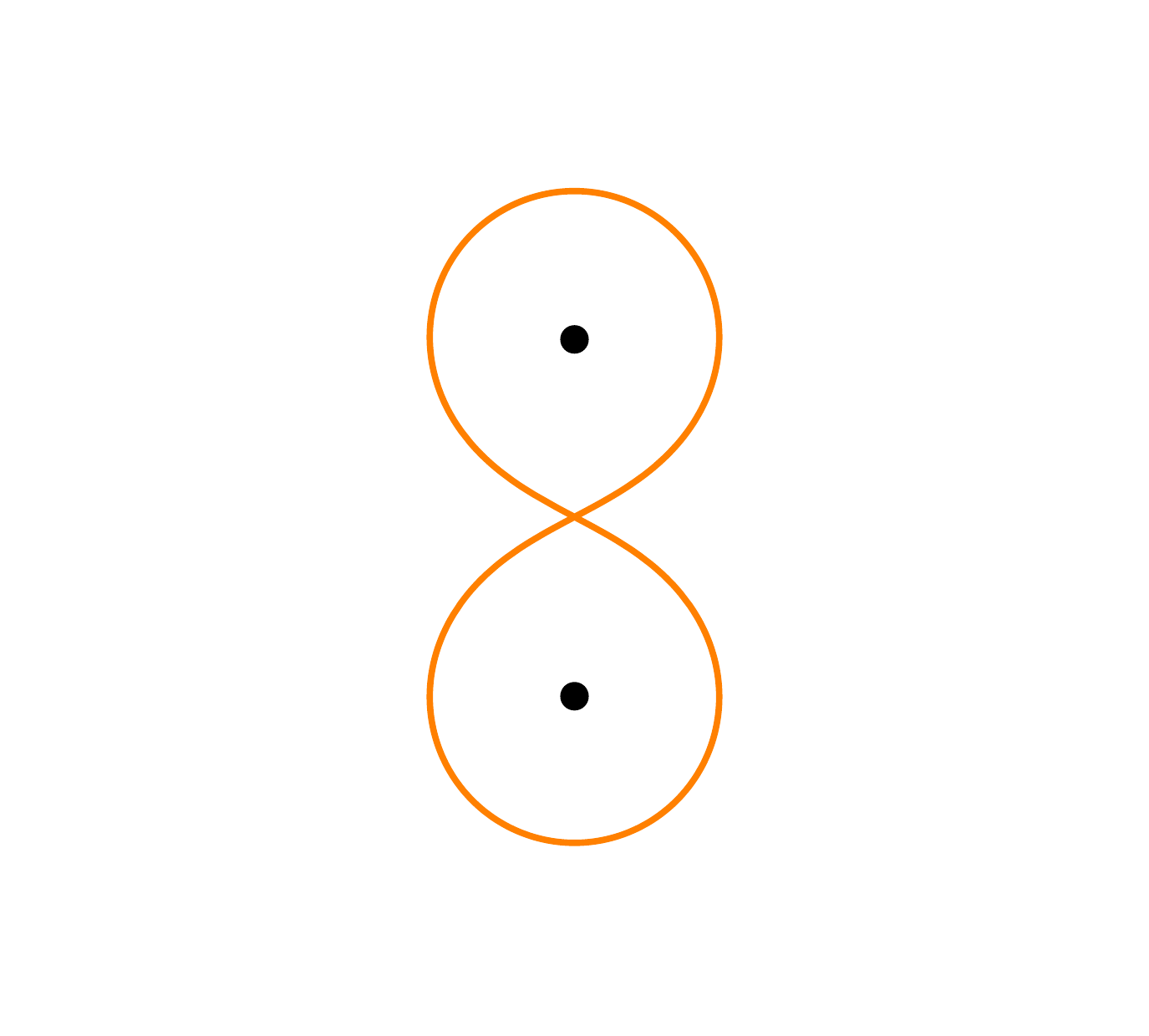}
 \includegraphics[width=3.8cm]{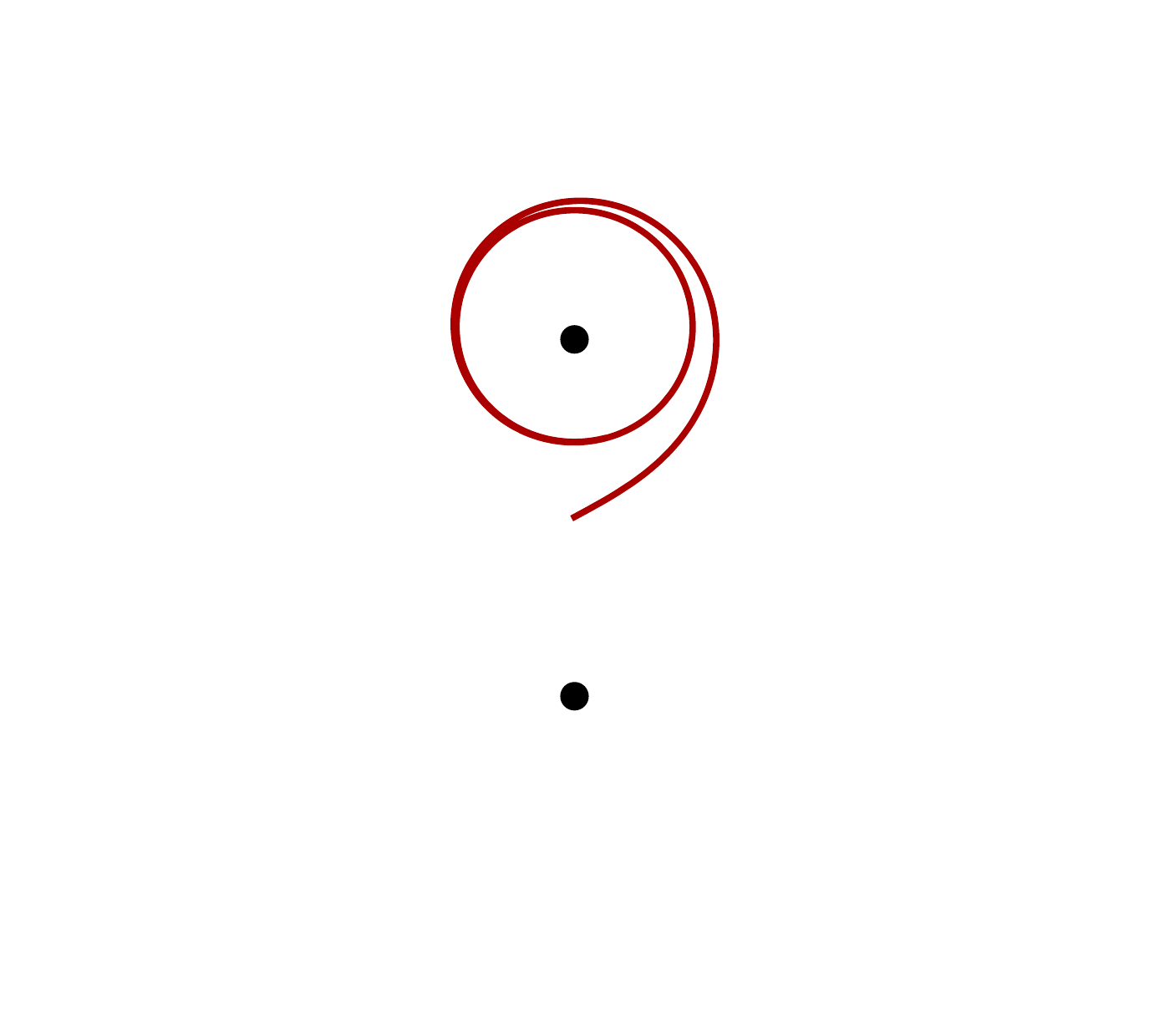} \\
 \includegraphics[width=8cm]{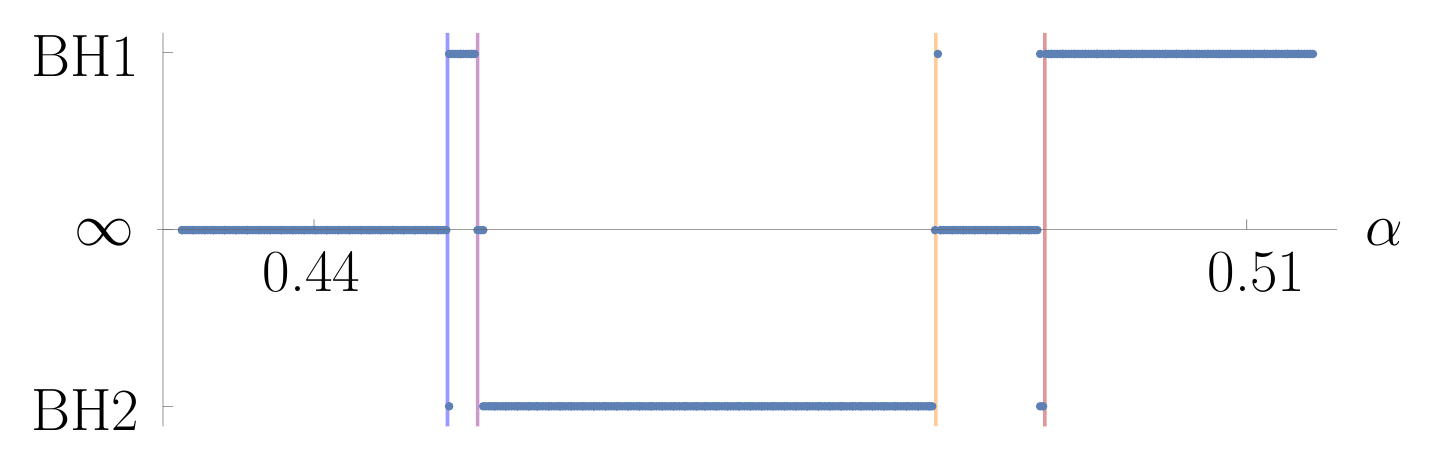}
 \caption{\emph{Above:} Examples of asymptotically periodic null orbits that start from the centre of mass. In our symbolic dynamics, these orbits are given by the recurring sequences $000\cdots$, $0202\cdots$, $222\cdots$, $444\cdots$ (left-to-right). \emph{Below:} The vertical lines indicate the critical values of the angle $\alpha$ corresponding to these asymptotically periodic orbits. Due to the Cantor-like distribution of perpetual orbits on the initial data, any open interval in $\alpha$ contains either zero or an infinite number of perpetual orbits. \label{fig:periodicorbits}}
\end{figure}

One can therefore describe planar null rays in the binary MP spacetime using a base-5 system. Moreover, it is clear that the \emph{perpetual} orbits can be described by an infinite sequence of digits from the symbolic alphabet which do not contain the digits $1$ or $3$, since they linger in the strong-field region without being absorbed by the black holes or escaping to infinity (by contrast, the rays which are absorbed or scattered are finite-length sequences which terminate in $1$ or $3$). For example, the sequence $000\cdots$ corresponds to a geodesic that orbits both black holes, whilst the sequence $222\cdots$ corresponds to the figure-of-eight orbit. Additional examples of perpetual orbits in the double black hole spacetime are presented in Fig.~\ref{fig:periodicorbits}. The \emph{periodic} orbits -- which form a subset of the perpetual orbits -- are described in the decision dynamics by recurring sequences; these sequences correspond to \emph{rational} numbers. However, a generic perpetual orbit need not be periodic. Thus, the representation of such orbits will not be a recurring sequence; non-periodic perpetual orbits correspond to \emph{irrational} numbers. The use of a base-5 symbolic alphabet in the decision scheme provides a natural map of perpetual orbits on to the (uncountably infinite) $5$-adic Cantor set.

(In an alternative approach, presented by Cornish and Gibbons in Ref.~\cite{Cornish:1996de}, a geodesic is described by a sequence of digits recording its passage through three `windows' on the symmetry axis. The three digits $-1$, $0$ and $1$ correspond to the open intervals $z \in (-\infty, z_-)$, $(z_-,z_+)$, and $(z_+, +\infty)$, respectively. Recalling the three-disc example \cite{Eckhardt:1988}, we classify this as a `collision dynamics', as repeated neighbouring digits are prohibited (see Fig.~\ref{fig:threedisc}). It is explored further in Appendix \ref{sec:translation}. We prefer to use `decision dynamics' as it provides insight into the ordering of perpetual orbits in initial data, as we shall describe.)


\subsection{Fractal structure of the one-dimensional shadow\label{subsec:1dfractal}}

\subsubsection{Ordering of perpetual orbits in the initial data\label{subsubsec:ordering}}

\begin{figure}[h]
 \subfigure[]{\includegraphics[width=8cm]{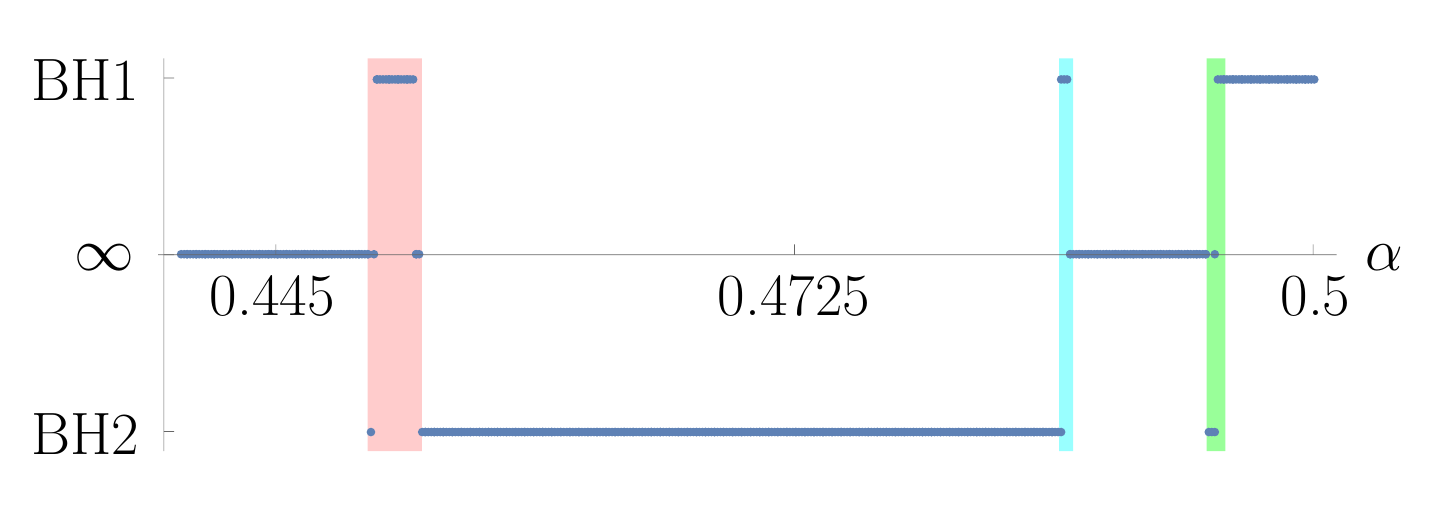}\label{fig:fractalplotregions}}
  \vspace{0.1cm}
\begin{tabular}{ccc}
 \subfigure[]{\includegraphics[width=5.5cm]{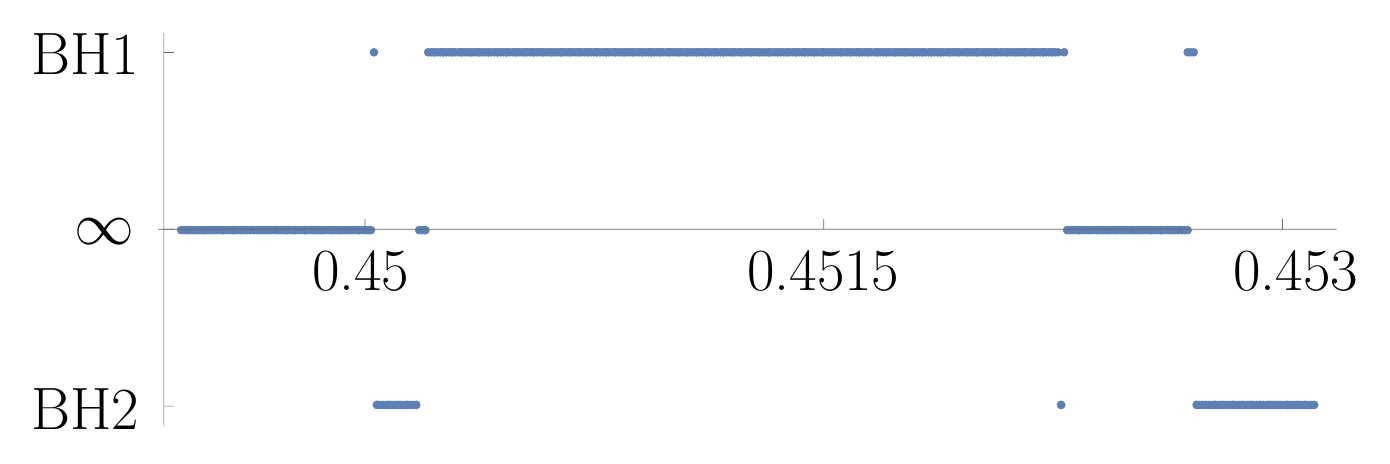}\label{fig:fractalplotlz1}} &
 \subfigure[]{\includegraphics[width=5.5cm]{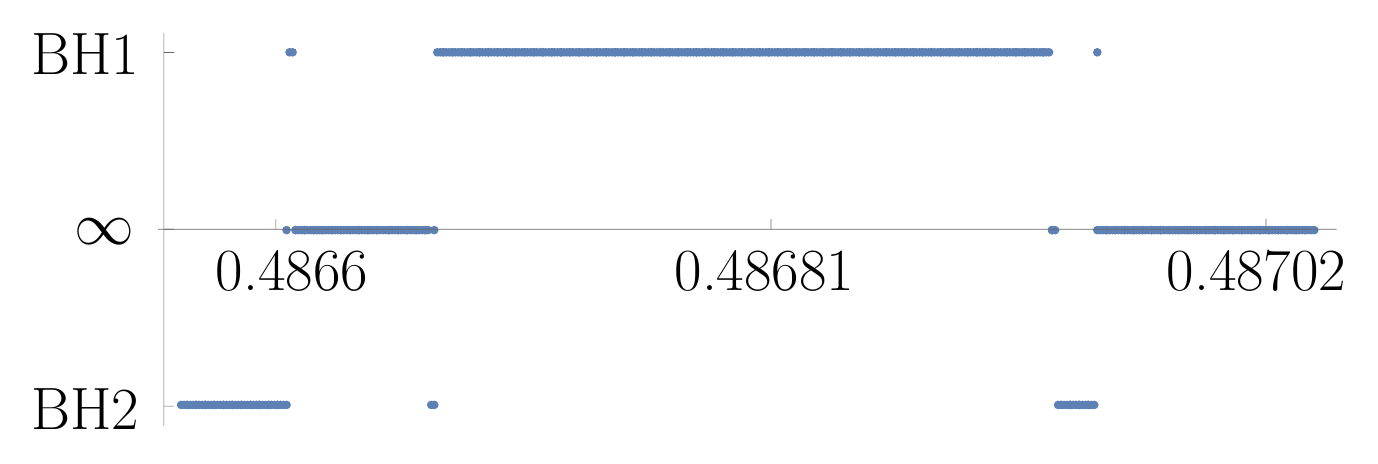}\label{fig:fractalplotmz1}} &
 \subfigure[]{\includegraphics[width=5.5cm]{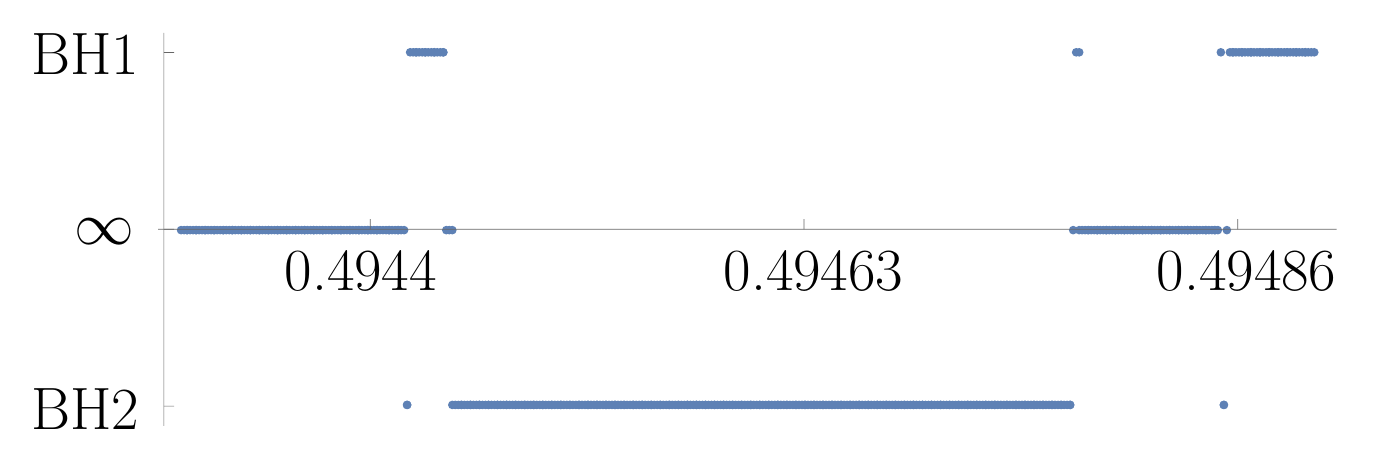}\label{fig:fractalplotrz1}} \\
 \vspace{0.1cm}
 \subfigure[]{\includegraphics[width=5.5cm]{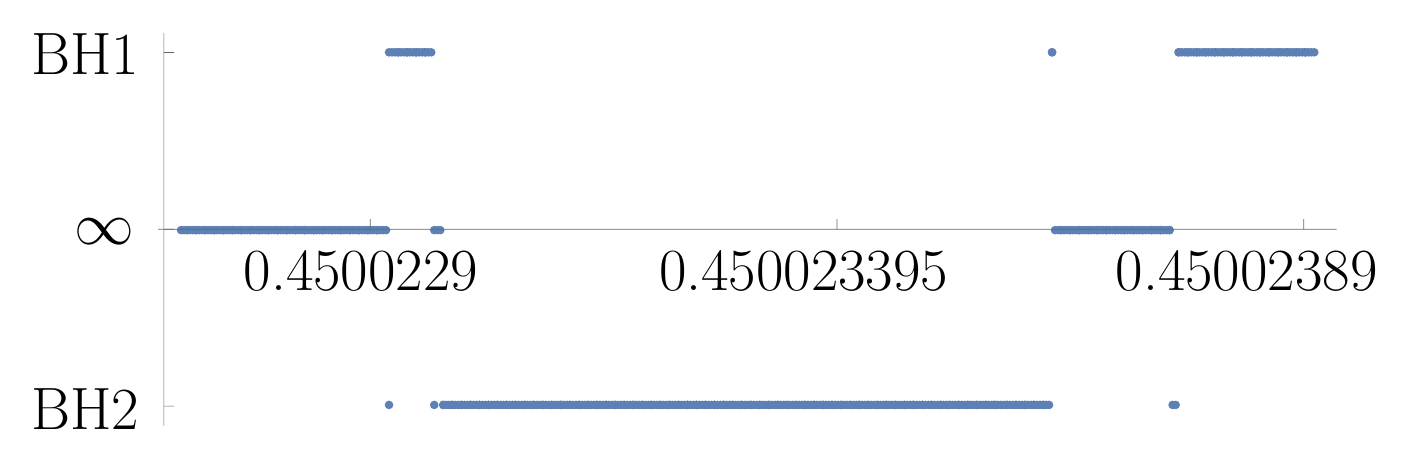}\label{fig:fractalplotlz2}} &
 \subfigure[]{\includegraphics[width=5.5cm]{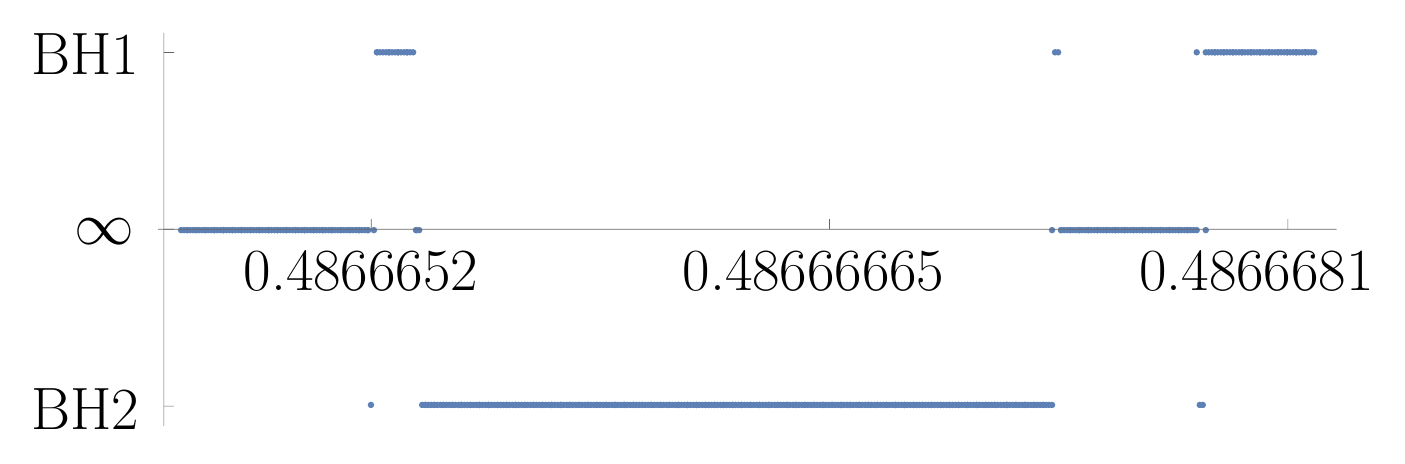}\label{fig:fractalplotmz2}} &
 \subfigure[]{\includegraphics[width=5.5cm]{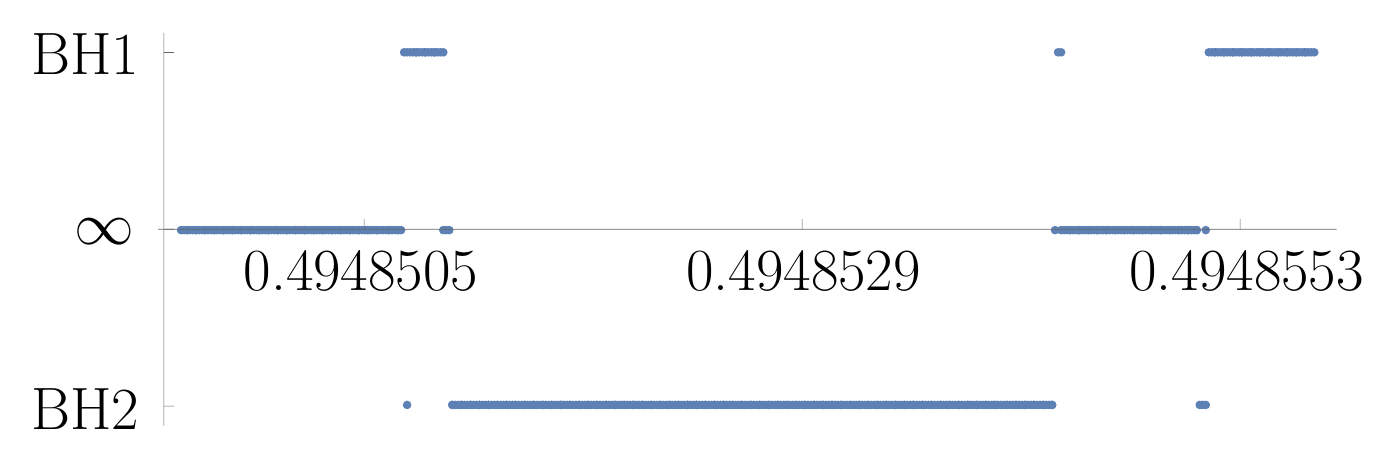}\label{fig:fractalplotrz2}} \\
  \vspace{0.1cm}
 \subfigure[]{\includegraphics[width=5.5cm]{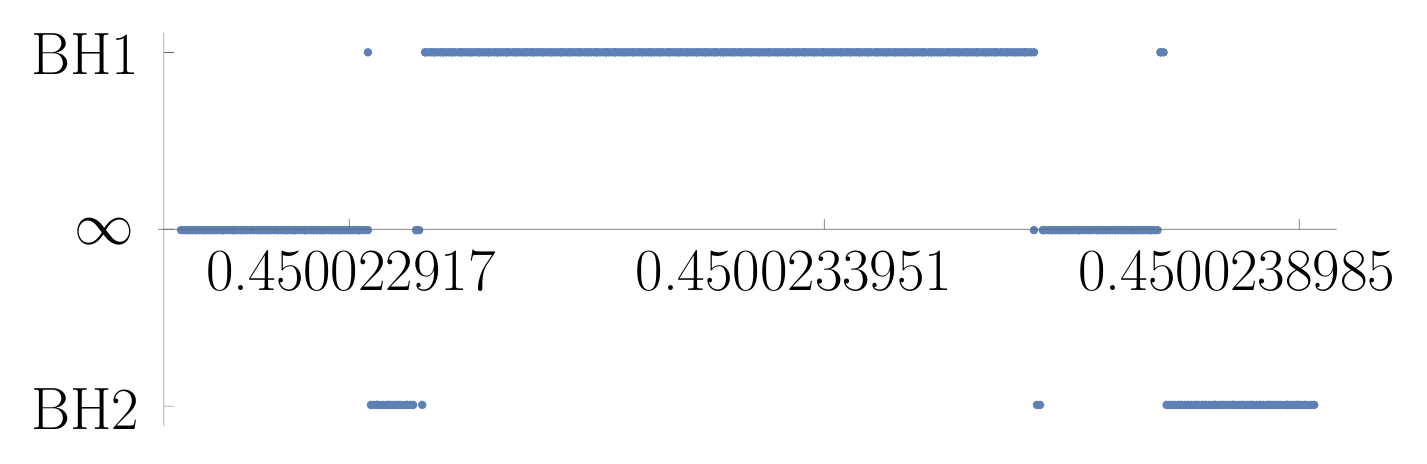}\label{fig:fractalplotlz3}} &
 \subfigure[]{\includegraphics[width=5.5cm]{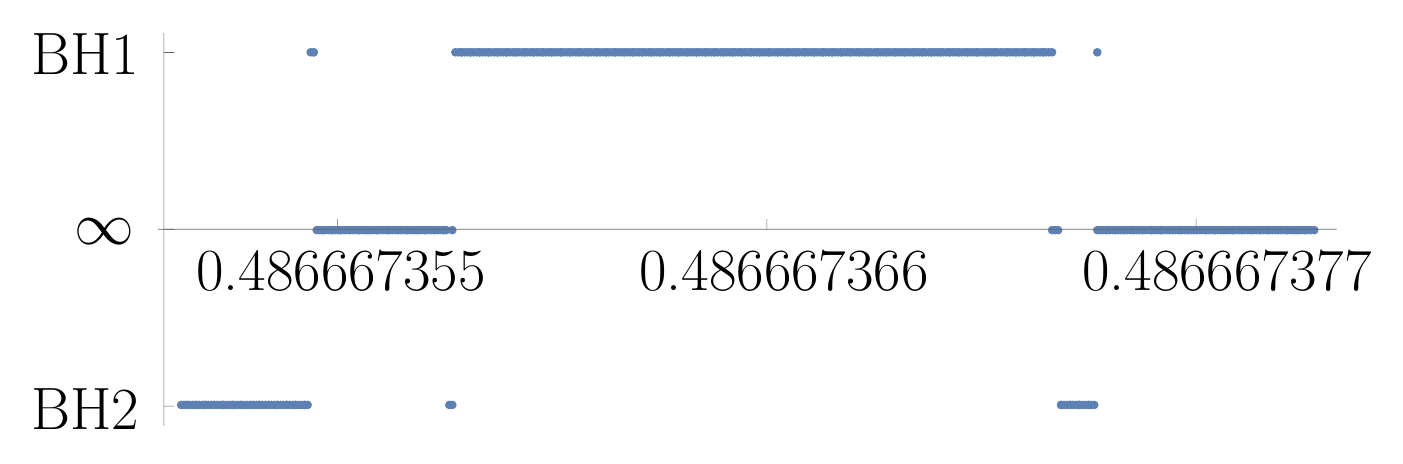}\label{fig:fractalplotmz3}} &
 \subfigure[]{\includegraphics[width=5.5cm]{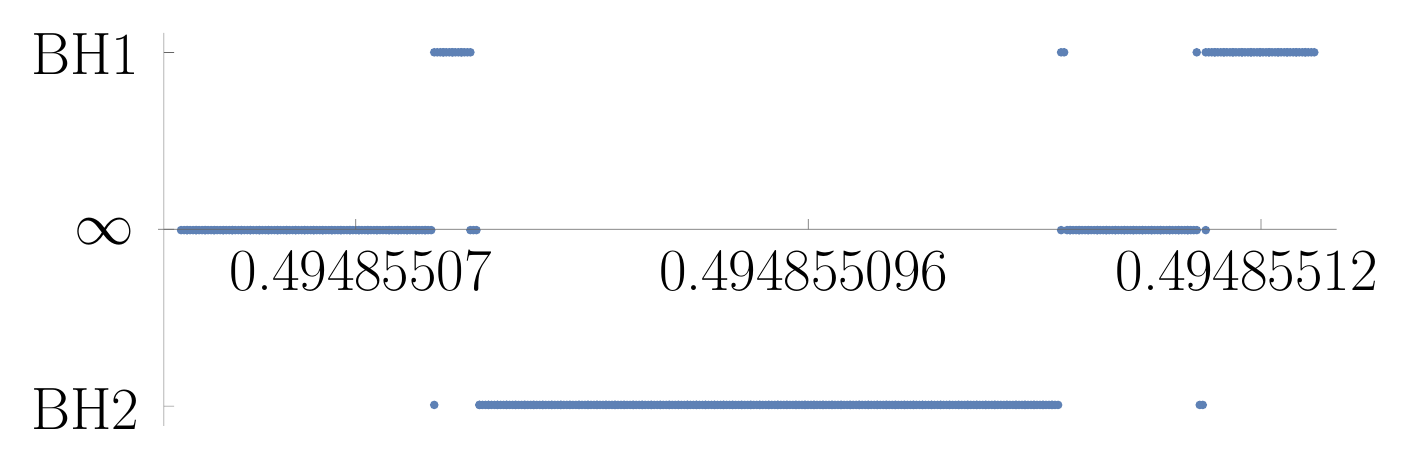}\label{fig:fractalplotrz3}} \\
\end{tabular}
\caption{Zooming in on the 1D shadow. These plots show the fate of a null geodesic starting at the centre of mass with initial angle $\alpha$ (cf.~Fig.~\ref{fig:trajectorytypes}) and $M_\pm = 1$, $a = 2$. With the exception of a measure-zero set (the perpetual orbits), all rays fall in to the upper black hole (BH1), the lower black hole (BH2), or escape to infinity ($\infty$). Between the pink and blue regions, and blue and green regions, are open intervals in which all rays share the same fate. Within the pink, blue and green regions, corresponding to `decisions' 0, 2, and 4 in Fig.~\ref{fig:decision}, there are an infinite number of perpetual orbits. In plots (b), (e) and (h) we repeatedly `zoom in' on the left (pink, 0) region; in plots (c), (f) and (i) on the middle (blue, 2) region; and in (d), (g) and (h) the right (green, 4) interval. Self-similar structure is apparent. A subtlety is that (d), (g) and (j) are similar to (a); whereas (b), (e) and (h) are repeatedly reflected in the $x$-axis; and (c), (f) and (i) are also repeatedly flipped in the horizontal sense. These features are explained in the text. }
\label{fig:cantorzoom}
\end{figure}

Let us now apply symbolic dynamics to understand the ordering and organisation of the perpetual orbits, (i.e.~scattering singularities) in an initial data set. Let us begin by examining the 1D shadow shown in Fig.~\ref{fig:fractalplotregions}, in domain $\alpha \in [0,\pi/2]$, in which we have highlighted three intervals of interest in pink (left), light blue (middle), and green (right). Outside of these regions there are no perpetual orbits, but within each of these three intervals are embedded infinitely many perpetual orbits.

To the left of the pink interval (in initial angle $\alpha$), all trajectories escape to infinity [Fig.~\ref{fig:trajectorytypes}(a)]. Between the pink and light blue intervals, all trajectories fall into the lower black holes [Fig.~\ref{fig:trajectorytypes}(b)], and between the light blue and green intervals, all trajectories again escape to infinity. Finally, to the right of the green interval, all of the trajectories pass into the upper black hole [Fig.~\ref{fig:trajectorytypes}(c)].

Now consider what happens when we increase $\alpha$, starting from the equator $\alpha = 0$. All trajectories escape to infinity, until we reach some critical value of $\alpha$ which corresponds to the \emph{first} perpetual orbit. This orbit is depicted on the left of Fig.~\ref{fig:periodicorbits}, and has symbolic representation $000\cdots$ in our decision dynamics. Similarly, let us consider what happens when we decrease $\alpha$, starting from the pole $\alpha = \pi/2$. As discussed in Sec. \ref{subsec:1dshadow}, all of the trajectories fall into the upper black hole, until we reach a critical value of $\alpha$ corresponding to the \emph{final} perpetual orbit. This orbit is depicted on the right of Fig.~\ref{fig:periodicorbits}, and it corresponds to the sequence $444\cdots$ in our symbolic code.

One might infer from the above examples that the ordering in the initial data exactly matches the ordering of the sequences in our decision dynamics, but this is not quite the case. To be precise, we now introduce a function $F$ that maps a `decision dynamics' sequence $X = X_1 X_2 X_3 \cdots$ (where $X_i$ are single digits in base-$5$, cf.~Fig.~\ref{fig:decision}) onto a real number in the unit interval. We demand that the ordering of the image of $F$ in the unit interval corresponds to the ordering of rays in the initial data $\alpha$, so that $F(X^{(1)}) < F(X^{(2)})$ if and only if $\alpha^{(1)} < \alpha^{(2)}$, where $X^{(i)}$ and $\alpha^{(i)}$ are the sequences and initial data for any null rays.

We take $F(X) = f( \widetilde{X} )$. Here, $f$ simply maps the digits of a sequence on to the corresponding digits of a real number base 5, in a straightforward way: $f(Y_1Y_2Y_3 \ldots) = 0.Y_1Y_2Y_3\cdots$. The tilde denotes a parity-reordering operation, in which each digit $X_i$ in $X$ is mapped to a digit $\widetilde{X}_i$ in $\widetilde{X}$ according to a procedure which keeps track of `parity' $P$. Working from left-to-right in the sequence, starting with $P=+1$, we examine each digit $X_i$ in turn, and (i) set
\beq
\widetilde{X}_i = \begin{cases} X_i , & P = +1, \\ 4 - X_i , & P = -1 ; \end{cases}
\eeq
(ii) reverse the parity if $X_i = 2$ ($P \rightarrow -P$); (iii) iterate, $i \rightarrow i+1$. We note that $\overset{\approx}{X} = X$, so the operation is its own inverse (i.e.~it is an involution). The geometrical explanation here is that decision $2$ describes a geodesic passing between the black holes (cf.~Fig.~\ref{fig:decision}), and this reverses the sense of the orbit, from clockwise to counter-clockwise and {\it vice versa}.

For example, consider the sequences $A = 20202020\cdots$ and $B = 24242424\cdots$. Applying the parity-reordering, $\widetilde{A} = 24202420\cdots$ and $\widetilde{B} = 20242024\cdots$. As $0.20242024\cdots < 0.24202420\cdots$ (i.e.~$F(B) < F(A)$) it is clear that the perpetual orbit represented by decision sequence $B$ will precede that represented by decision sequence $A$ in the initial data $\alpha$.


\subsubsection{Constructing a Cantor-like set on initial data\label{subsec:cantor}}

It is well-known that the standard Cantor set may be constructed by an iterative procedure. Starting with the (closed) unit interval $\left[0, 1\right]$, one removes the open middle third $(\frac{1}{3},\frac{2}{3})$, leaving two closed intervals $\left[0, \frac{1}{3}\right]$ and $\left[\frac{2}{3}, 1\right]$. Next, one removes the open middle third of the remaining closed intervals, continuing in this fashion \emph{ad infinitum} until left with a set of points which were not removed from the unit interval at any step. These points make up the Cantor set: a one-dimensional self-similar fractal. One may construct other types of Cantor sets by removing more than one interval; and/or by changing the proportionate width of the interval(s) removed.

\begin{figure}
 \includegraphics[width=0.6\textwidth]{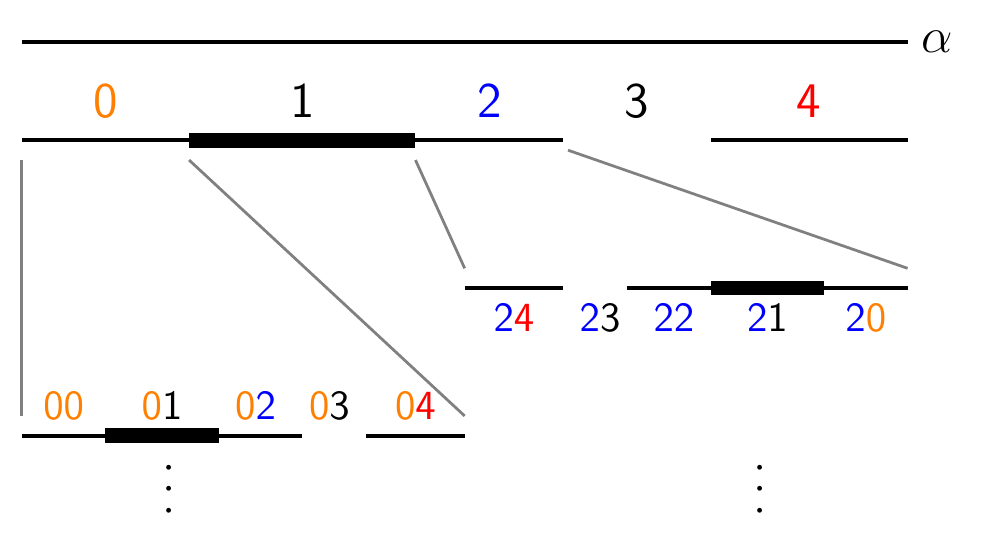}
 \caption{An iterative procedure for the construction of the shadow. First, from a region of initial data $\alpha$, we remove two open intervals, corresponding to decision 1 (ray capture by a BH) and decision 3 (ray escape to infinity). The former interval [thick black] lies in the shadow, and the latter does not. Next, we iterate on the remaining closed intervals $0$, $2$ and $4$, removing two open intervals from each (note parity-reversal for case $2$). Repeating this procedure \emph{ad infinitum} leads to (i) two infinite sets of disjoint open intervals, corresponding to shadow and non-shadow regions, respectively, and (ii) an uncountably-infinite number of distinct points of measure zero, corresponding to perpetual orbits. The latter set is isomorphic to the $5$-adic Cantor set. \label{fig:cantorconstruction}}
\end{figure}

With symbolic dynamics as our guide, we may develop a similar iterative procedure to construct the 1D shadow on the initial data set $\alpha$. Let $\alpha_{\widetilde{X}}$ denote an initial value corresponding to a perpetual orbit, with $X$ an infinite sequence of `decisions'  and $\widetilde{X}$ the parity-reordered sequence which determines its order in the initial data (see Sec.~\ref{subsubsec:ordering}). We begin by focussing on the `interesting' interval $C = [\alpha_{\dot{0}}, \alpha_{\dot{4}}]$, where an overdot denotes infinite recurrence (i.e.~$\alpha_{\dot{0}} = \alpha_{000\cdots}$). From this interval we may remove two open intervals, $O_1 = (\alpha_{0\dot{4}}, \alpha_{2\dot{0}})$ and $O_3 = (\alpha_{2\dot{4}}, \alpha_{4\dot{0}})$, corresponding to geodesics that immediately fall into the (lower) black hole or escape to infinity, respectively. $O_1$ forms part of the shadow. Now we may iterate this procedure on each closed interval that remains. Iterating is equivalent to following the geodesics that linger in the vicinity of the black holes until they reach the next decision point. A schematic diagram which demonstrates this iterative process is presented in Fig.~\ref{fig:cantorconstruction}.

Suppose we have iterated $k$ times, and are now considering the closed interval $C_{\widetilde{X}} \equiv [\alpha_{\widetilde{X}\dot{0}}, \alpha_{\widetilde{X}\dot{4}}]$, where $\widetilde{X}$ is a (parity-reordered) sequence with $k$ digits. To iterate we next remove the open intervals $O_{\widetilde{X}1}$ and $O_{\widetilde{X}3}$, where $O_{\widetilde{X}1} \equiv (\alpha_{\widetilde{X}0\dot{4}},\alpha_{\widetilde{X}2\dot{0}})$ and $O_{\widetilde{X}3} \equiv (\alpha_{\widetilde{X}2\dot{4}}, \alpha_{\widetilde{X}4\dot{0}})$, We add $O_{\widetilde{X}1}$ to the shadow, and continue.

We note that $O_{\widetilde{X}1}$ and $O_{\widetilde{X}3}$ are the open intervals of initial data corresponding to null geodesics that go through a decision sequence $X$ before falling into the black hole, or escaping to infinity, respectively. In the former case, to determine which black hole is selected, we count the number of digits $n$ in the sequence $\widetilde{X}$ that are not equal to $4$; if $n$ is even (odd), the geodesic falls into the lower (upper) black hole.

After $k$ iterations, we have partitioned the initial data into $3^k - 1$ open intervals $\{\{O_1, O_3\}, \{O_{01}, O_{03}, O_{21}, O_{23}, O_{41}, O_{43}\}, \cdots \}$, corresponding to geodesics that make up to $k-1$ decisions before falling into the black hole or escaping, and $3^k$ closed intervals corresponding to geodesics that linger long enough around the black holes to make $k$ decisions, and yet whose ultimate fate has not been determined at this level of precision. As $k \rightarrow \infty$, one is left with an infinite number of closed sets of zero measure -- a structure isomorphic to the $5$-adic Cantor set -- and an infinite number of open sets which are isomorphic to its complement. The black hole shadow is the union of the open sets $ O_{\widetilde{X}1}$, where $X$ is \emph{any} sequence without the digits $1$ or $3$.

In principle, at each stage the ratio of the widths of the two open intervals removed to the closed interval in which they are embedded will depend on the entire prior history of the geodesic motion; and thus, on all the preceding digits in its symbolic representation. In practice, the ratios will depend principally on the most recent decision taken (i.e.~the previous digit in the sequence), and the effect of dependence on previous decisions (earlier digits) is exponentially suppressed. Thus, we expect the structure that arises to be clearly self-similar, as we confirm in the next section.

The MP di-hole system is therefore an example of a chaotic scattering system, due to the existence of this Cantor-like set on the initial data. Recalling Eckhardt's definition \cite{Eckhardt:1988} stated in Sec.~\ref{sec:introduction}, scattering in a Hamiltonian system is chaotic if there exists an infinity of distinct scattering singularities of measure zero, on some manifold of initial data, which are typically arranged into a fractal set (here shown to be a generalized $5$-adic Cantor set). These scattering singularities are discussed further in Sec.~\ref{subsec:chaotic} (see Fig.~\ref{fig:timedelay}).

\subsubsection{Demonstrating self-similarity in the one-dimensional shadow\label{subsec:selfsimilar}}

Having developed a symbolic code based on decision dynamics, we now use it to understand the fractal properties of the 1D shadow. It is clear from the 1D shadow plot presented in Fig.~\ref{fig:fractalplotregions} that there are three open intervals in which we see rich fractal structure. By zooming in on these intervals successively, we will confirm that the shadow is self-similar.

Let us first examine the leftmost interval of the 1D shadow plot, which is highlighted in pink, to obtain Fig.~\ref{fig:fractalplotlz1}. Zooming in on this region once is equivalent to making decision $0$. We see that this image is qualitatively very similar to Fig.~\ref{fig:fractalplotregions}. In fact, the image is identical except for the trajectories which fall into the upper and lower black holes are exchanged. Zooming in on the left-hand interval again -- i.e. making decision $0$ once more -- we obtain Fig.~\ref{fig:fractalplotlz2}, which is similar to Fig.~\ref{fig:fractalplotregions}. Finally, at the next level of zoom, presented in Fig.~\ref{fig:fractalplotlz3}, we see that again upper and lower black holes are exchanged; we obtain an image which is identical to Fig.~\ref{fig:fractalplotlz1}.

Similarly, we may zoom in on the middle interval (highlighted in light blue in Fig.~\ref{fig:fractalplotregions}). This is equivalent to repeatedly taking decision $2$. Three successive levels of zoom are displayed in Fig.~\ref{fig:fractalplotmz1}, Fig.~\ref{fig:fractalplotmz2} and Fig.~\ref{fig:fractalplotmz3}, respectively. We observe that, at each level, the trajectories which fall into the upper and lower black holes are again interchanged. Furthermore, the image is mirrored in $\alpha$ about the centre of the interval (N.B.~decision $2$ reverses the sense of the orbit).

Finally, we may zoom in on the right-hand interval (green) by repeatedly taking decision $4$; this yields Fig.~\ref{fig:fractalplotrz1}, Fig.~\ref{fig:fractalplotrz2} and Fig.~\ref{fig:fractalplotrz3}. We see that each plot is similar to the original. 

In Fig.~\ref{fig:cantorzoom}(a)--(j) we have zoomed in on the \emph{same} region repeatedly, that is, followed geodesics which make the same decision at each stage. Zooming in on the intervals in this fashion yields a particular class of self-similar images: those in which the proportions of the intervals corresponding to each trajectory type are the similar at each level of zoom. However, if we were to take different decisions at each stage, we would still observe self-similarity, but with the relative proportions of the intervals depending, principally, on the previous digit in the decision sequence.

\subsection{The strange repellor and chaotic scattering\label{subsec:chaotic}}

The set of all (unstable, unbound) trajectories that remain confined in the scattering region as $t \rightarrow \infty$ constitutes the \emph{repellor} $\Omega_R$ of the scattering system \cite{Gaspard:1989}. 
In a \emph{two}-disc model, the repellor consists of a unique trajectory. The associated dynamics are regular, and the Kolmogorov--Sinai (KS) entropy is zero; thus, the repellor is \emph{regular}. 
By contrast, the repellor for the three-disc system forms a Cantor-like set (i.e.~an uncountably infinite set with fractal properties) \cite{Eckhardt:1988, Gaspard:1989}; this is called an \emph{irregular} (or \emph{strange}) repellor. In the previous sections, we have shown that a strange repellor $\Omega_R$ also exists for null geodesics on the double black hole MP spacetime, and it is also a Cantor-like set.

Gaspard \& Rice \cite{Gaspard:1989} showed that the repellor $\Omega_R$, and the natural measure it supports, are characterized by quantities such as the Lyapunov exponents ($\lambda_i$), the KS entropy per unit time ($h_{KS}$), the Hausdorff dimension ($D_H$) and the information dimension ($D_I$), the escape rate ($\gamma$) and the time-delay function ($T$). Here we briefly illustrate the latter.

\begin{figure}[h]
 \includegraphics[width=8cm]{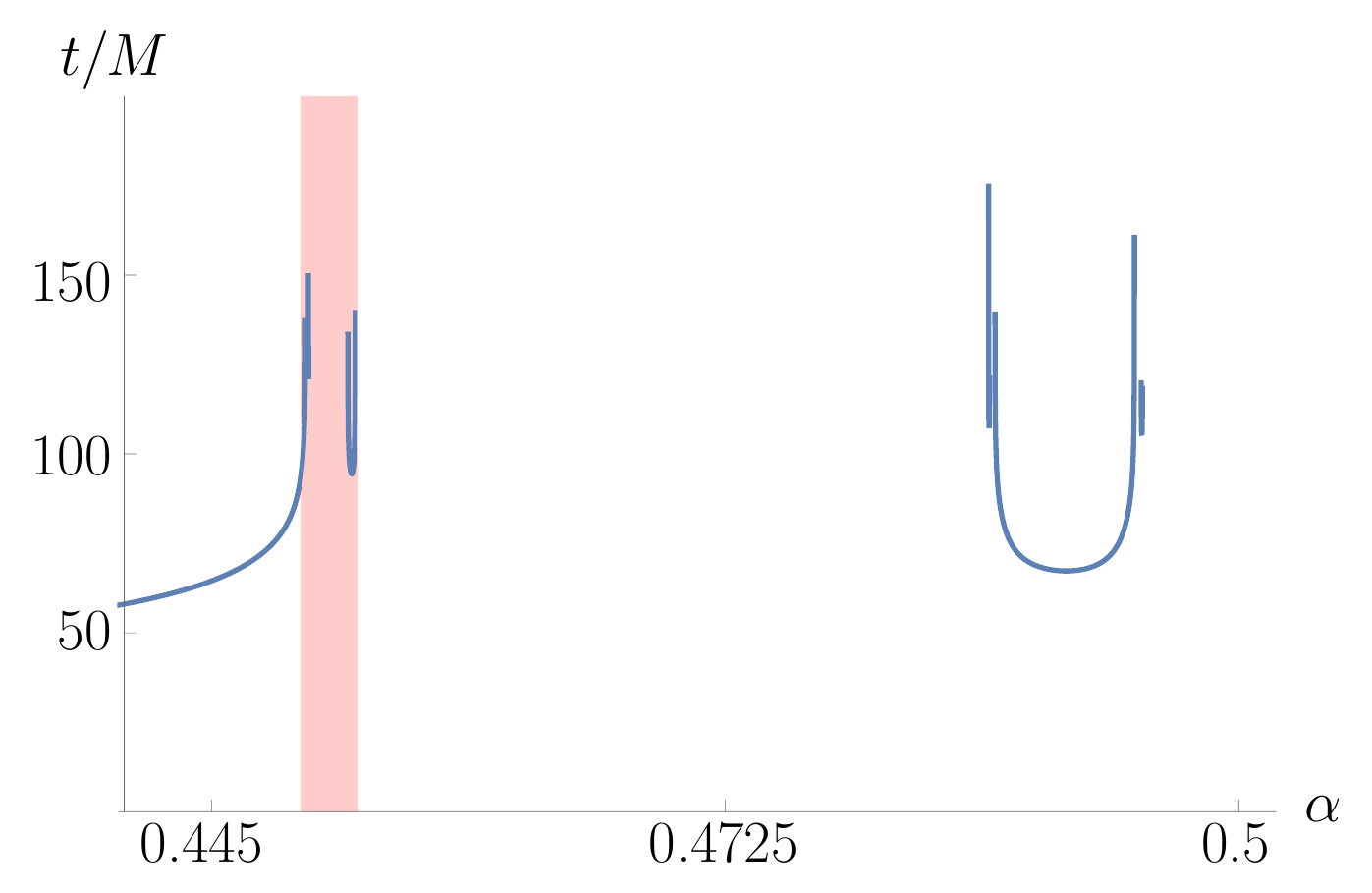}
 \includegraphics[width=8cm]{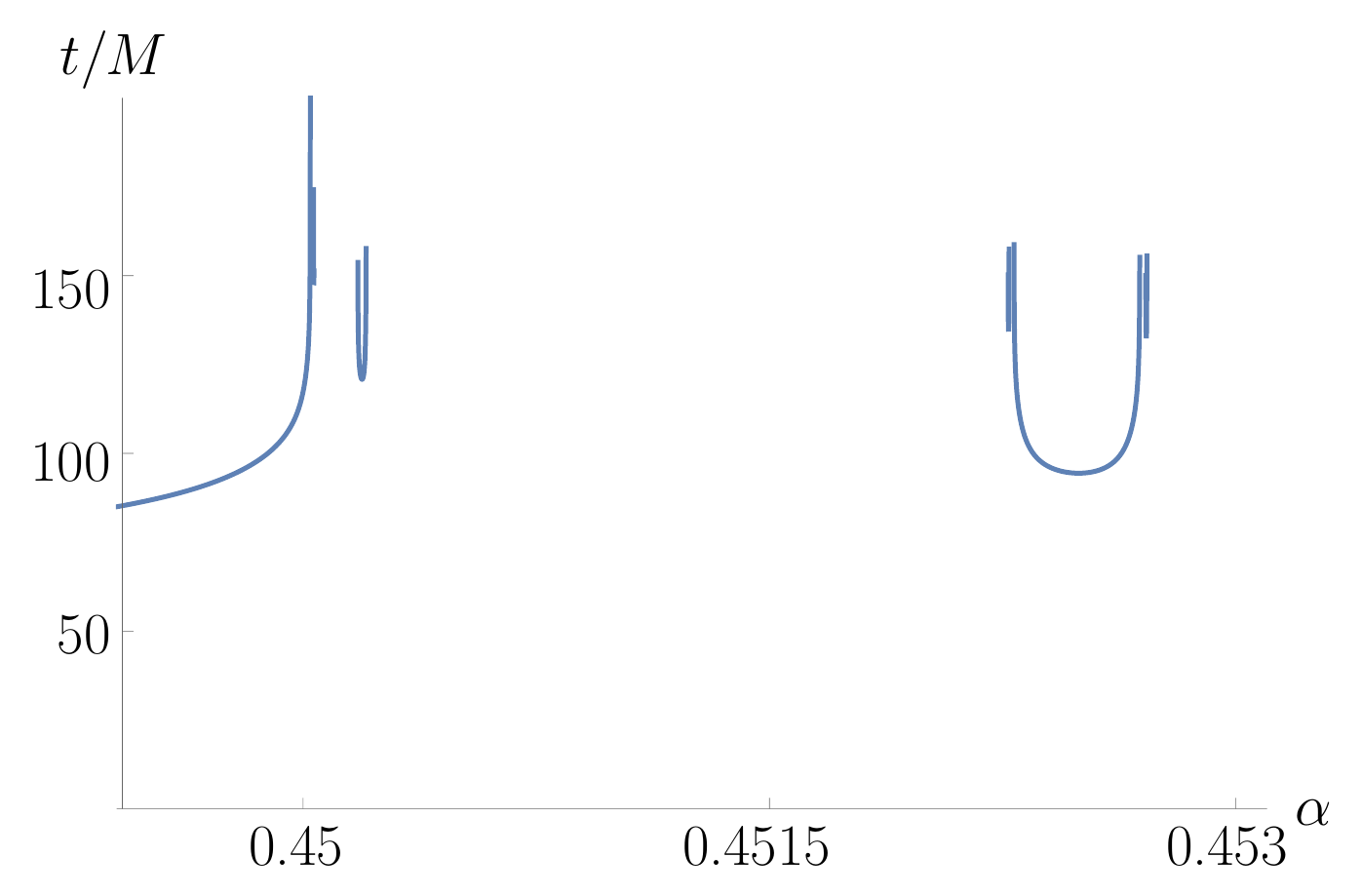}
  \caption{Time delay for scattering from the two black hole system. The plot shows the coordinate time it takes a null geodesic starting at the centre of mass to escape to large radius. In the blank regions, the geodesics fall into the black holes and the function is not defined. The right plot shows a close-up of the red region in the left-plot, highlighting the self-similarity of this function.}
 \label{fig:timedelay}
\end{figure}

Figure \ref{fig:timedelay} shows the time delay function $T(\alpha)$, defined here as the coordinate time it takes for a null geodesic starting at the centre of mass to reach some large radius $r_1$. Note that $T(\alpha)$ is not defined for trajectories that fall into the black holes, as $t$ diverges in the approach to an horizon.
In addition, $T$ diverges in the approach to a scattering singularity -- not because it approaches an horizon, but because it (asymptotically) approaches a perpetual orbit. Like the 1D shadow, the time-delay function has a self-similar geometry. In these respects, it is similar to the time delay function for the three-disc model shown in Fig.~2 of Ref.~\cite{Gaspard:1989} (see also Fig.~2 in Ref.~\cite{Ott:Tel:1993} for the deflection angle).

\section{Non-planar rays and binary shadows\label{sec:2Dshadows}}
In Sec.~\ref{subsec:nonplanar} we study rays with non-zero angular momentum about the symmetry axis. In Sec.~\ref{subsec:2D} we present a gallery of two-dimensional shadows, which we analyze by examining 1D slices.

 \subsection{Non-planar rays\label{subsec:nonplanar}}

Let us now consider non-planar motion governed by system (\ref{eq:nonplanar}), with a non-zero conserved azimuthal angular momentum $p_\phi$. Once again, it is rewarding to consider the fundamental perpetual null orbits (cf.~Fig.~\ref{fig:fundamental}). Where two or more (distinct-but-connected) fundamental null orbits exist, we expect chaotic scattering to occur (i.e.~a Cantor-like set of scattering singularities in initial data). However, we shall see that there are other possibilities.

\subsubsection{Fundamental orbits with angular momentum}
\begin{figure}[h]
\begin{tabular}{cccc}
 \subfigure[]{
  \includegraphics[width=5cm,trim={2.5cm 2.5cm 2.5cm 2cm},clip]{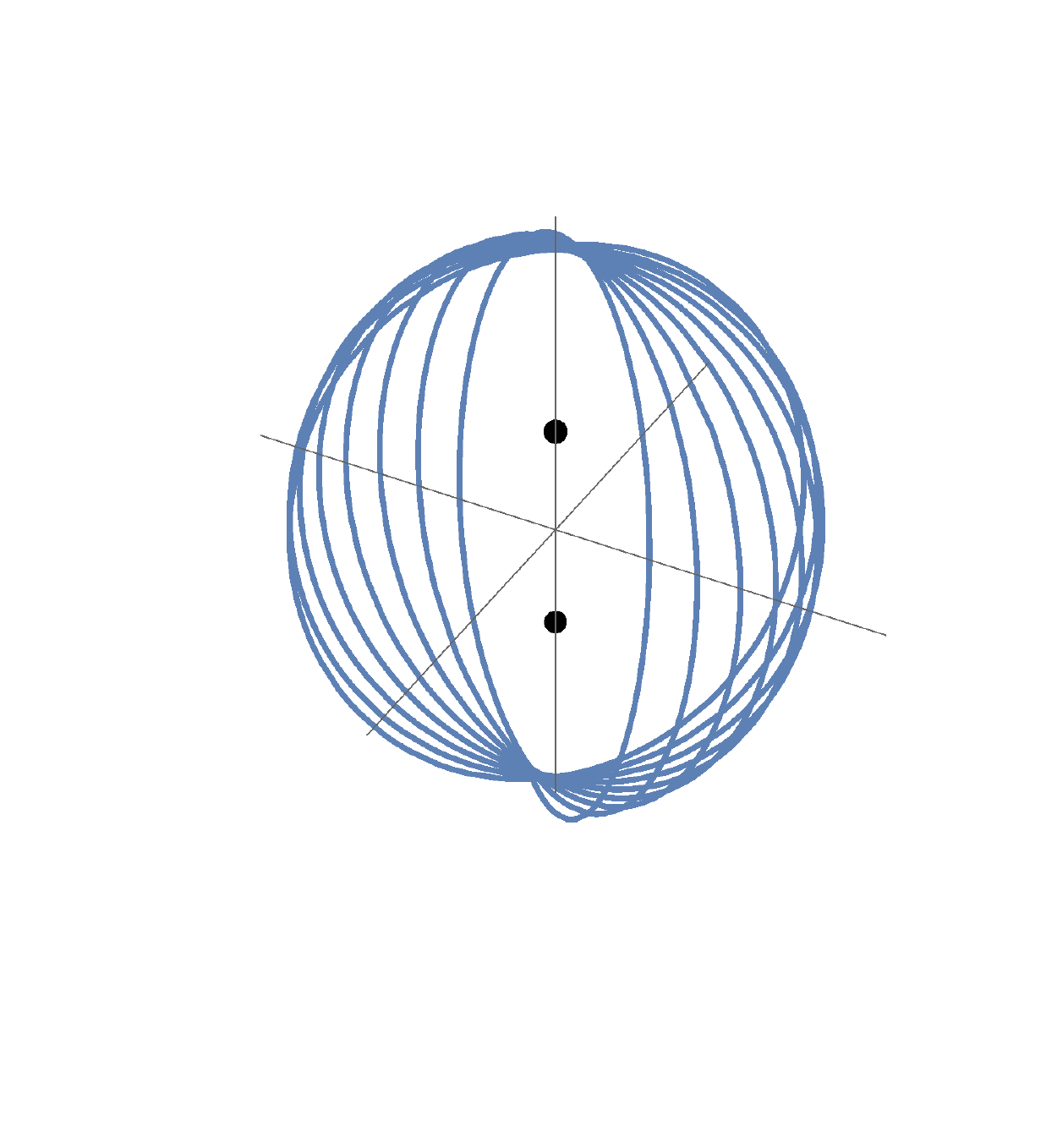}
  \includegraphics[width=3.0cm]{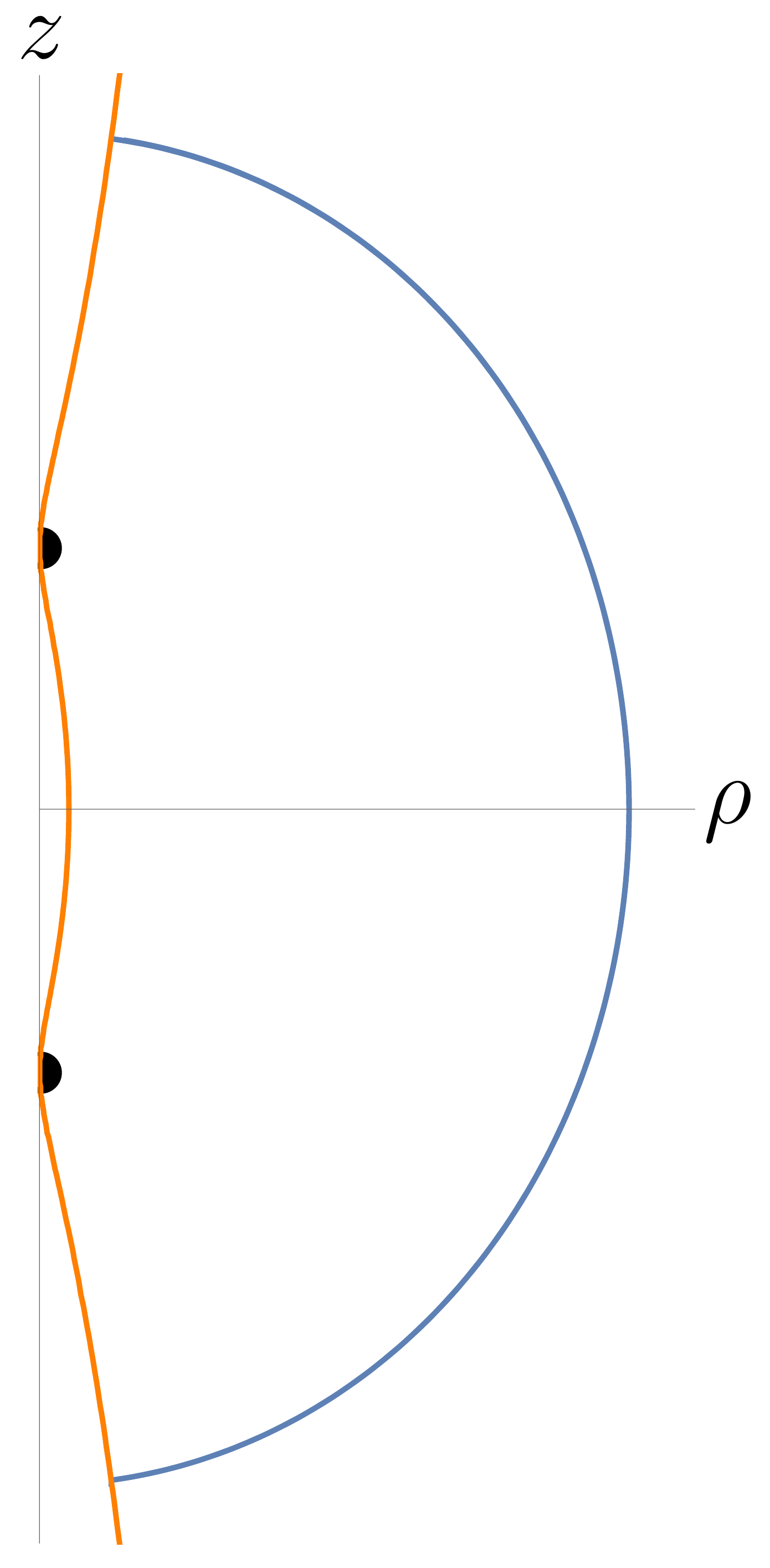}
  } &
 \subfigure[]{
  \includegraphics[width=5cm,trim={2.5cm 2.5cm 2.5cm 2cm},clip]{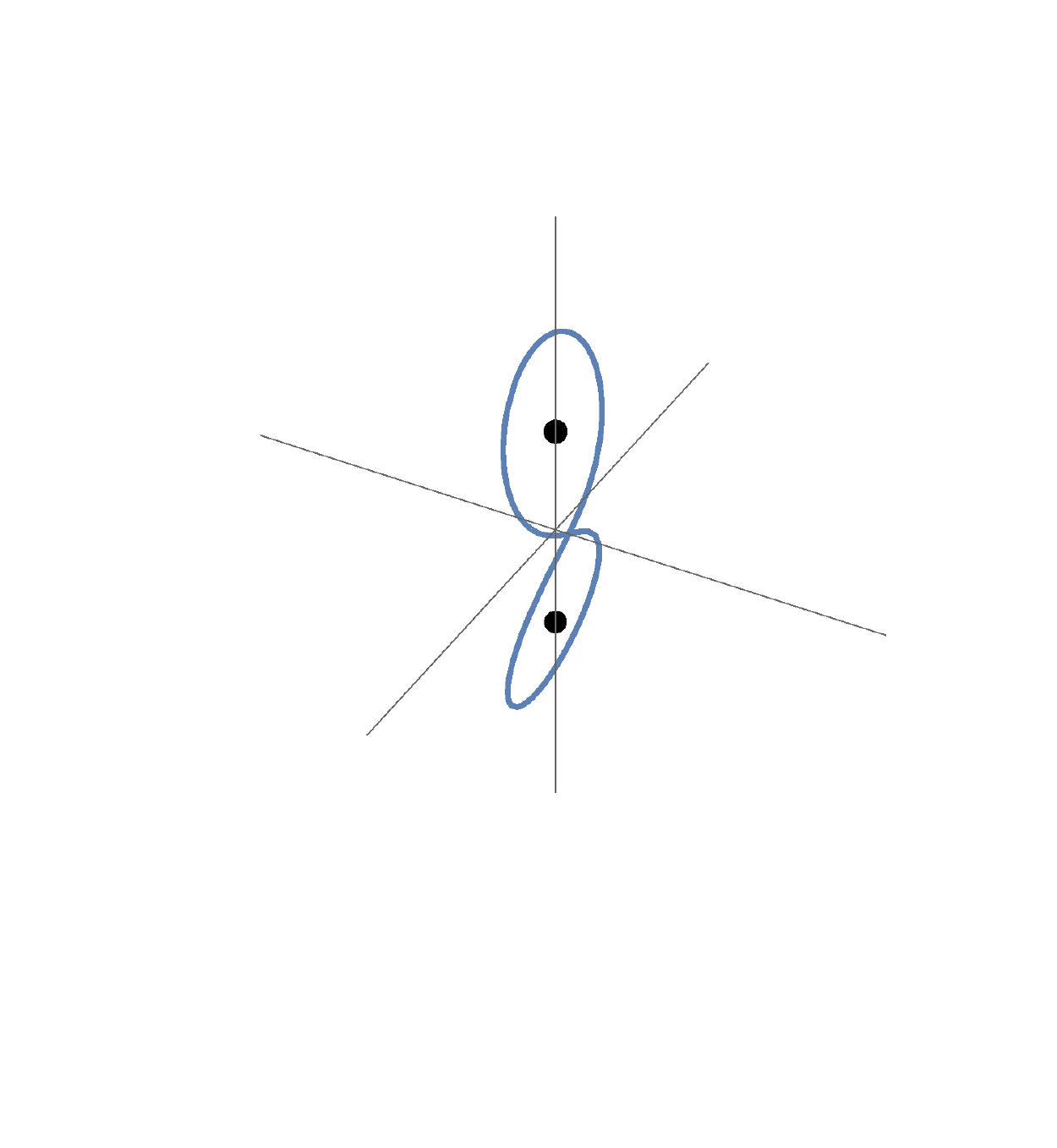}
  \includegraphics[width=3.0cm]{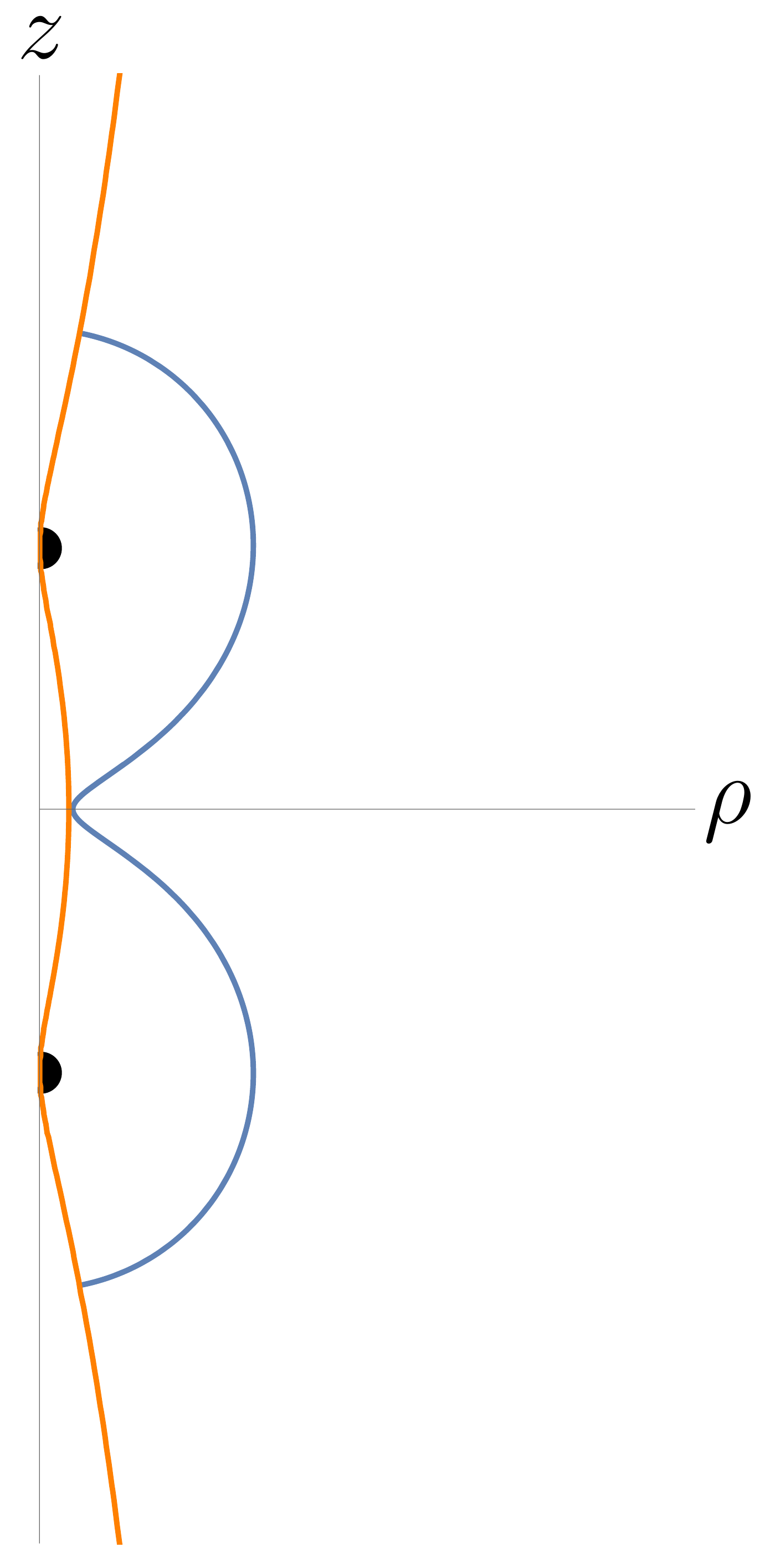}
 }
 \\ \vspace{0.1cm}
 \subfigure[]{
  \includegraphics[width=5cm,trim={2.5cm 2.5cm 2.5cm 2cm},clip]{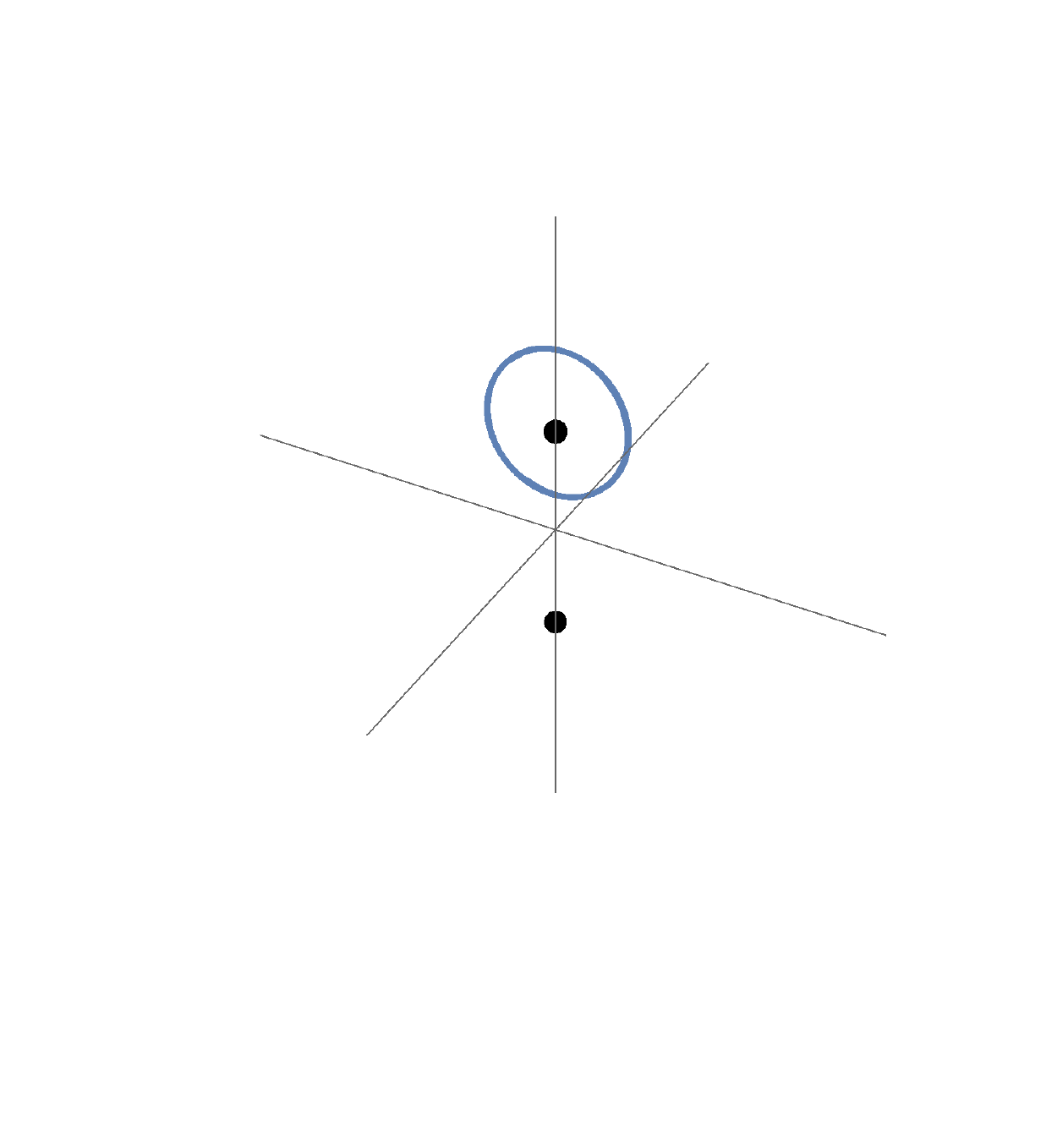}
  \includegraphics[width=3.0cm]{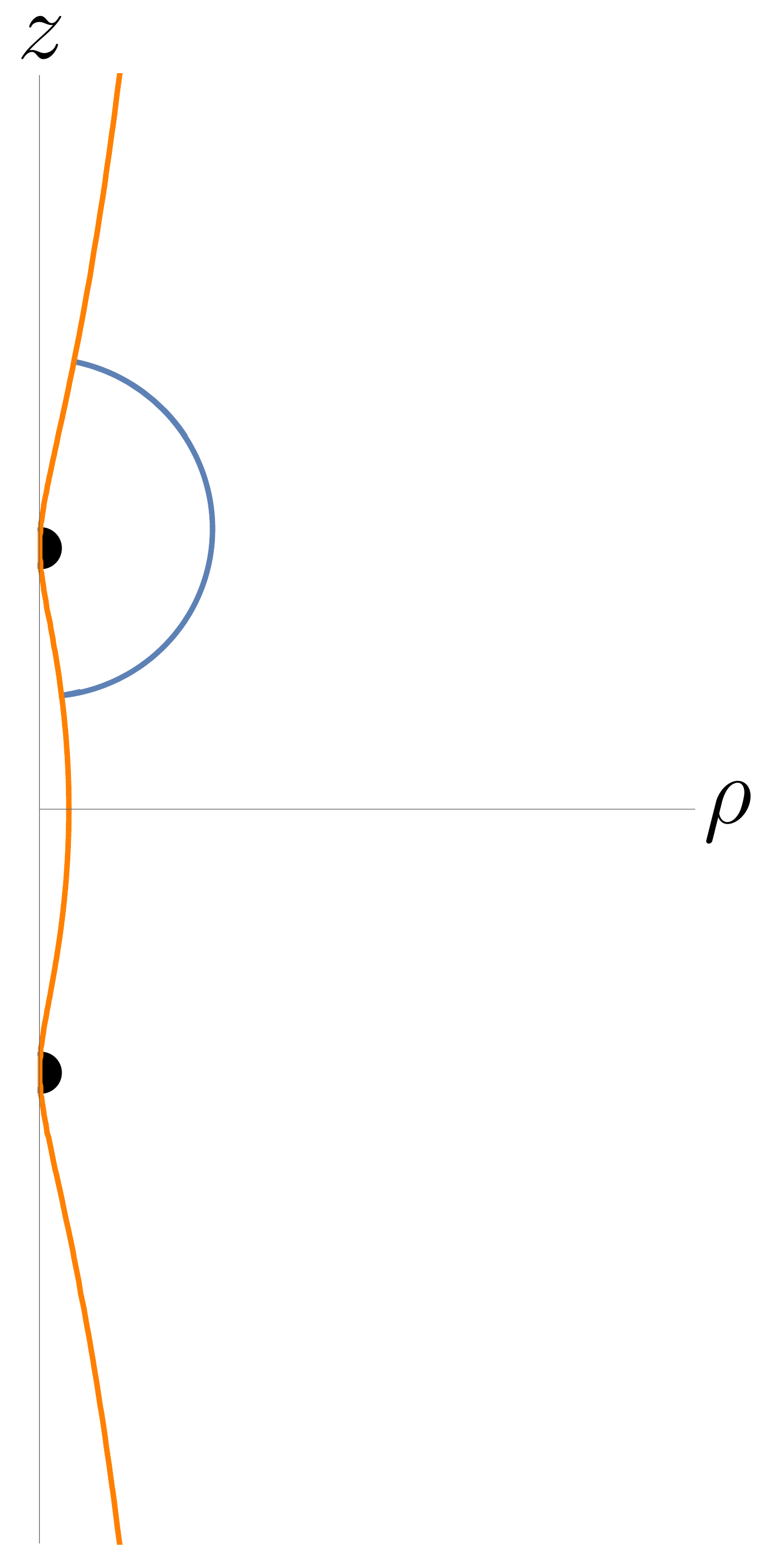}
 } &
 \subfigure[]{
  \includegraphics[width=5cm,trim={2.5cm 2.5cm 2.5cm 2cm},clip]{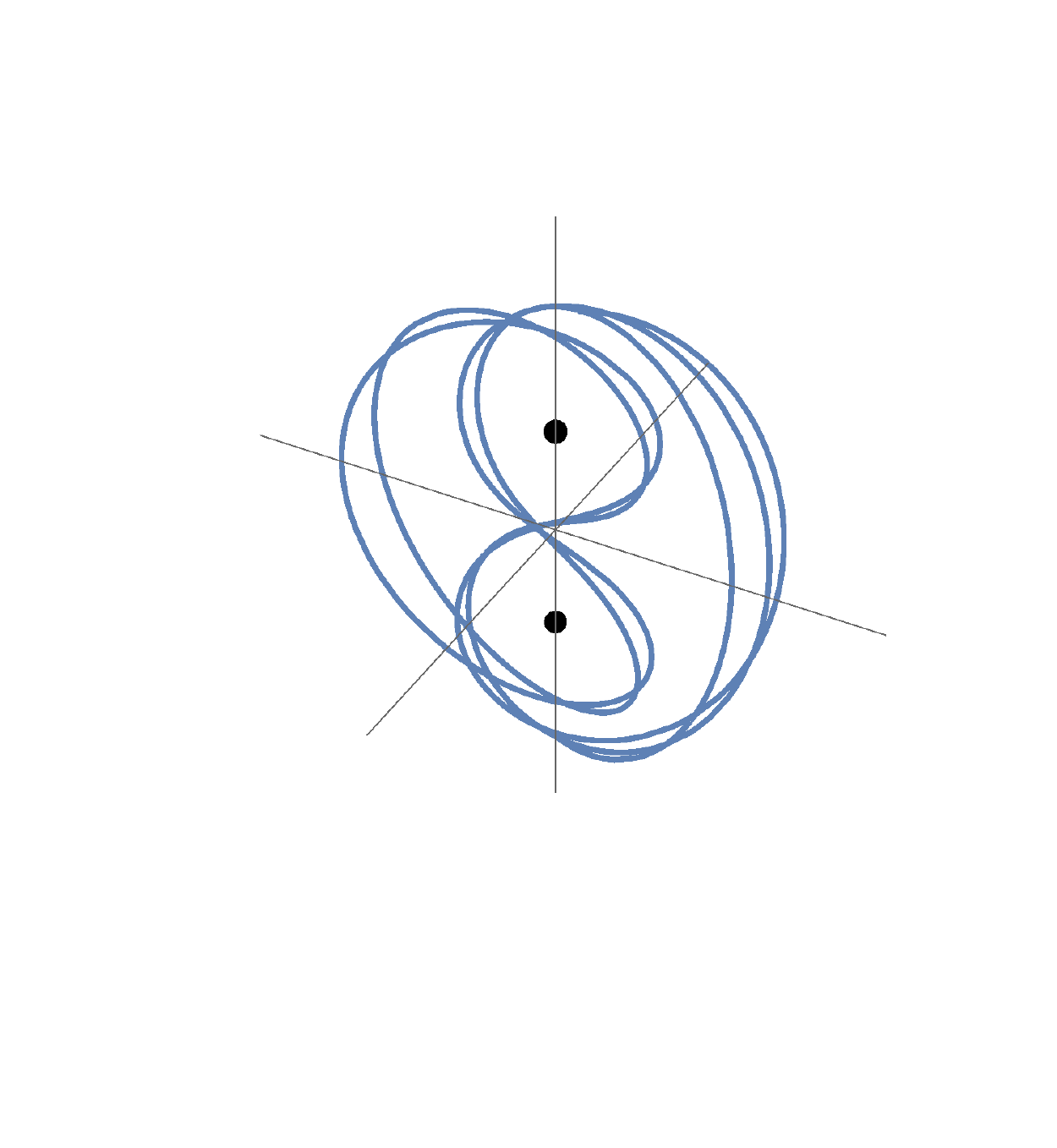}
  \includegraphics[width=3.0cm]{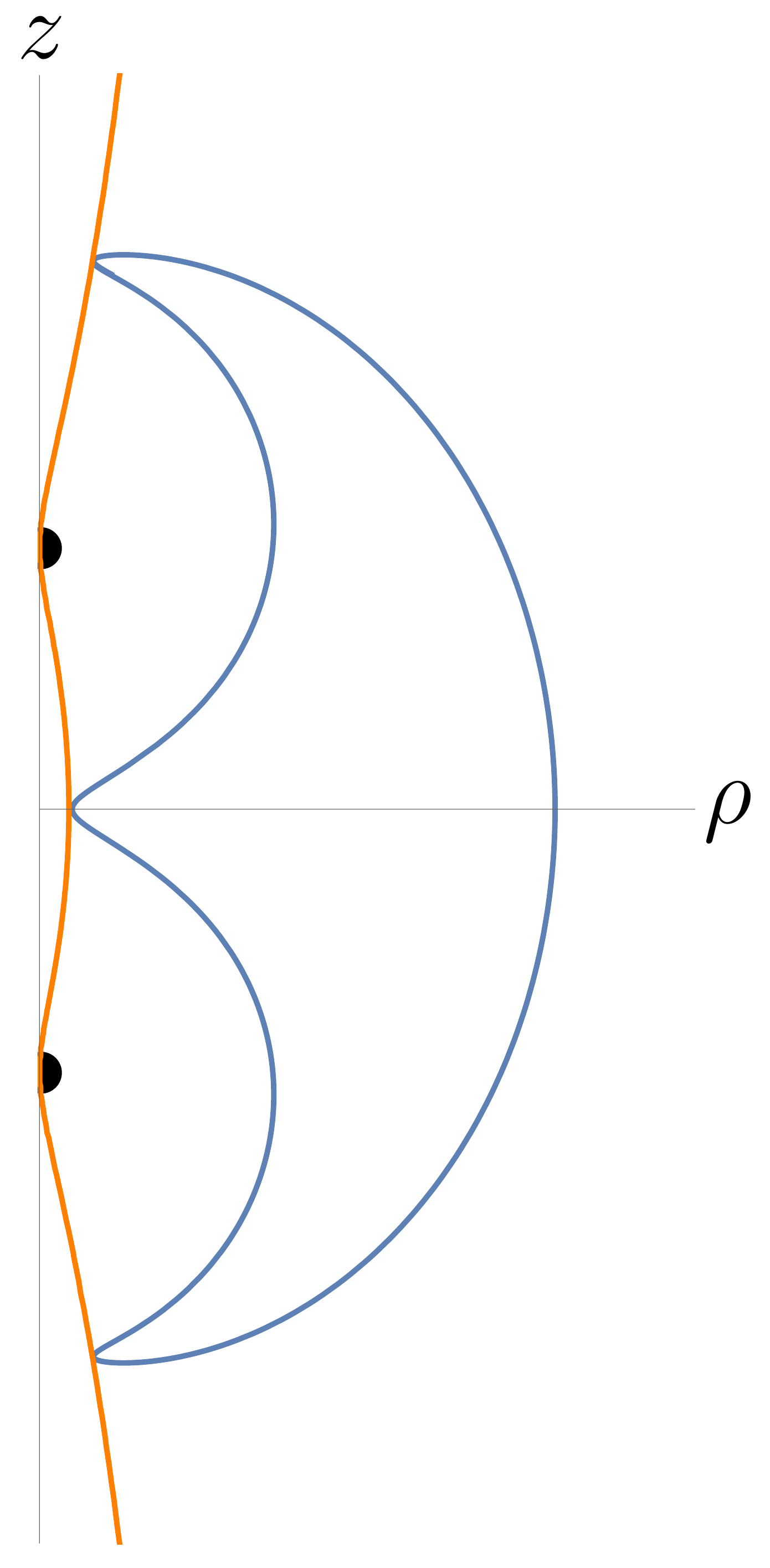}
 }
\end{tabular}
\caption{
 Examples of non-planar perpetual null geodesics ($p_\phi = 1$, $M_\pm = 1$, $a = 2$). The left plots show trajectories in 3D with $x,y,z$ axes; the right plots show the projection in the $(\rho, z)$-plane. Plots (a), (b) and (c) show the non-planar versions of the fundamental planar orbits shown in Fig.~\ref{fig:fundamental}, corresponding to decision sequences (a) $000\cdots$, $222\cdots$ and $444\cdots$. Plot (d) shows the non-planar version of the second plot in Fig.~\ref{fig:periodicorbits}, with decision sequence $0202\cdots$. The yellow line is the contour $\rho U^2 = p_\phi$ (see text).}
\label{fig:nonplanar}
\end{figure}

Figure \ref{fig:nonplanar} shows examples of non-planar fundamental null orbits for $p_\phi = 1$, $a=2$. The fundamental orbits of Fig.~\ref{fig:fundamental} persist, keeping their distinct character, even though the motion is no longer planar. The three fundamental orbits (with decision sequences (a) $000\cdots$, (b) $222\cdots$, (c) $444\cdots$) are shown, along with the alternating case (d) $0202\cdots$. The latter case (d) indicates that transitions between the fundamental orbits are possible, just as in the planar case, and thus there will exist an infinite family of perpetual orbits. Though the motion in the $(\rho, z)$-plane is periodic, the non-commensurate motion in $\phi$ means that these geodesics are not closed in 3D; instead they trace out a two-surface [left plots].

\subsubsection{A potential function}
The null condition $H = 0$ [Eq.~(\ref{eq:hamiltonian})] yields an `energy equation',
\beq
p_\rho^2 + p_z^2 = U^4 - \frac{p_\phi^2}{\rho^2} = \frac{1}{\rho^2} \left( h(\rho,z) + p_\phi \right) \left[ h(\rho, z) - p_\phi  \right] . \label{eq:energy}
\eeq
Here we have introduced the function
\beq
h(\rho,z) \equiv \rho U^2  \label{eq:h}
\eeq
that determines the sign of the right-hand side of Eq.~(\ref{eq:energy}), via the term in square brackets. The contours of $h$ (i.e.~the solutions to $h(\rho,z) = p_\phi$) are curves on which a ray may be instantaneously stationary in the $(\rho, z)$-plane (as $p_\rho = 0 = p_z$ and thus $\dot{\rho} = 0 = \dot{z}$). For a given $p_\phi$, the contour $h(\rho,z) = p_\phi$ demarcates a `forbidden' region that a null ray cannot access; for $p_\phi \neq 0$ this region includes all of the symmetry axis, with the possible exception of $z = z_\pm$.

For a ray which `touches' a contour $C$ (defined by $h = p_\phi$), with $\dot{\rho} = 0 = \dot{z}$, Hamilton's equations are $\dot{p}_\rho|_C = \frac{1}{\rho} h_{,\rho}$, $\dot{p}_z|_C = \frac{1}{\rho} h_{,z}$. Consequently, rays that `touch' a contour of $h$ must be incident parallel to $\nabla h$, and thus orthogonal to the contours of $h$. This feature can be seen in Fig.~\ref{fig:nonplanar} (note that in case (d) we have confirmed that the ray does not quite touch the contour).

\begin{figure}[h]
\begin{tabular}{cccc}
 \subfigure[$p_{\phi}=4$]{
  \includegraphics[width=3.5cm]{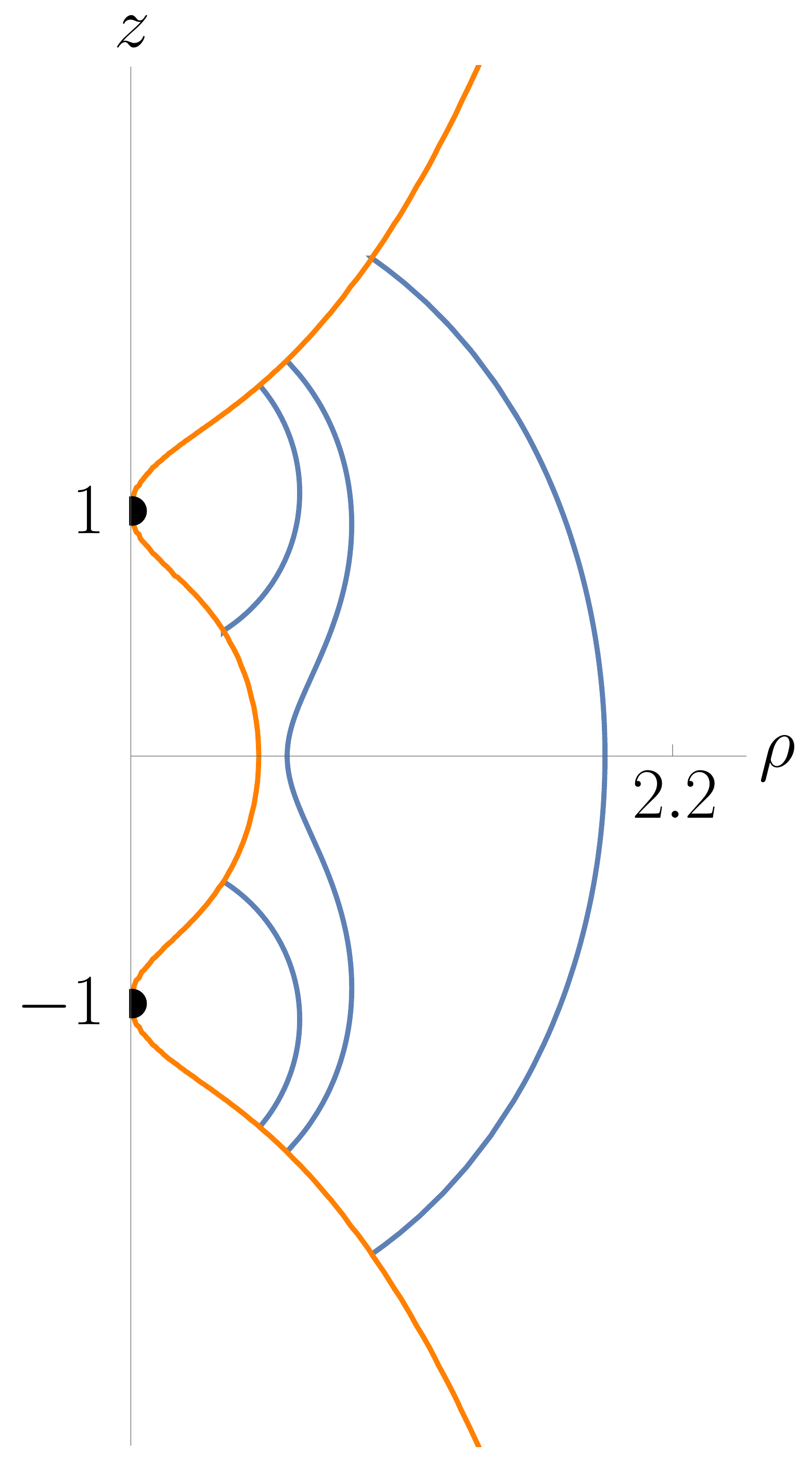}
 }  &
 \subfigure[$p_{\phi}=5$]{
  \includegraphics[width=3.5cm]{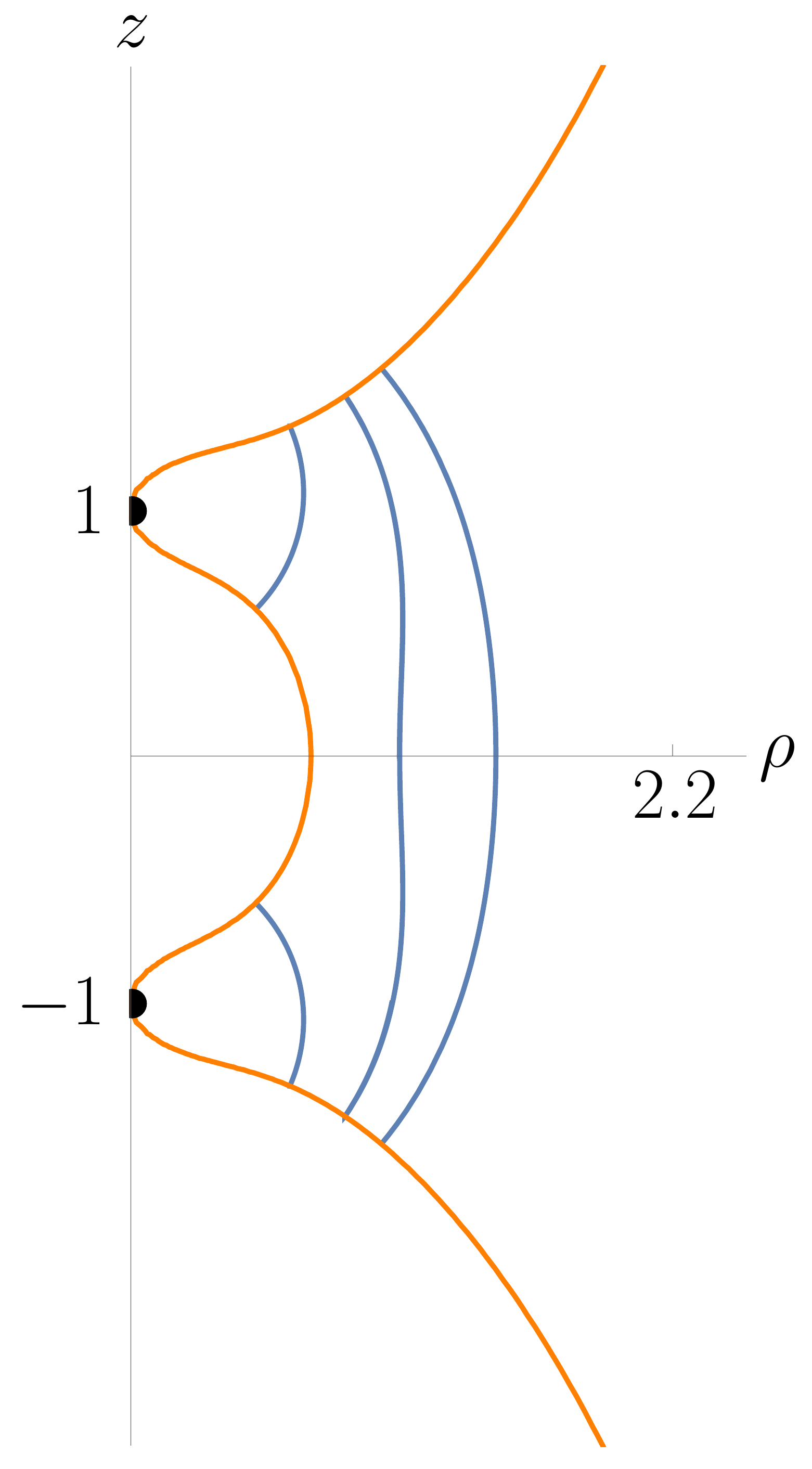}
 } &
 \subfigure[$p_{\phi}=5.08$]{
  \includegraphics[width=3.5cm]{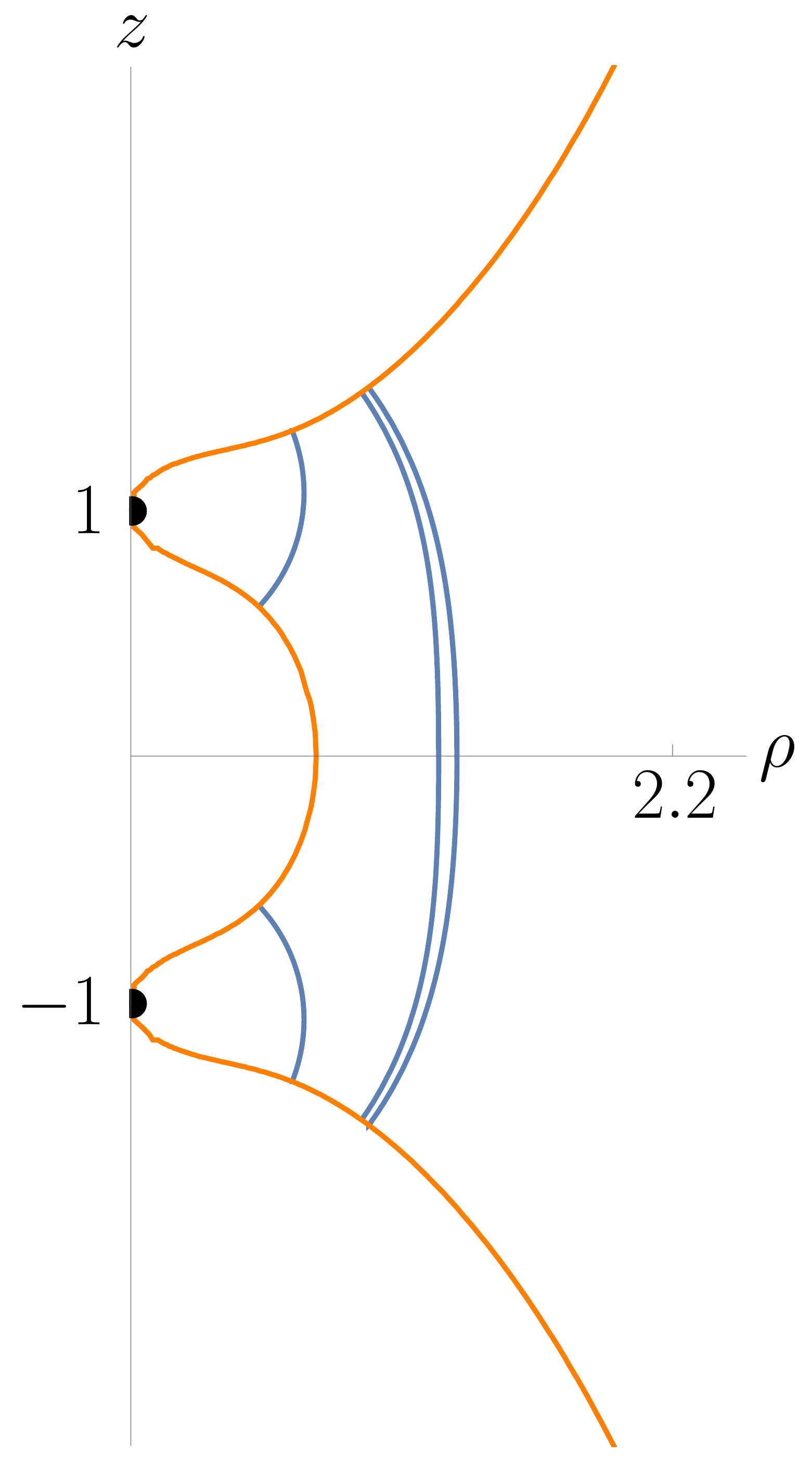}
 } &
 \subfigure[$p_{\phi}=5.9$]{
  \includegraphics[width=3.5cm]{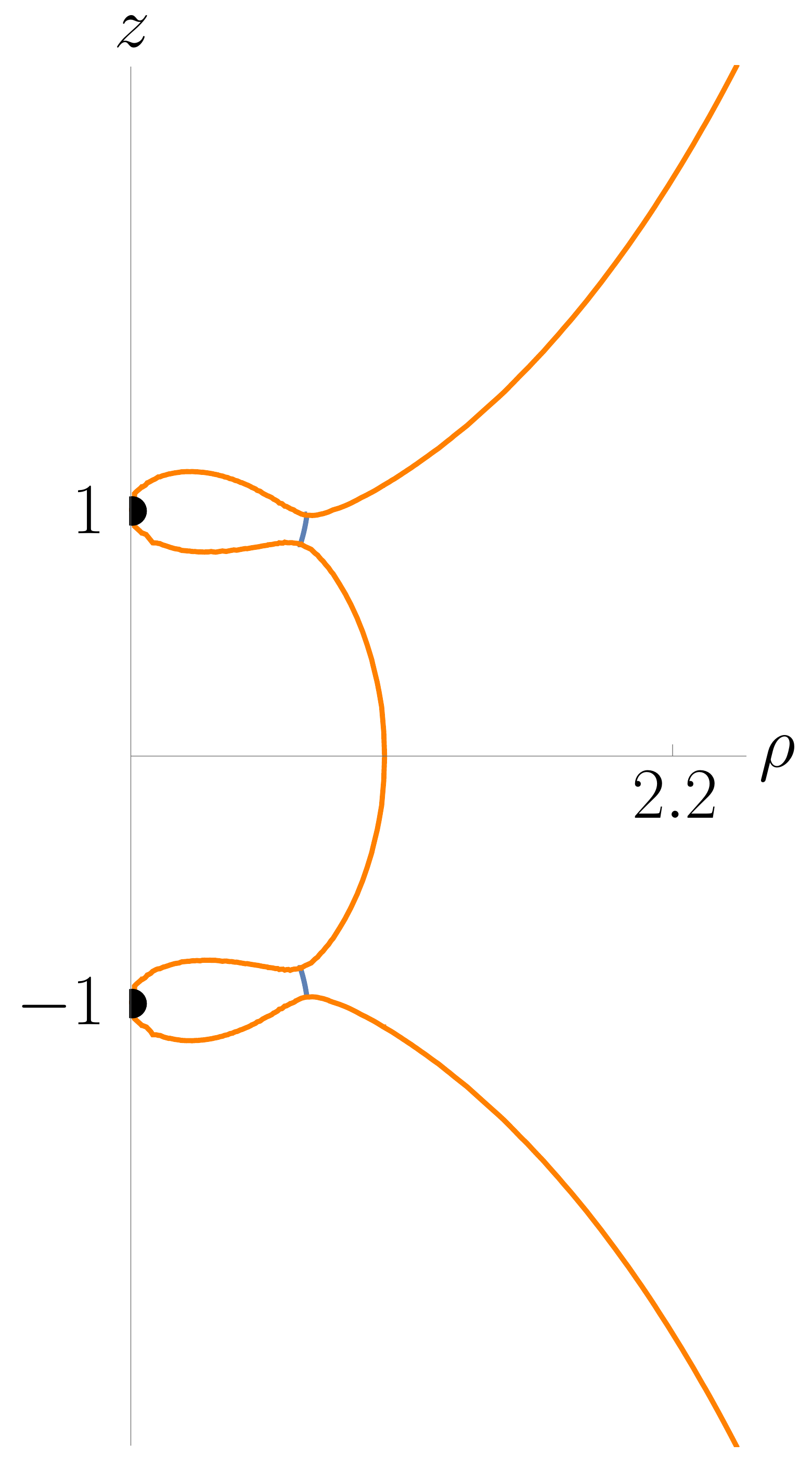}\label{fig:nonplanar-fundamental:d}
 }
\end{tabular}
\caption{Examples of non-planar fundamental orbits for $a=2$ and $p_\phi > 0$. The blue curves are null trajectories, and the yellow curve shows $h = p_\phi$. As $p_\phi$ increases, the $\dot{0}$ and $\dot{2}$ orbits move closer together. Beyond $p_\phi \approx 5.08$, these orbits do not exist. The $\dot{4}$ orbit [without symmetry in the equatorial plane] exists up to $p_\phi \approx 5.92214$. The consequences for chaotic scattering are explored in the text.}
\label{fig:nonplanar-fundamental}
\end{figure}

Figure \ref{fig:nonplanar-fundamental} shows the effect of increasing $p_\phi$ on the fundamental orbits, for the case $a=2$. We see that the $\dot{0}$ and $\dot{2}$ trajectories move closer together as $p_\phi$ is increased. At a critical value, $p_\phi = p_{\phi}^A$  ($p_\phi^A \approx 5.08\ldots$ for $a=2$) the orbits intersect, and beyond this value the `outer' fundamental orbits are not possible. However, the $\dot{4}$ orbits, around the upper and lower black holes, remain possible until the contour `pinches off' at $p_\phi^B$ ($p_\phi^B \approx 5.92214$ for $a=2$), where $p_\phi^B > p_\phi^A$.

\subsubsection{Stationary points of $h$\label{subsec:stationarypoints}}
\begin{figure}
\begin{tabular}{cc}
 \subfigure[$a=2$]{
  \includegraphics[width=7.5cm]{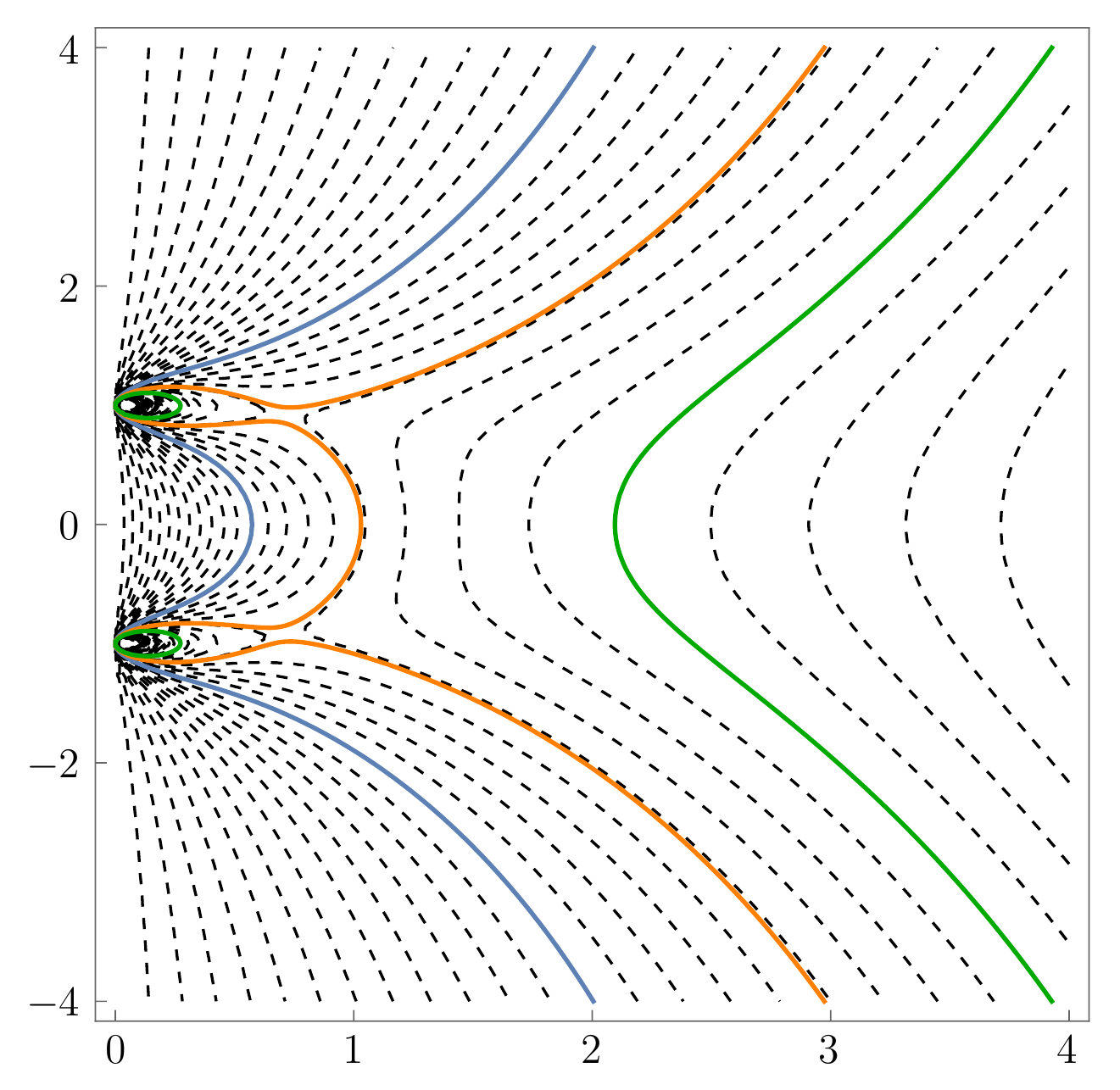}\label{fig:morphology:a}
  } &
 \subfigure[$a=0.5$]{
  \includegraphics[width=7.5cm]{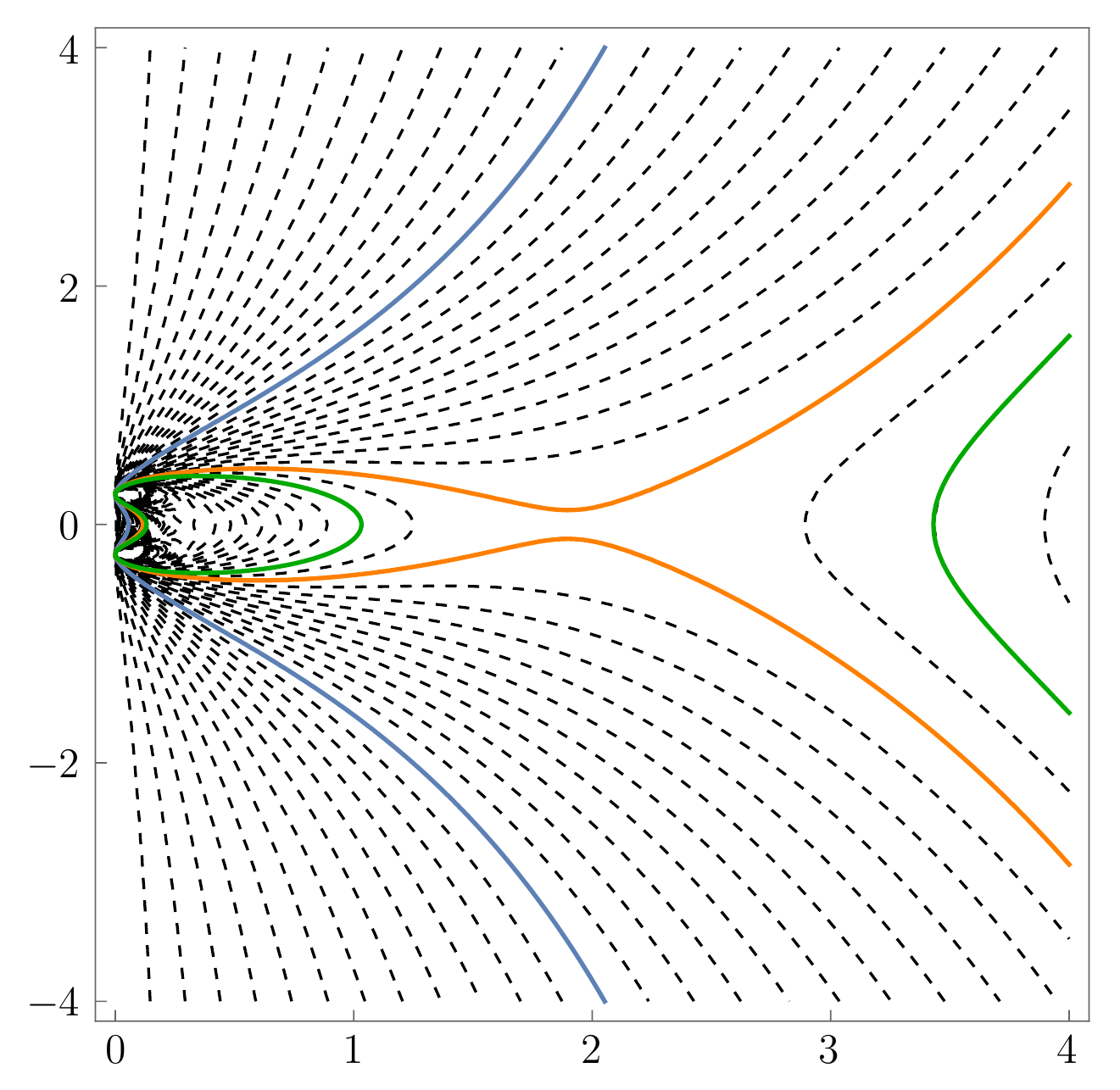}\label{fig:morphology:b}
 }
\end{tabular}
\caption{Contours of $h \equiv \rho U^2$ in the $(\rho, z)$-plane for black holes separated by $a = 2$ [left] and $a = 0.5$ [right].
Each contour $h(\rho,z) = p_\phi$ represents an impassable barrier for a ray with angular momentum $p_\phi$. Three typical cases are indicated: $p_\phi^{(1)}$ [green], $p_\phi^{(2)}$ [yellow], $p_\phi^{(3)}$ [blue], where $p_\phi^{(1)} > p_\phi^{(2)} > p_\phi^{(3)}$. A ray with $p_\phi^{(1)}$, approaching from infinity, is prevented from reaching the black holes by the green contour; whereas rays with $p_\phi^{(2,3)}$ may pass through. In the case $p_\phi^{(2)}$, absorption by a black hole would require the ray to pass through a narrow throat.  
}
\label{fig:morphology}
\end{figure}

Figures \ref{fig:morphology:a} and \ref{fig:morphology:b} indicate that the morphology of the contours of $h$ depends on the separation of the black holes $a$. For large $a$, the system behaves as two distinct black holes; whereas for small $a$ the system can effectively to resemble a single, distorted black hole. To better understand this, we should consider the stationary points of $h$. In Fig.~\ref{fig:morphology:a} [$a=2$], $h$ has a pair of saddle points associated with each black hole separately, above and below the equatorial plane; whereas in Fig.~\ref{fig:morphology:b} [$a=0.5$], $h$ has saddle points only in the equatorial plane. Perpetual orbits of type $\dot{4}$ are clearly not possible in the latter case.

We may understand the system more fully by classifying the stationary points of $h$ [Eq.~(\ref{eq:h})] that lie in the equatorial plane ($z = 0$). Let us define
\beq
a_1 = 4M / \sqrt{27} \approx 0.769800 M, \quad \quad
a_2 = \sqrt{2} \, a_1 \approx 1.088662 M .  \label{eq:a1a2}
\eeq
For $a > a_2$, there are no stationary points on the equatorial plane. For $a = a_2$ there is a cusp along the $\rho$-axis (pointing inwards) at $\rho = \sqrt{5}a/2$ (N.B.~the cusp corresponds to a point of inflexion in the $\rho$ direction and a maximum in the $z$ direction). For $a_1 < a < a_2$ there are two stationary points: a saddle point at $\rho = \rho_+$ and a maximum at $\rho = \rho_-$, where $\rho_+ > \rho_- > a/\sqrt{2}$. For $a < a_1$, the stationary points at $\rho_+$ and $\rho_-$ are saddle points, such that $\rho_+ > a/\sqrt{2} > \rho_-$.

The separation $a=1$ is a special case: one contour connects three saddle points, as shown in Fig.~\ref{fig:nonplanar-fundamental:d}. Remarkably, the contour value and the saddle point positions can be written in closed form in terms of the Golden Ratio, $\varphi \equiv \frac{1}{2}(1 + \sqrt{5})$. The saddle points at $\rho = \frac{1}{2} 5^{1/4} \varphi^{3/2}$, $z=0$ and $\rho = \frac{1}{2} 5^{1/4} \varphi^{-1/2}$, $z = \pm 1/(2 \varphi)$ are connected by a single contour of height $h = \frac{1}{2} 5^{5/4} \varphi^{3/2}$. This contour encloses a maximum at $\rho = \sqrt{3} / 2$, $z=0$ with $h = 9 \sqrt{3}/2$. More detail on the derivation of these results is given in Appendix \ref{appendix:derivation}.

The saddle points of $h$ may be thought of as unstable `Lagrange points' for null rays in the system. We note that, wherever there are saddle points, there are also neighbouring contours (of equal height) on either side of the saddle which are almost parallel. As null rays intersect the contours orthogonally, it seems that,  generically, (unstable) null orbits will occur between such neighbouring contours. Figure \ref{fig:nonplanar-fundamental:d} shows a specific example.

\begin{figure}[h]
\begin{tabular}{ccccc}
 \subfigure[$a=2$]{\hspace{-1em}
  \includegraphics[width=3cm]{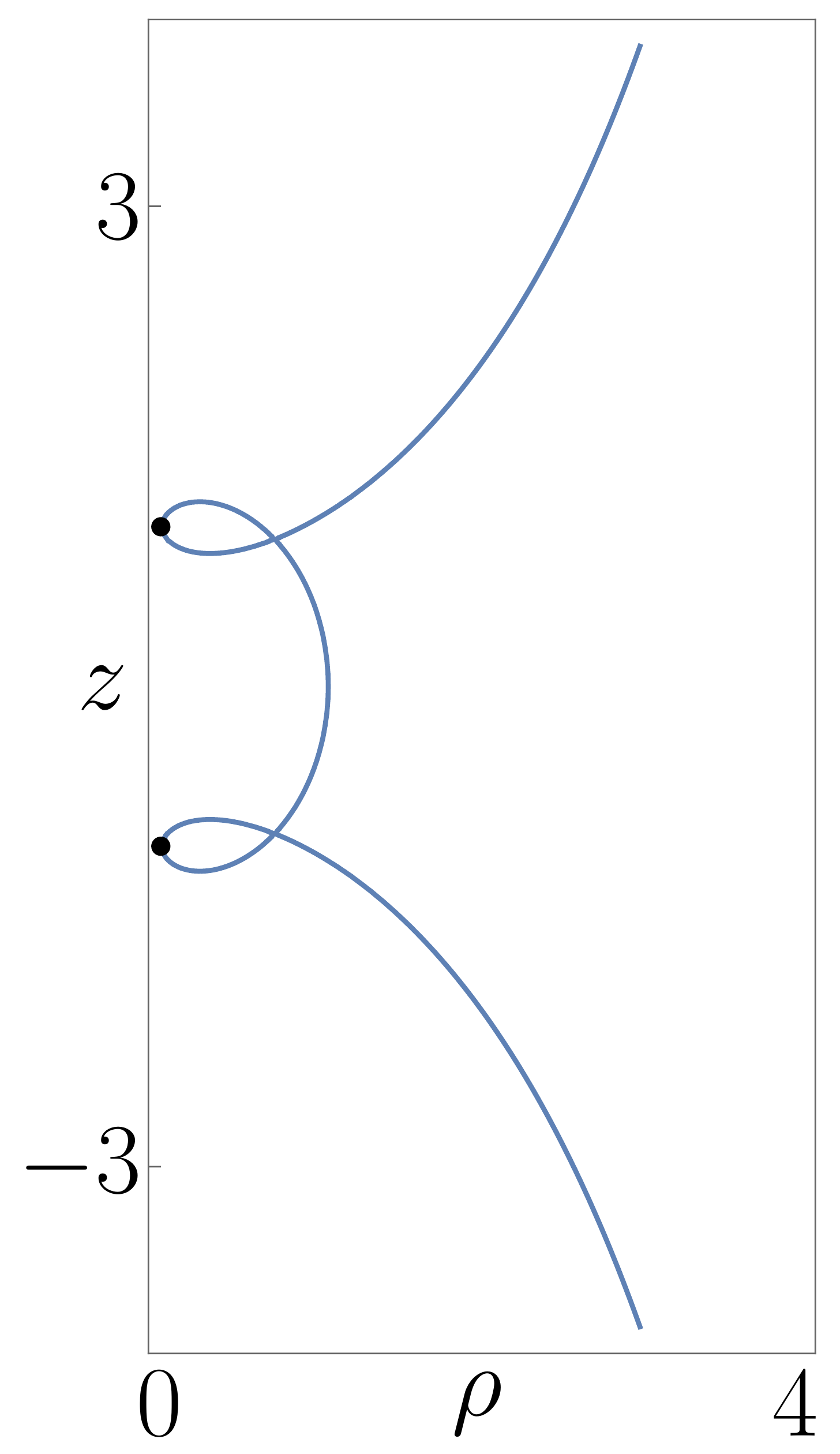}
  } &
 \subfigure[$a=\sqrt{32 / 27}$]{
  \includegraphics[width=3cm]{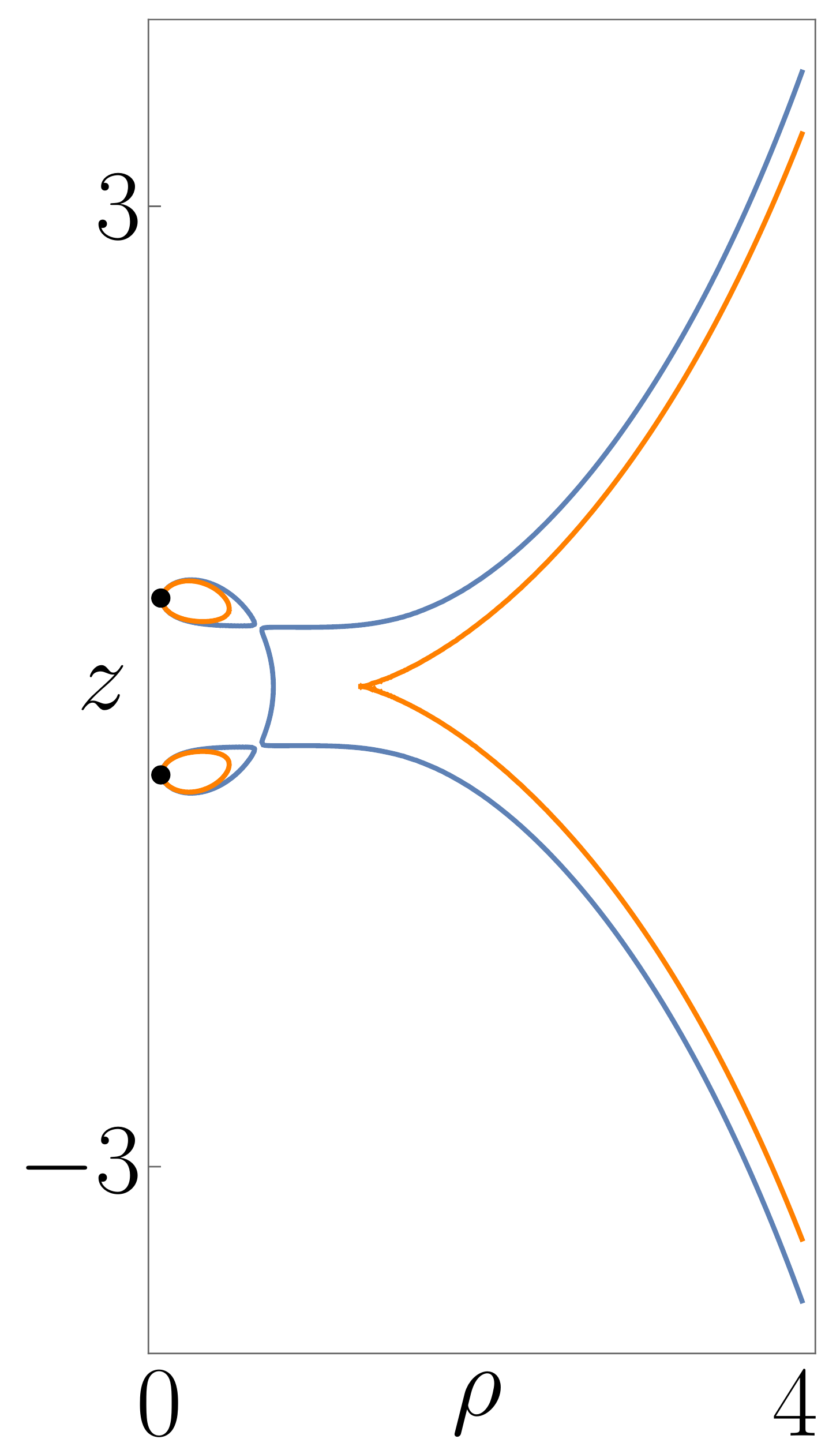}
 } &
 \subfigure[$a=1$]{
  \includegraphics[width=3cm]{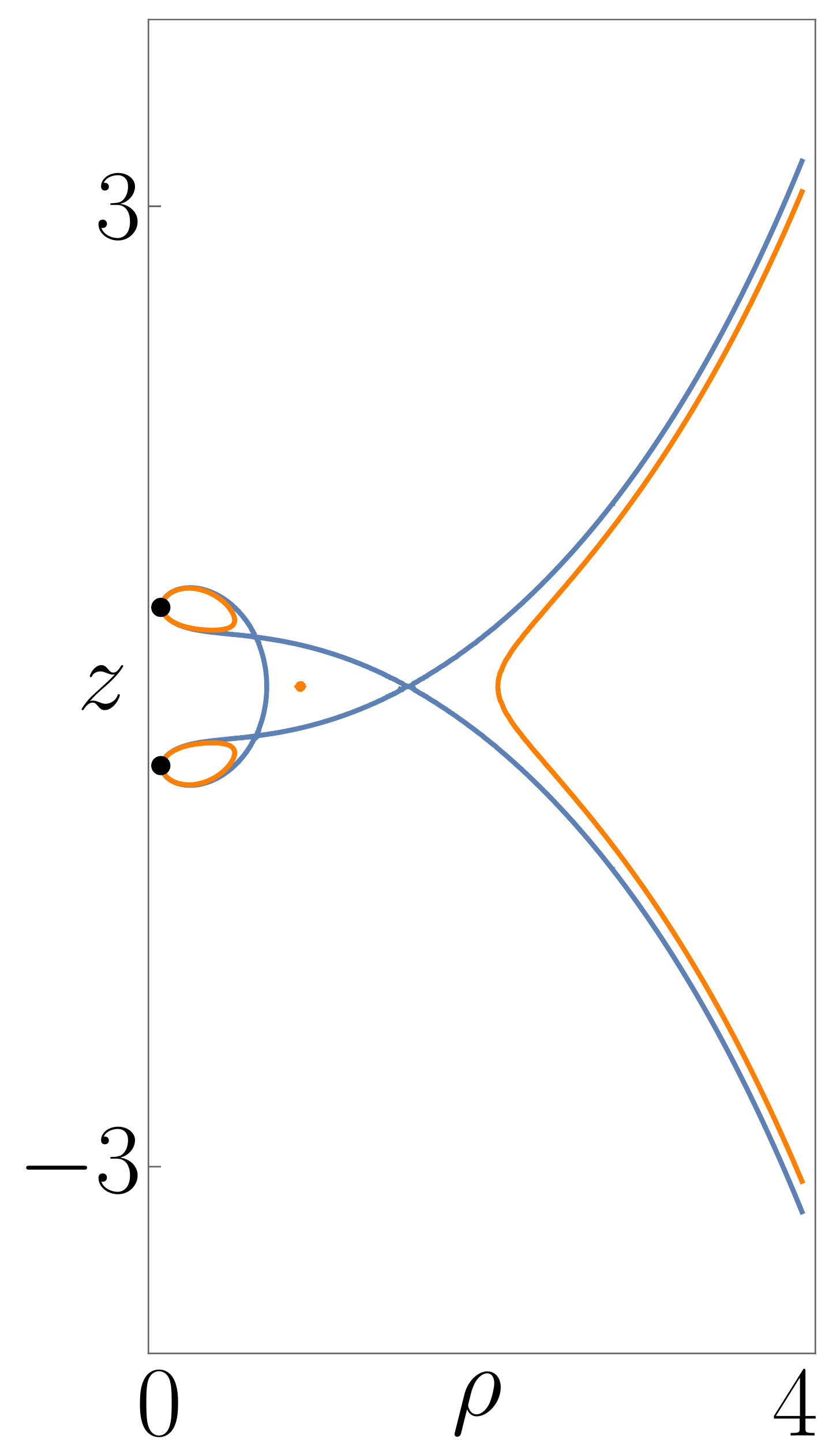}\label{fig:contour:c}
 } &
 \subfigure[$a=\sqrt{16 / 27}$]{
  \includegraphics[width=3cm]{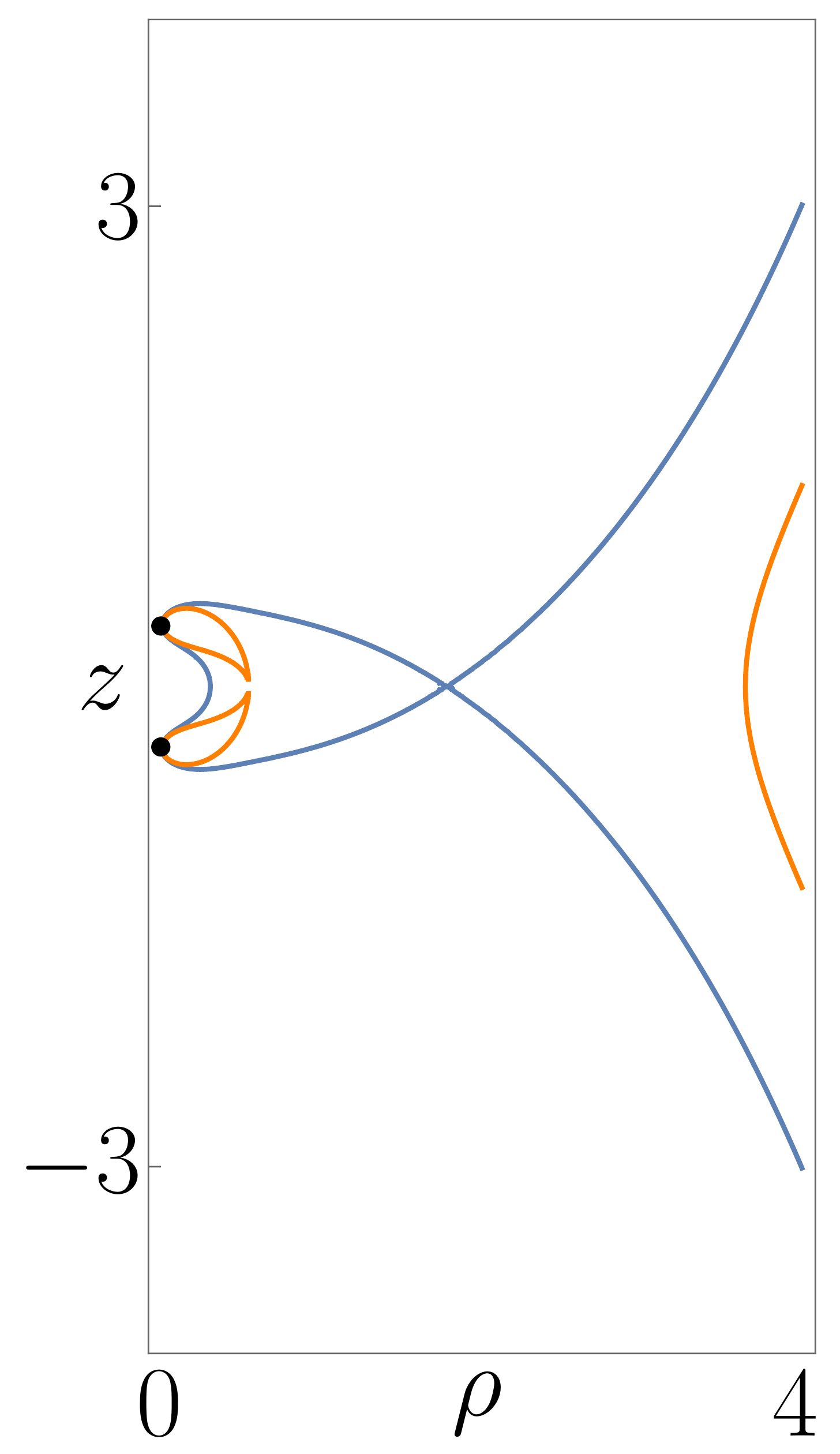}
 } &
 \subfigure[$a=0.5$]{
  \includegraphics[width=3cm]{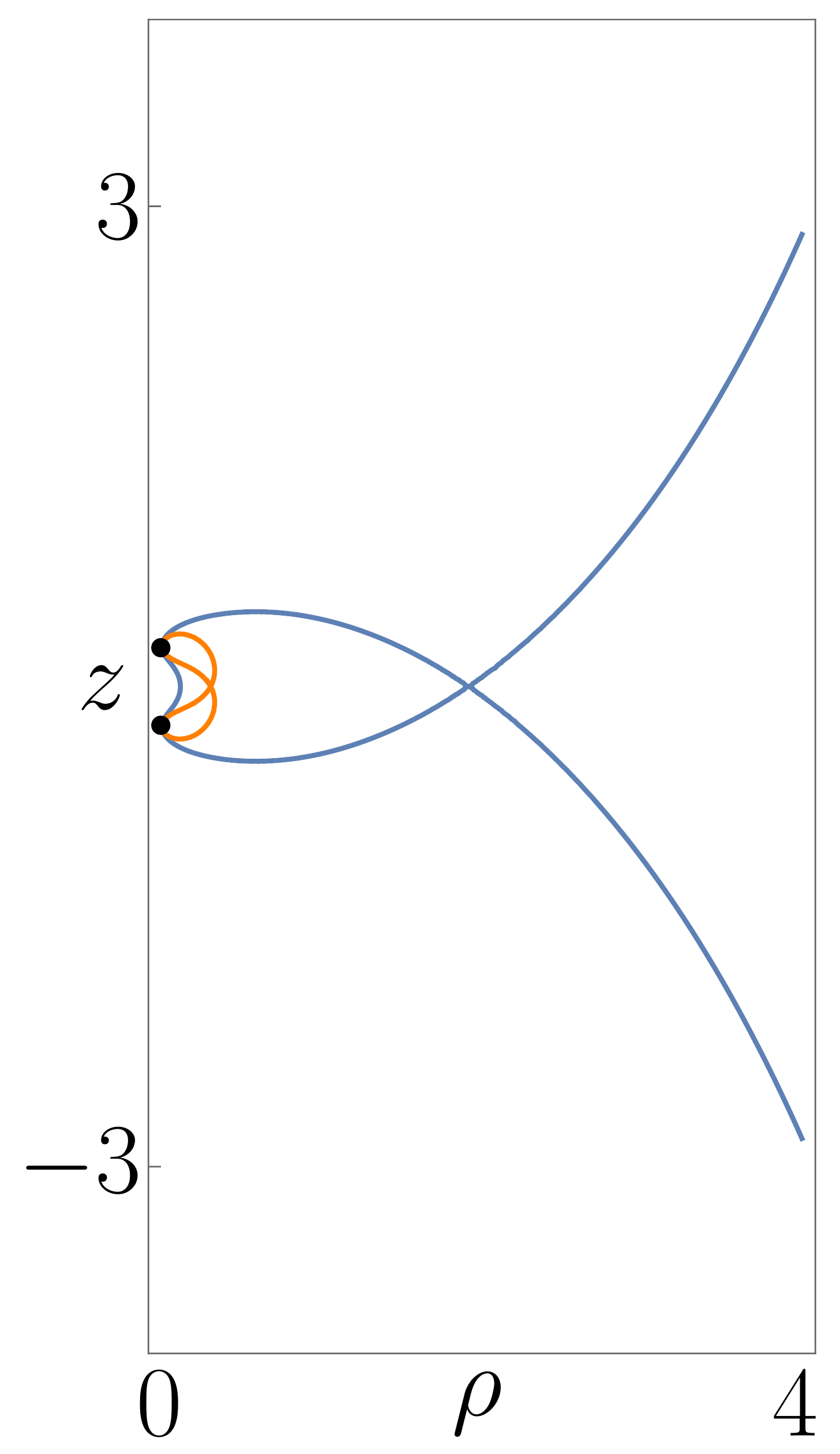}
 }
\end{tabular}
\caption{Contours that intersect the stationary points of $h(\rho, z) \equiv \rho U^2$, in the cylindrical-coordinate plane $(\rho, z)$, for a selection of values of the coordinate separation of the black holes $a$ (with $M_\pm = 1$). 
Plot (c) shows the intriguing case of \emph{bounded} null geodesics that neither fall into either black hole, nor escape to infinity. Rays starting in the vicinity of $\rho \approx 1$, $z=0$ are bounded by the blue contour that completely encloses this region. The yellow dot indicates a local maximum of $h$ (i.e.~a \emph{stable} circular orbit in the equatorial plane). On the equatorial plane, a maximum exists for $a_1 < a < a_2$; for $a < a_1$ there are two distinct saddle points; and for $a > a_2$ there are no stationary points in the equatorial plane [cf.~plot (a)]. \\
Parameters: (a) $a = 2$, $p^{(1)}_\phi \approx 5.92214$; (b) $a = a_2 = \sqrt{32/27} \approx 1.08866$, $p^{(1)}_\phi \approx 7.41479$ [blue] and $p^{(2)}_\phi \approx 7.60726$ [yellow]; (c) $a=1$, $p^{(1)}_\phi \approx 7.69421$ [blue] and $p^{(2)} \approx 7.79423$ [yellow]; (d) $a = a_1 = \sqrt{16/27} \approx 0.76980$, $p^{(1)}_\phi \approx 7.83645$ [blue] and $p^{(2)}_\phi \approx 8.7093$ [yellow]; (e) $a = 0.5$, $p^{(1)}_\phi \approx 7.93511$ [blue] and $p^{(2)}_\phi \approx 11.24300$ [yellow].}
\label{fig:contour}
\end{figure}

\subsubsection{Bounded orbits\label{subsec:bounded}}

\begin{figure}[h]
 \includegraphics[height=6cm,trim={2.5cm 2.5cm 2.5cm 2.5cm},clip]{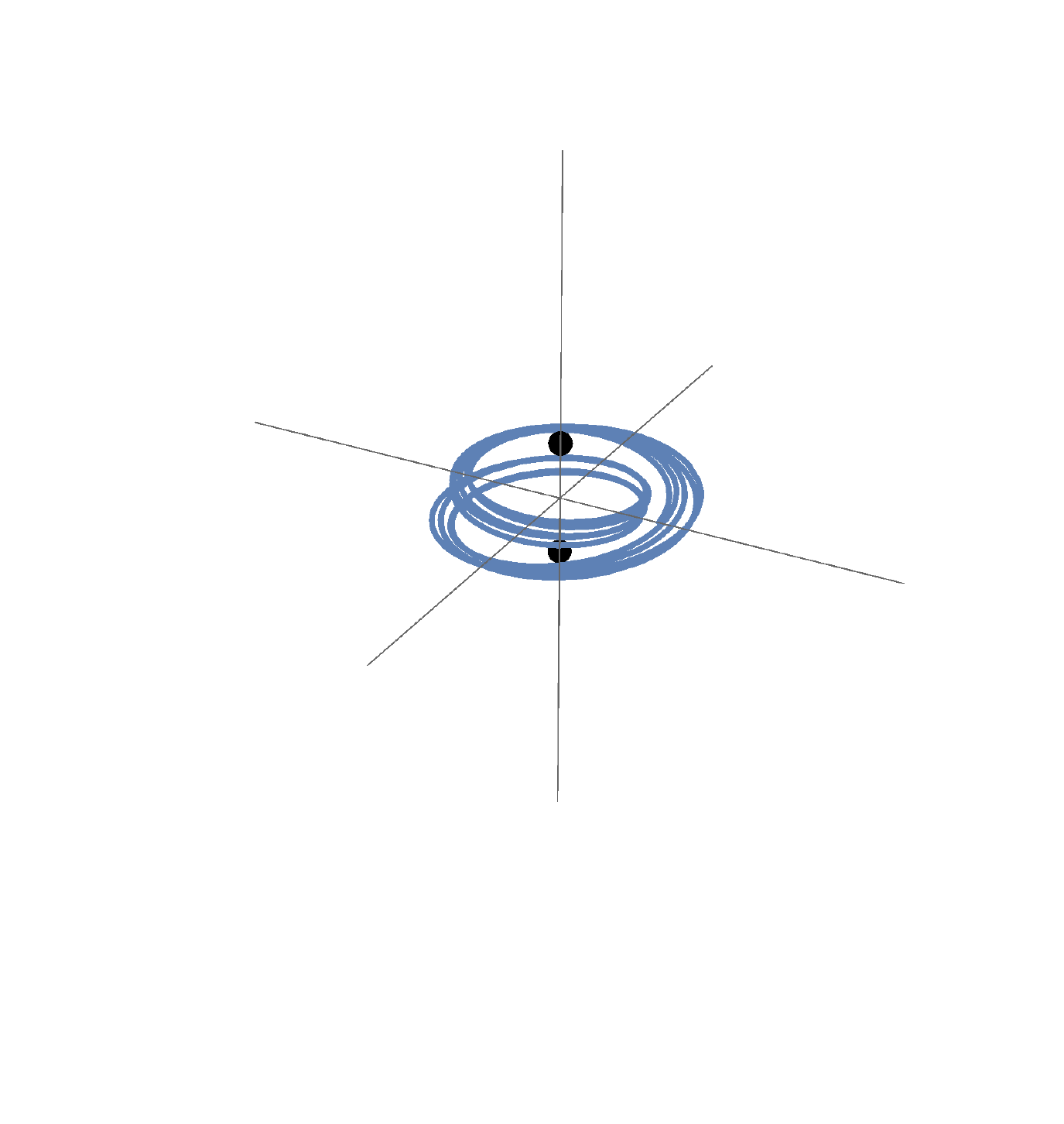}
 \hspace{1cm}
 \includegraphics[height=6cm]{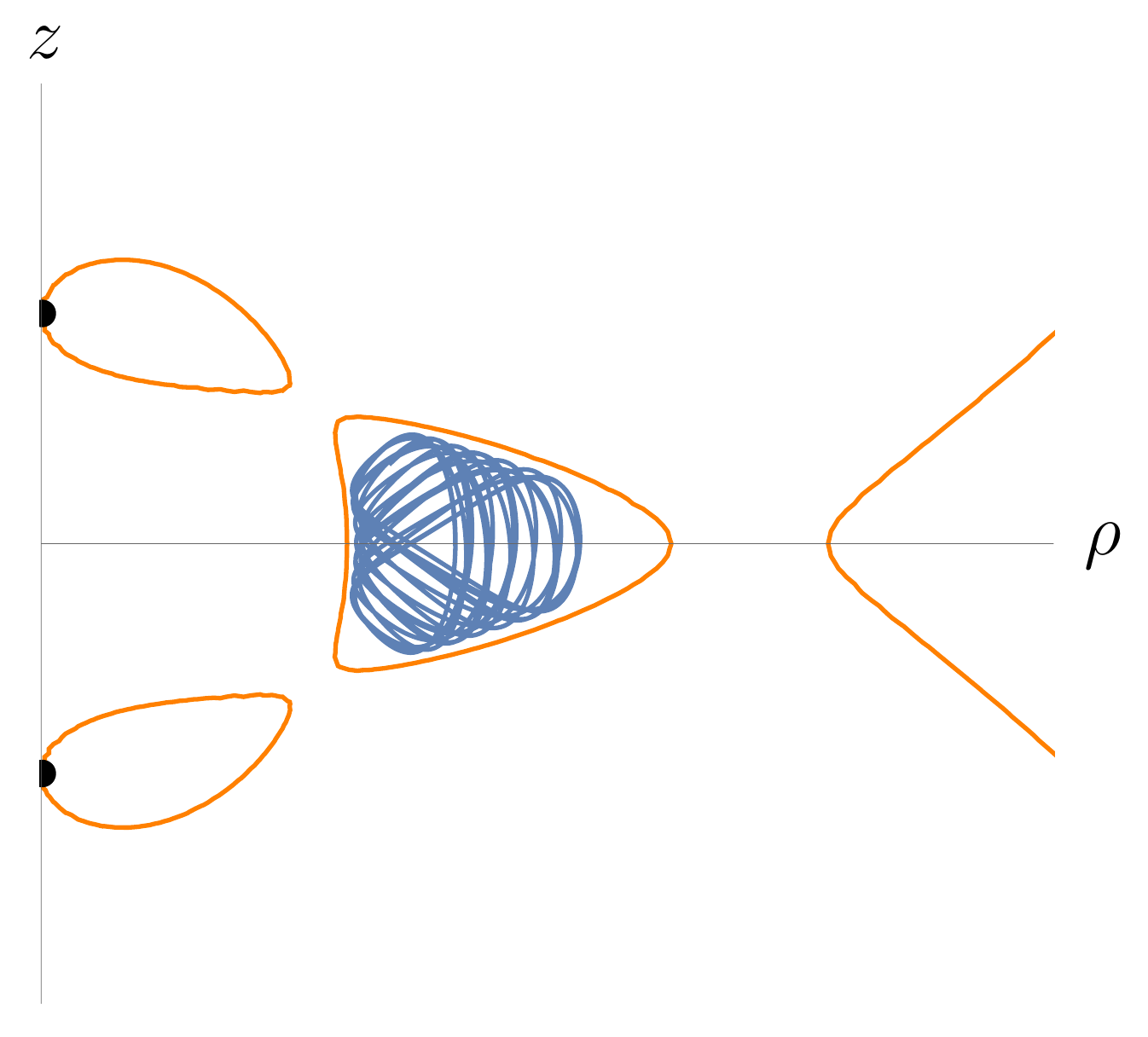}
  \caption{Example of a bounded null geodesic ($p_{\phi} = 7.70421$, $M = 1$, $a = 1$). The left-hand image is a three-dimensional plot of the null geodesic on $x$, $y$, $z$ axes. The right-hand plot shows the projection of the bounded geodesic in the $(\rho, z)$-plane [blue], with the energy surface $h \equiv \rho U^{2} = p_{\phi} = 7.70421$ shown in yellow.}
 \label{fig:bounded}
\end{figure}

Intriguingly, the existence of a maximum of $h \equiv \rho U^2$ for $a_1 < a < a_2$ implies the existence of stable \emph{bounded} null geodesics which are confined to a compact region of the $(\rho, z)$-plane. For $p_\phi > p_\phi^c$, this region is completely inaccessible to scattering trajectories that encroach from $\rho \rightarrow \infty$. An example of a bounded null geodesic is shown in Fig.~\ref{fig:bounded}.

\begin{figure}[h]
 \includegraphics[width=6.5cm]{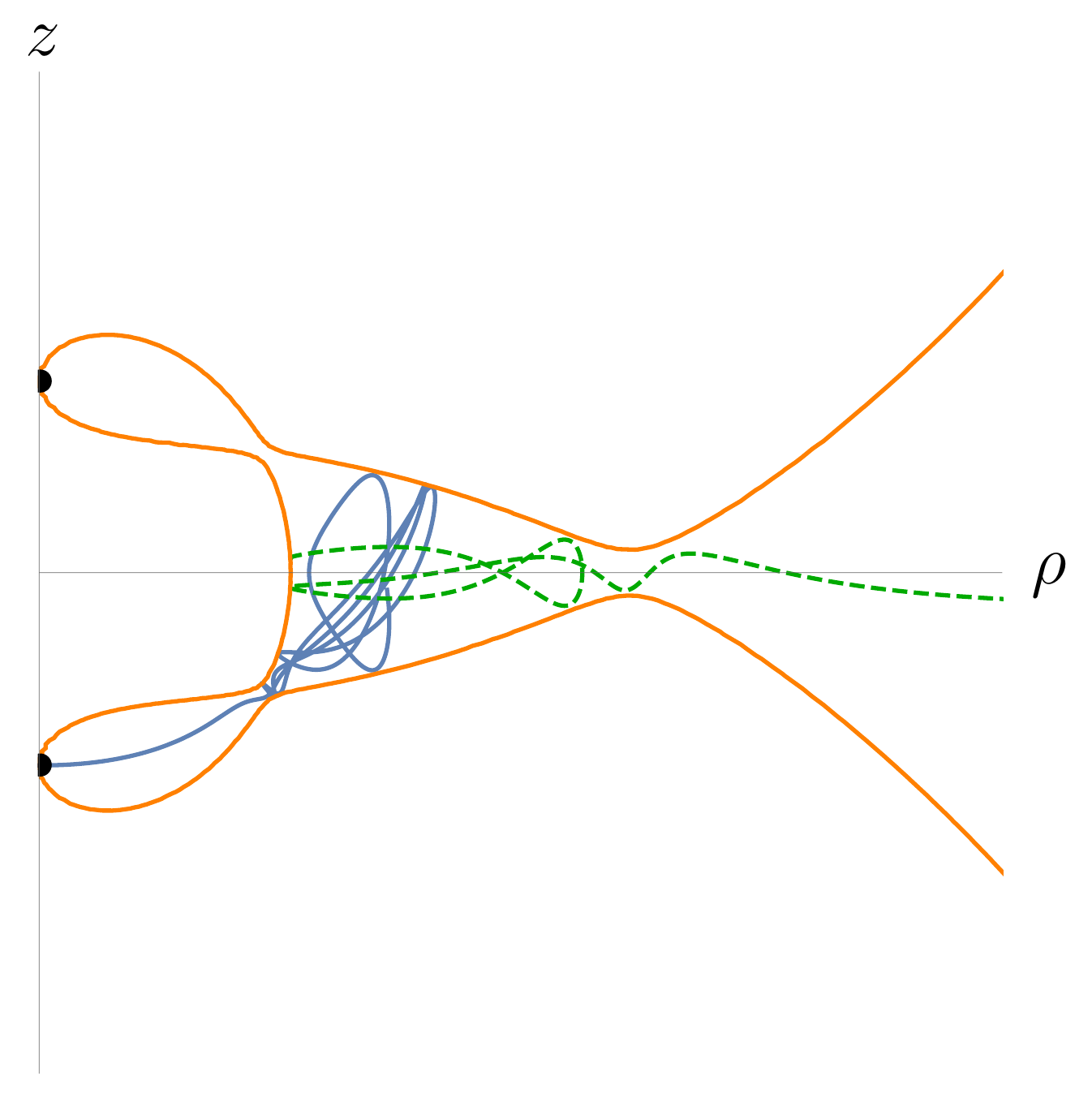}
  \caption{Null rays in the `pocket' with three throats for $a = 1$, $p_\phi = 7.69$. This unexpected feature generates qualitatively different chaotic behaviour (see text).}
 \label{fig:threethroats}
\end{figure}

By decreasing $p_\phi$ slightly from $p_\phi^c$, one may construct a `pocket' in the $(\rho, z)$-plane which is connected to the black holes horizons and to spatial infinity by narrow throats. An example of this case for $a=1$ is shown in Fig.~\ref{fig:threethroats}. Qualitatively different chaotic dynamics is associated with this feature, and the `decision dynamics' of Sec.~II is not a suitable framework. Furthermore, for $a < 1$, it is possible to connect the pocket to the black holes, without also connecting to infinity. For $a > 1$, the pocket may be connected to infinity, without connecting to the black holes.

We recall that the potential for the H\'enon--Heiles system \cite{Henon:1964}, $V(x,y) = \frac{1}{2}(x^2+y^2) +x^2 y - \frac{1}{3} y^3$, also has a single contour $V_c=1/6$ connecting three saddle points. Thus, for Hamiltonian values $H = V_c + \eps$, the H\'enon--Heiles system also has a pocket with three throats. It has been shown \cite{Aguirre:2001} that this system exhibits the Wada property (cf.~the Lakes of Wada, described by K.~Yoneyama in 1917 \cite{Yoneyama:1917}): any point on the boundary of one escape basin  is also on the boundaries of the other two basins \cite{Kennedy:1991, Poon:1996, Sweet:1999}. It has been argued that the Wada property is a general feature associated with compact regions of phase space with three or more escapes \cite{Poon:1996, Sweet:1999, Aguirre:2001}. We anticipate that our `pocket' system will inherit this property; however, this remains to be investigated.

To our knowledge, bounded null geodesics have not been explored in a binary model before (though see Refs.~\cite{CoelhoHerdeiro2009, Wunsch:2013st}). However, bounded null geodesics are known to arise in certain `singleton' contexts: for instance, inside Kerr--Newman black holes and around naked singularities \cite{Liang:1974, Stuchlik:1981, Balek:1989}; in non-asymptotically flat black hole spacetimes \cite{Stuchlik:2002}; around 5D black rings \cite{Igata:2013be}; in the exterior of hairy black holes solutions \cite{Cunha:2015yba}; and in ultra-compact horizonless systems \cite{Schee:2015nua, Cardoso:2014sna}. Such orbits will be investigated in further detail in Ref.~\cite{Dolan:2016bxj}.

 \subsection{Two-dimensional shadows\label{subsec:2D}}

Let us now turn our attention to the 2D shadow cast by a pair of black holes in the MP spacetime. We define the shadow with respect to a null congruence passing orthogonally through a planar surface with its central point at $\rho_0$, $z_0$, where $\sqrt{\rho_0^2 + z_0^2} = r_{\text{max}}$ (typically, we use $r_{\text{max}} = 50M$). We define the angle of incidence via $\sin \theta = \rho_0 / r_{\text{max}}$, $\cos \theta = -z_0 / r_{\text{max}}$.

\subsubsection{A gallery of shadows\label{subsec:gallery}}
\begin{figure}[h]
 \vspace{-0.2cm}
 \includegraphics[width=3.8cm]{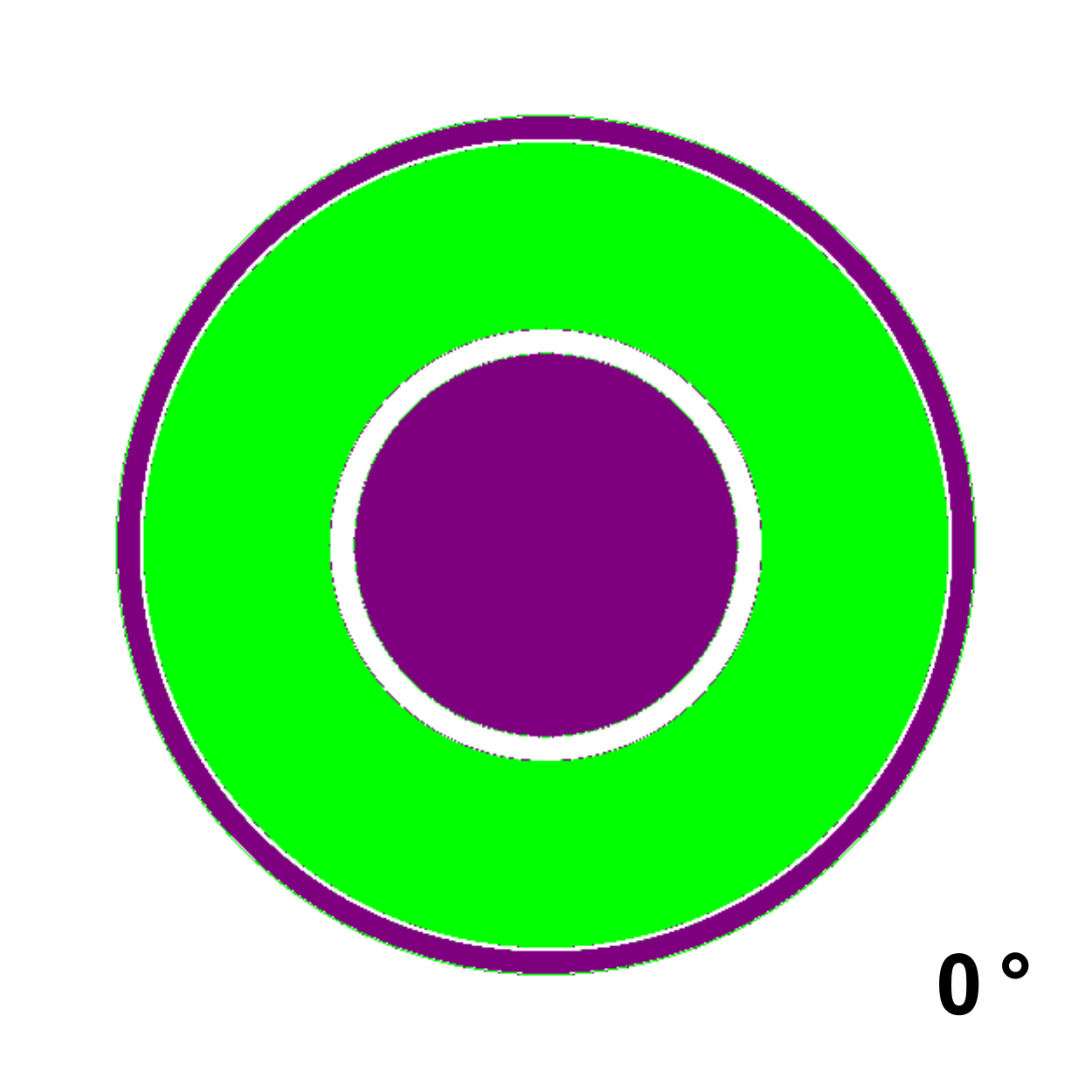}
 \includegraphics[width=3.8cm]{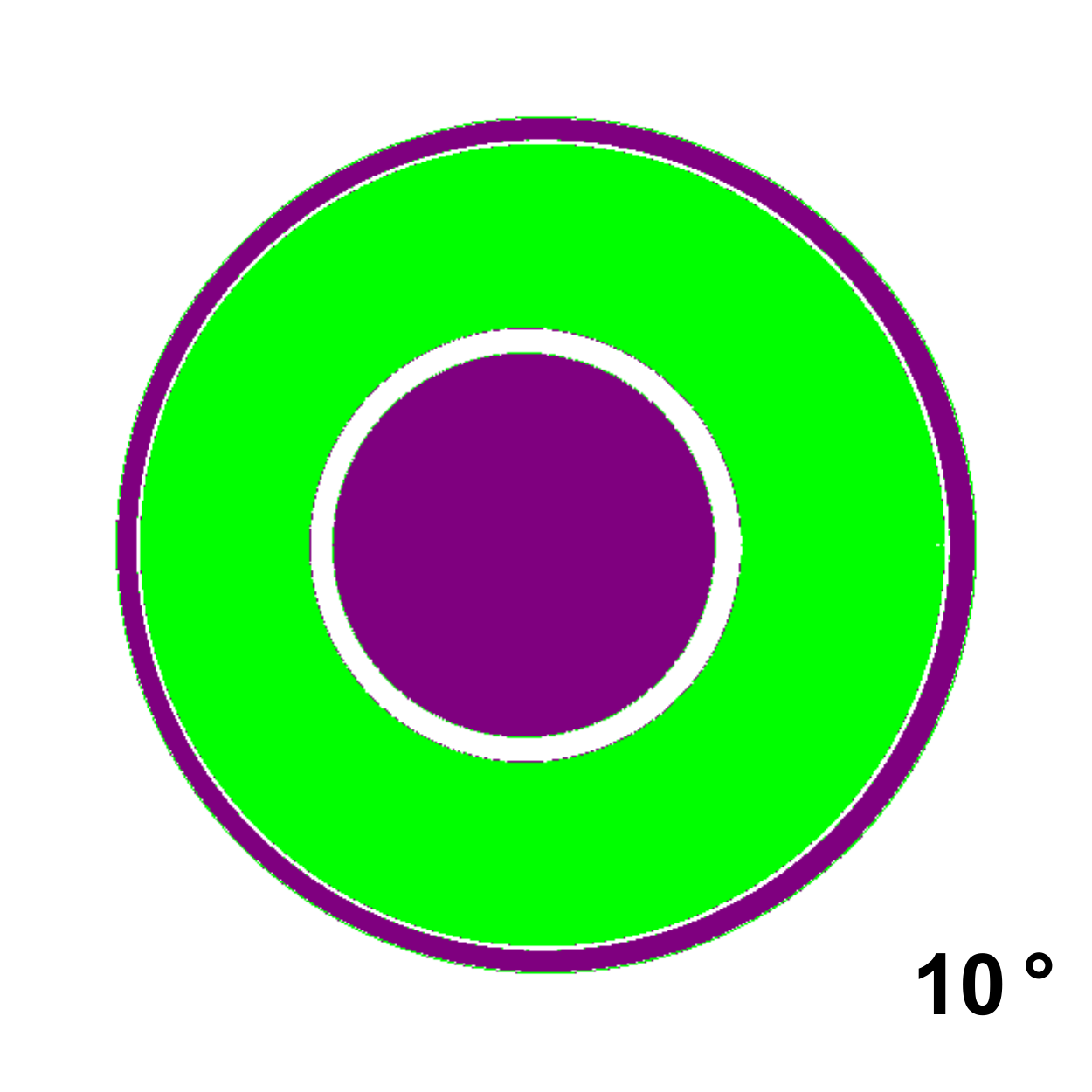}
 \includegraphics[width=3.8cm]{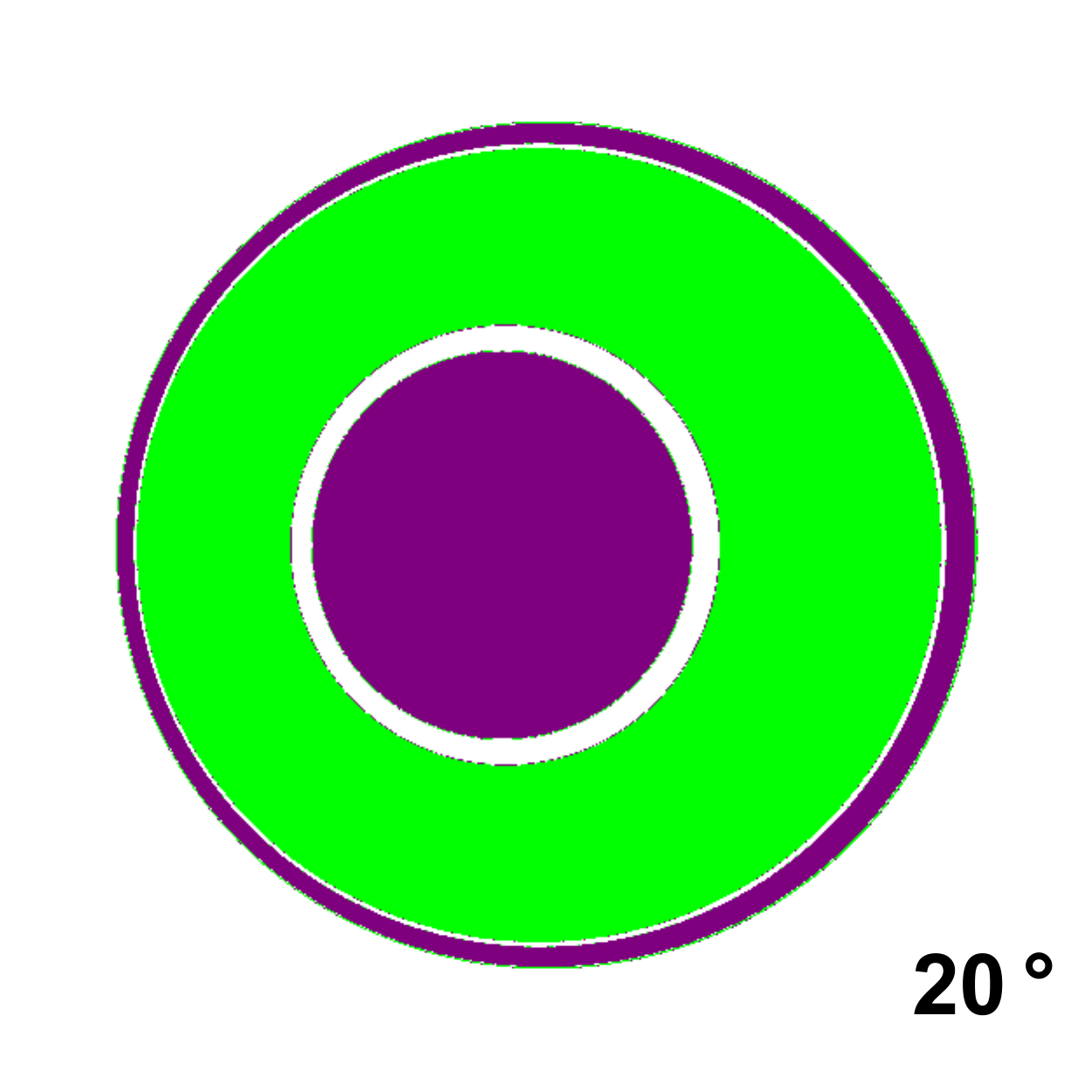}
 \includegraphics[width=3.8cm]{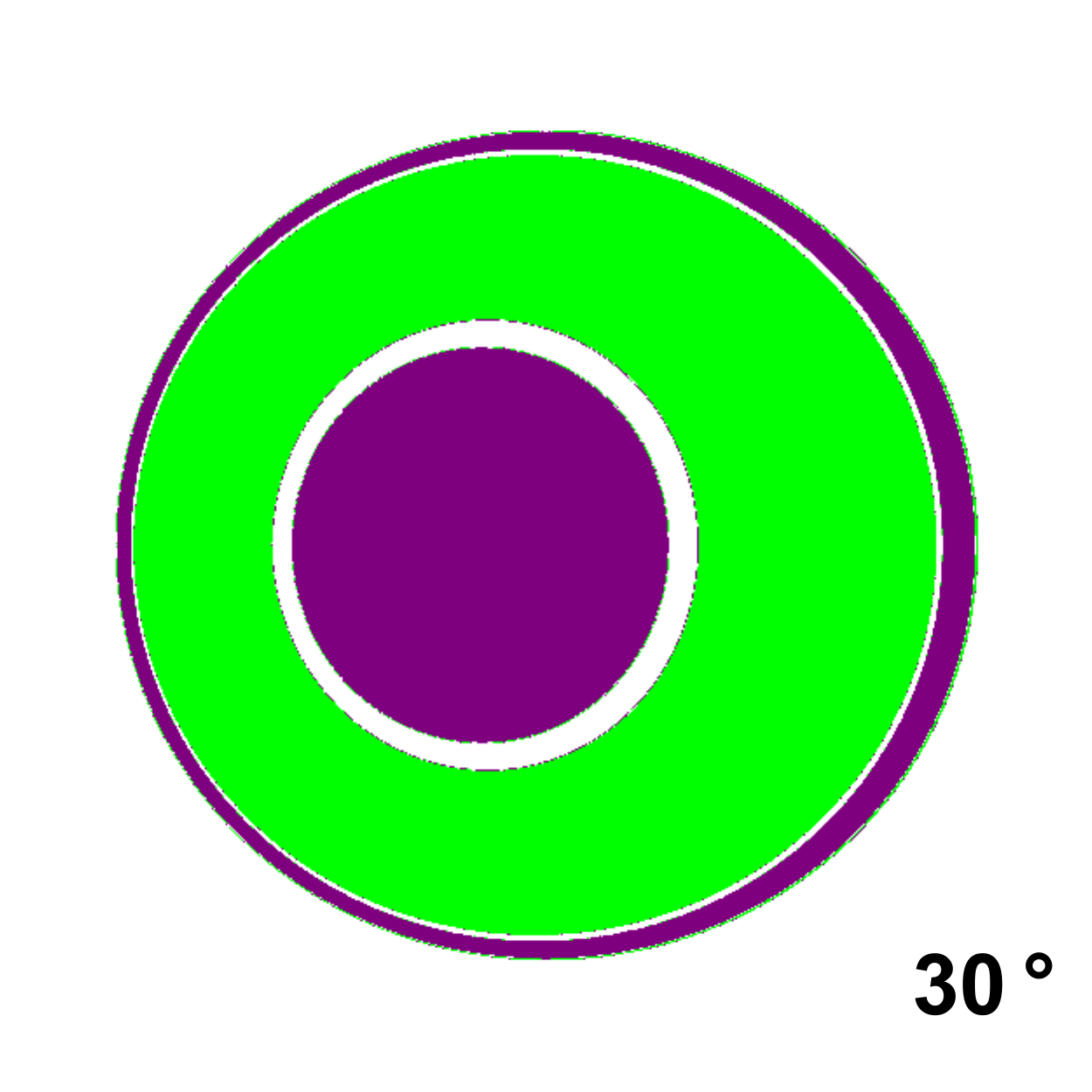}
  \vspace{-0.2cm}
 \includegraphics[width=3.8cm]{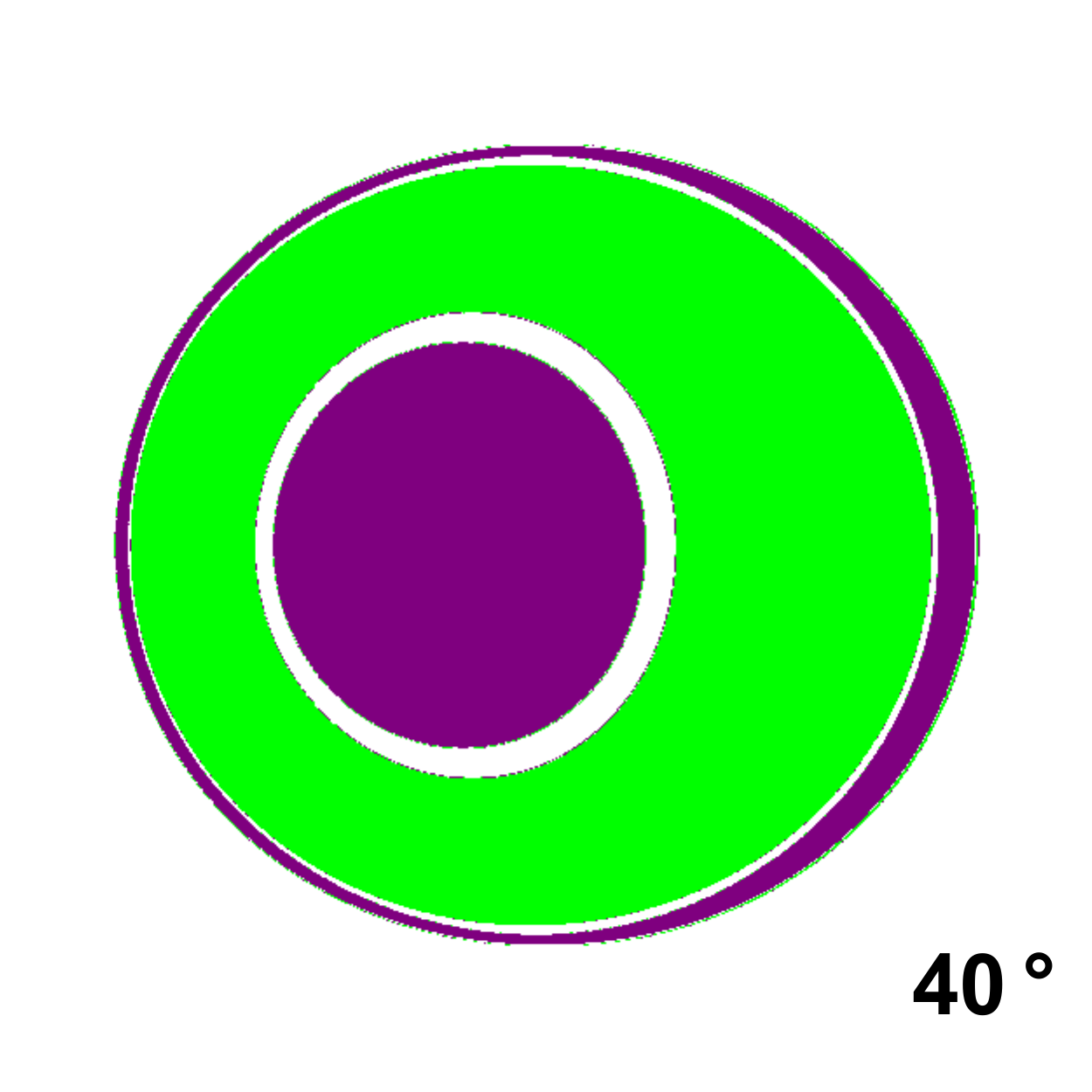}
 \includegraphics[width=3.8cm]{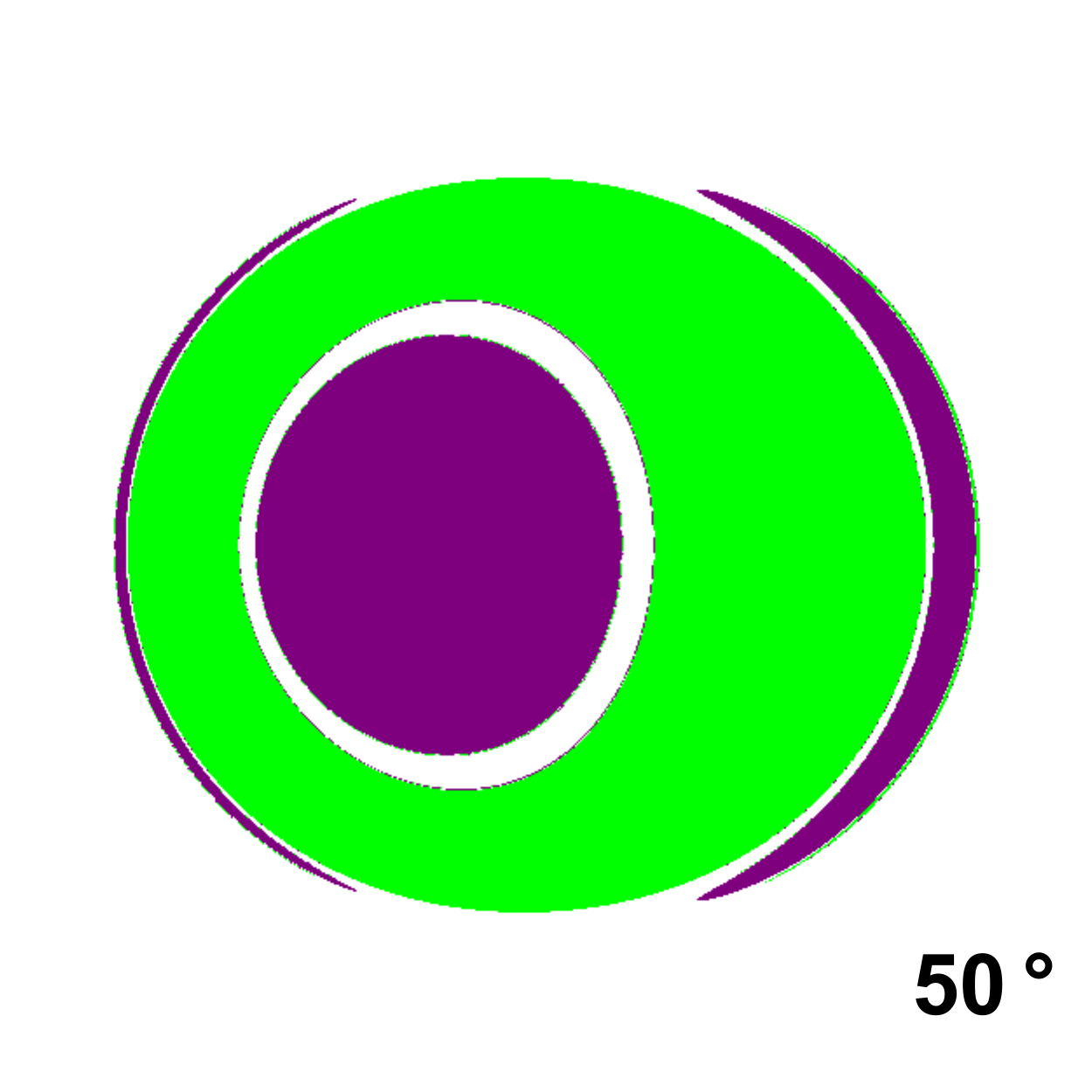}
 \includegraphics[width=3.8cm]{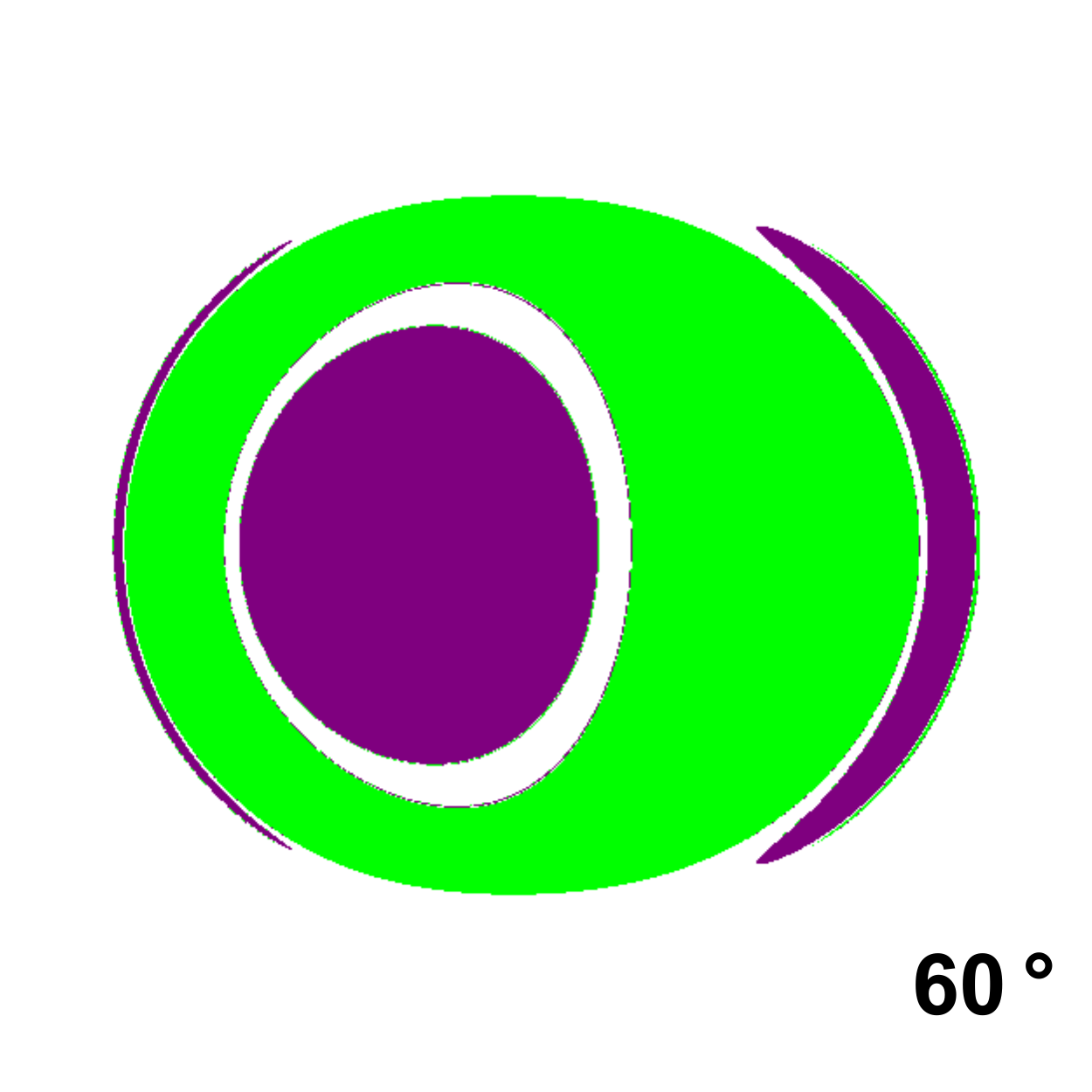}
 \includegraphics[width=3.8cm]{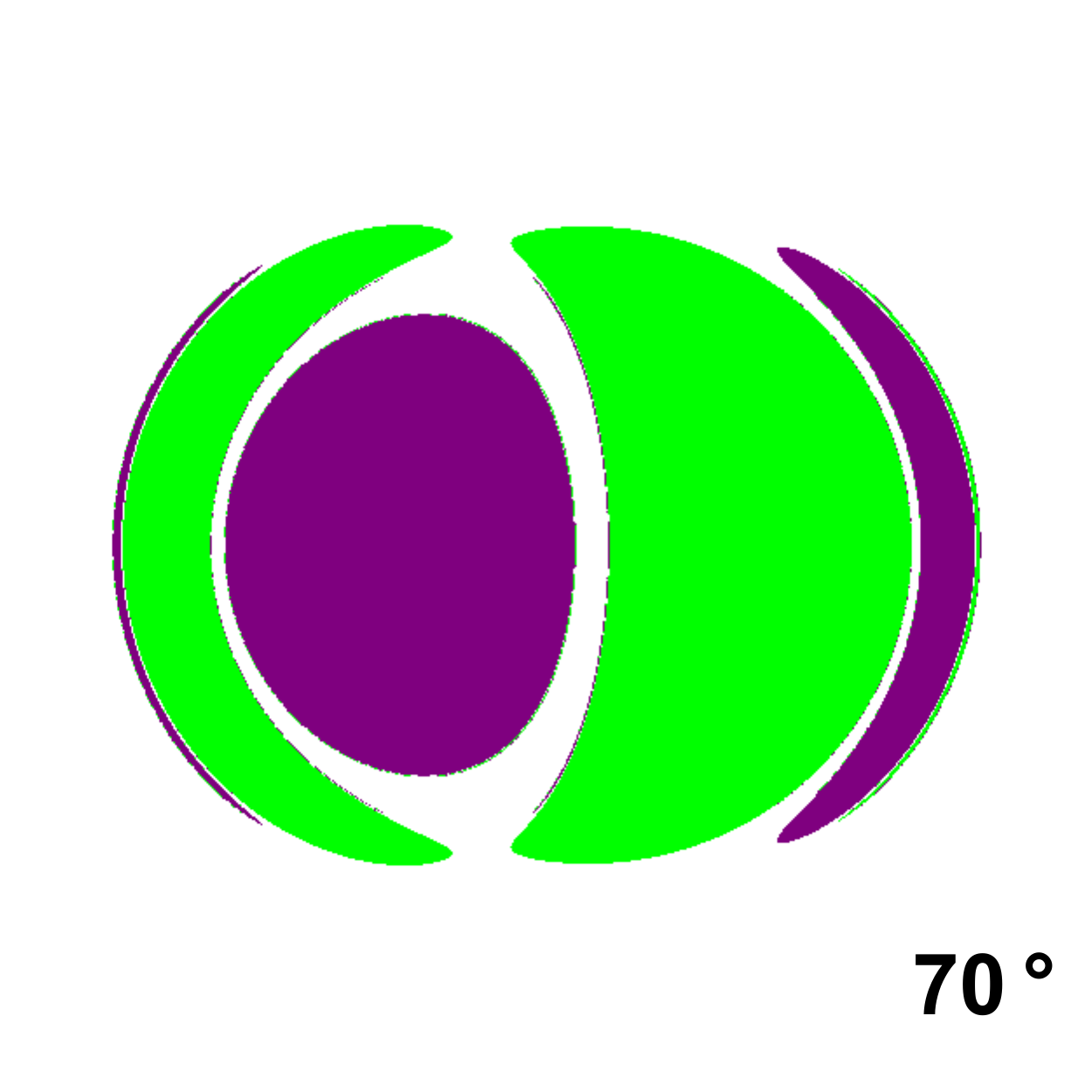}
  \vspace{-0.2cm}
 \includegraphics[width=3.8cm]{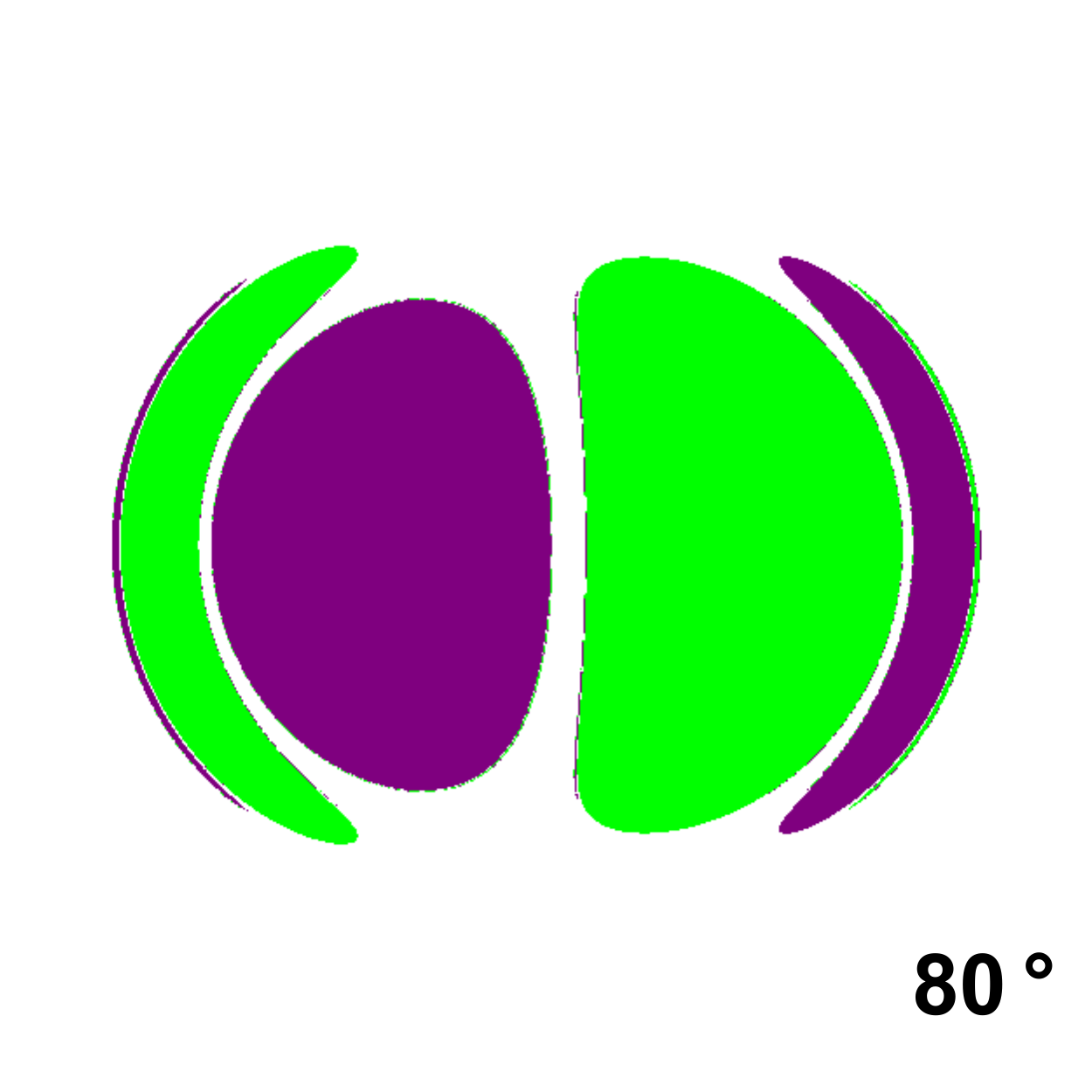}
 \includegraphics[width=3.8cm]{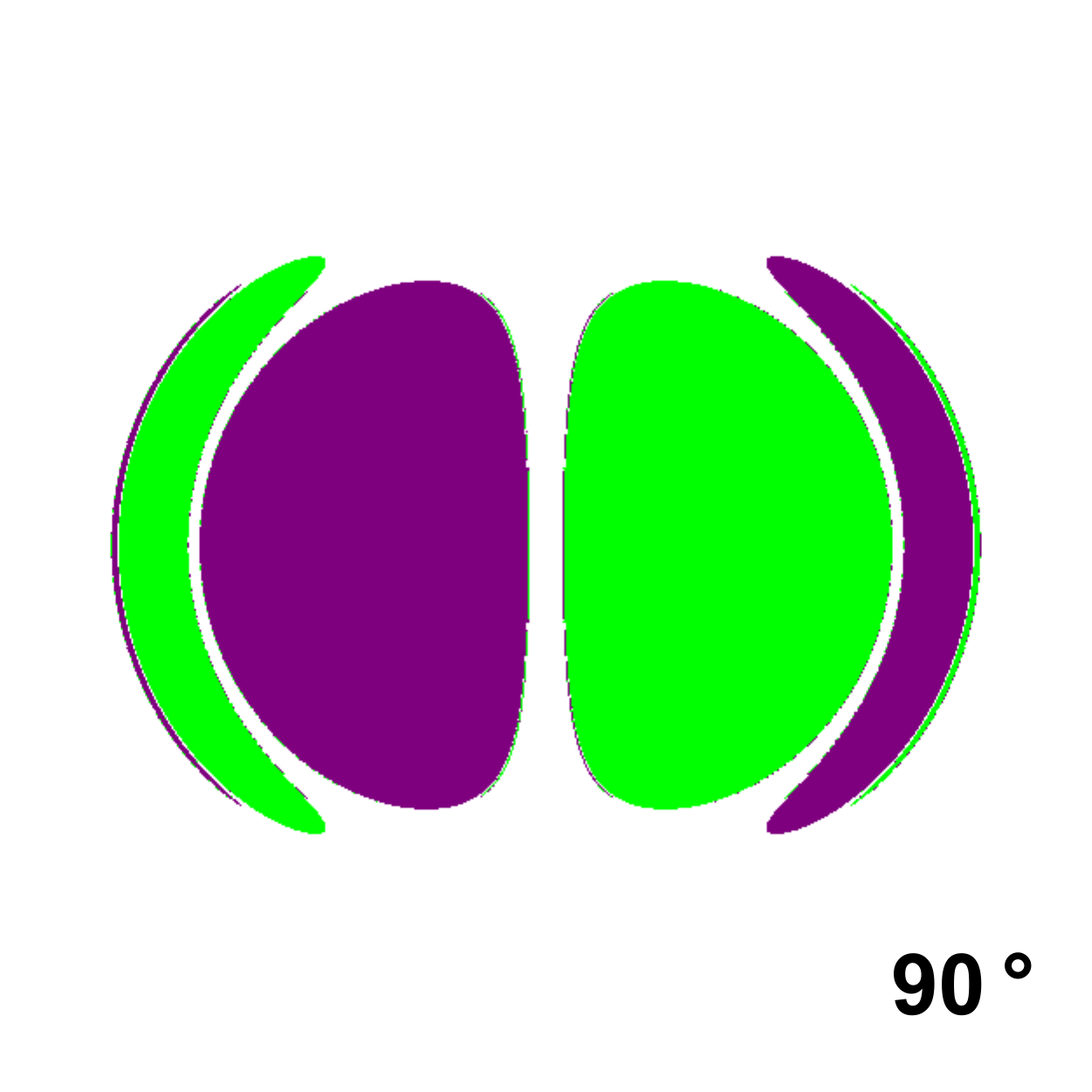}
 \includegraphics[width=3.8cm]{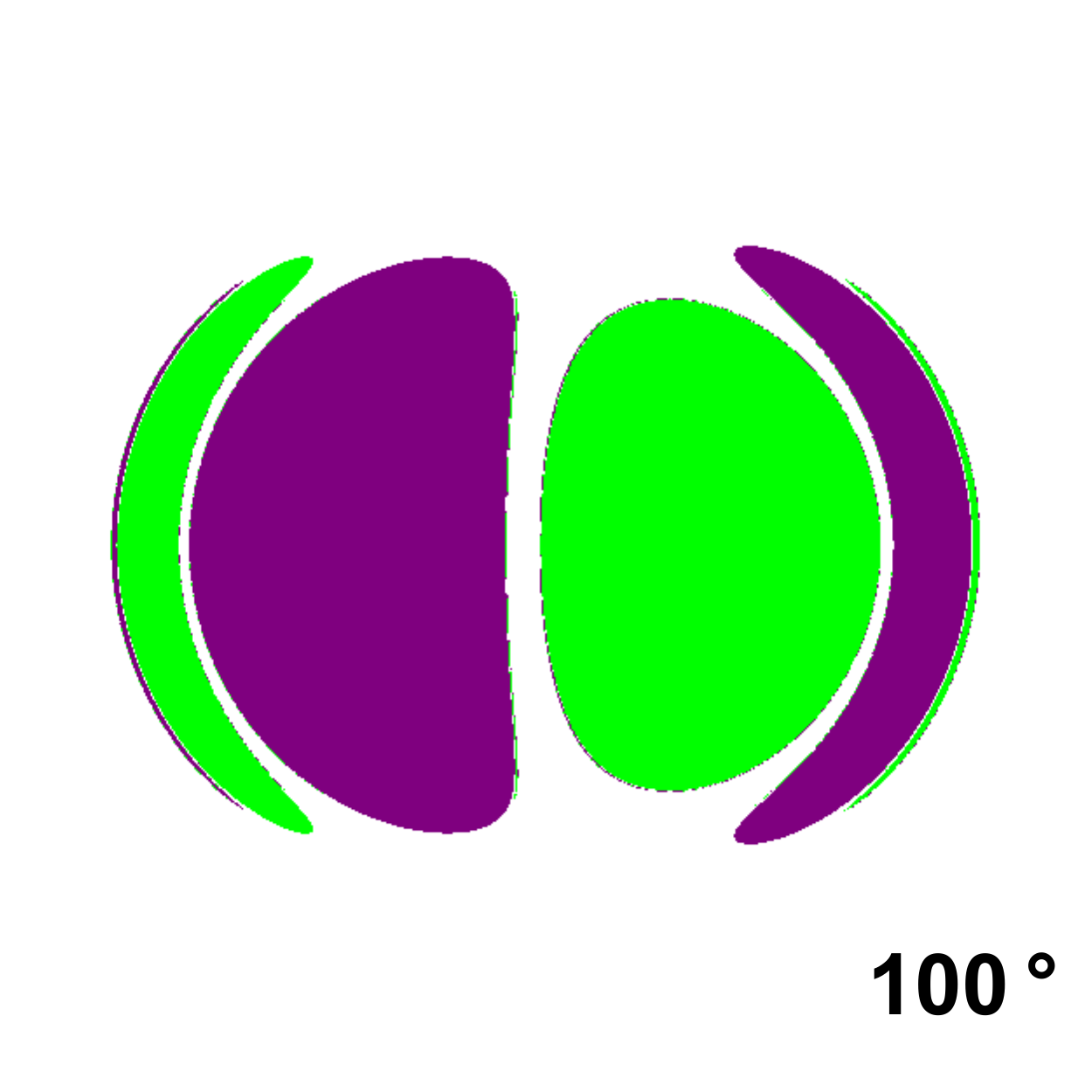}
 \includegraphics[width=3.8cm]{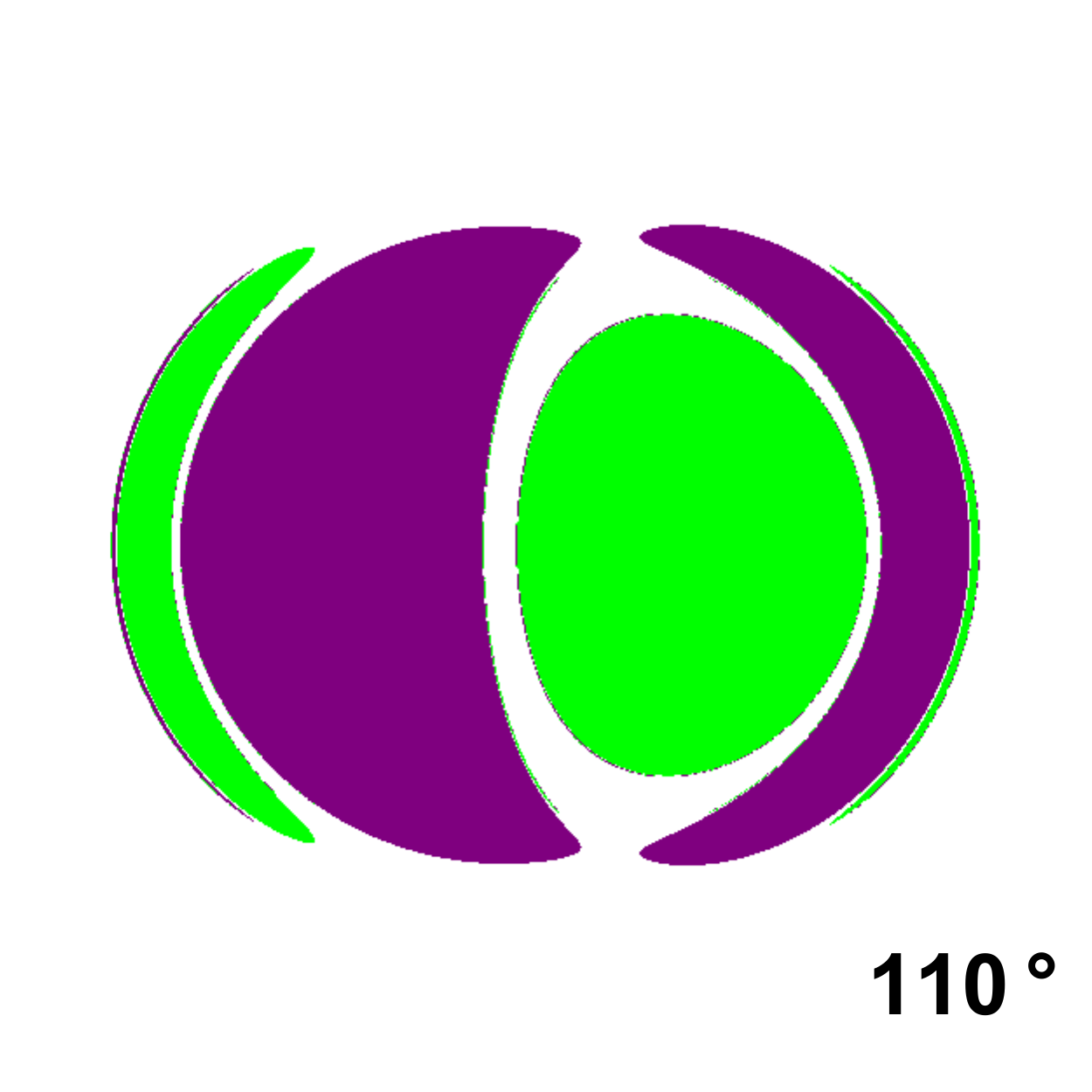}
  \vspace{-0.2cm}
 \includegraphics[width=3.8cm]{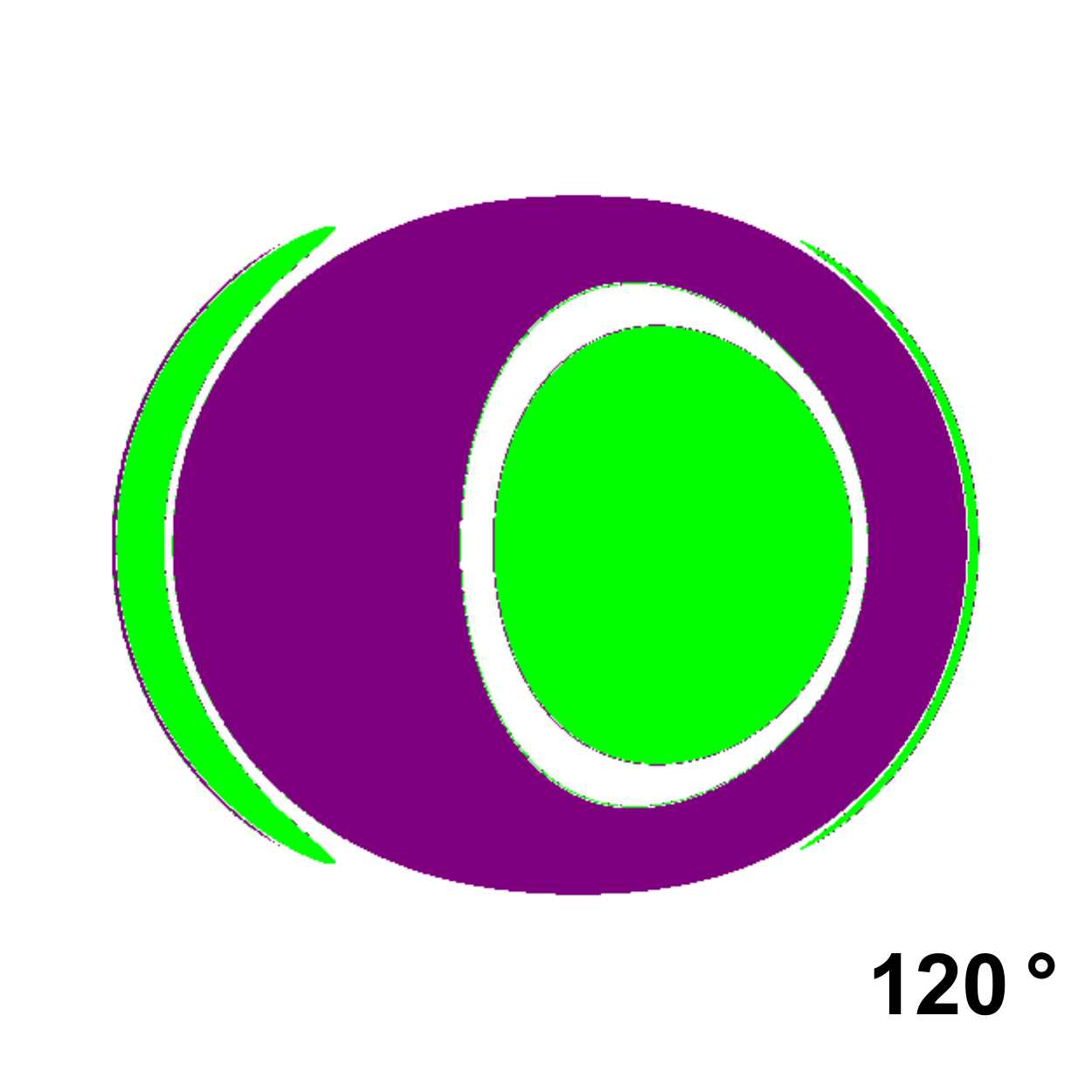}
 \includegraphics[width=3.8cm]{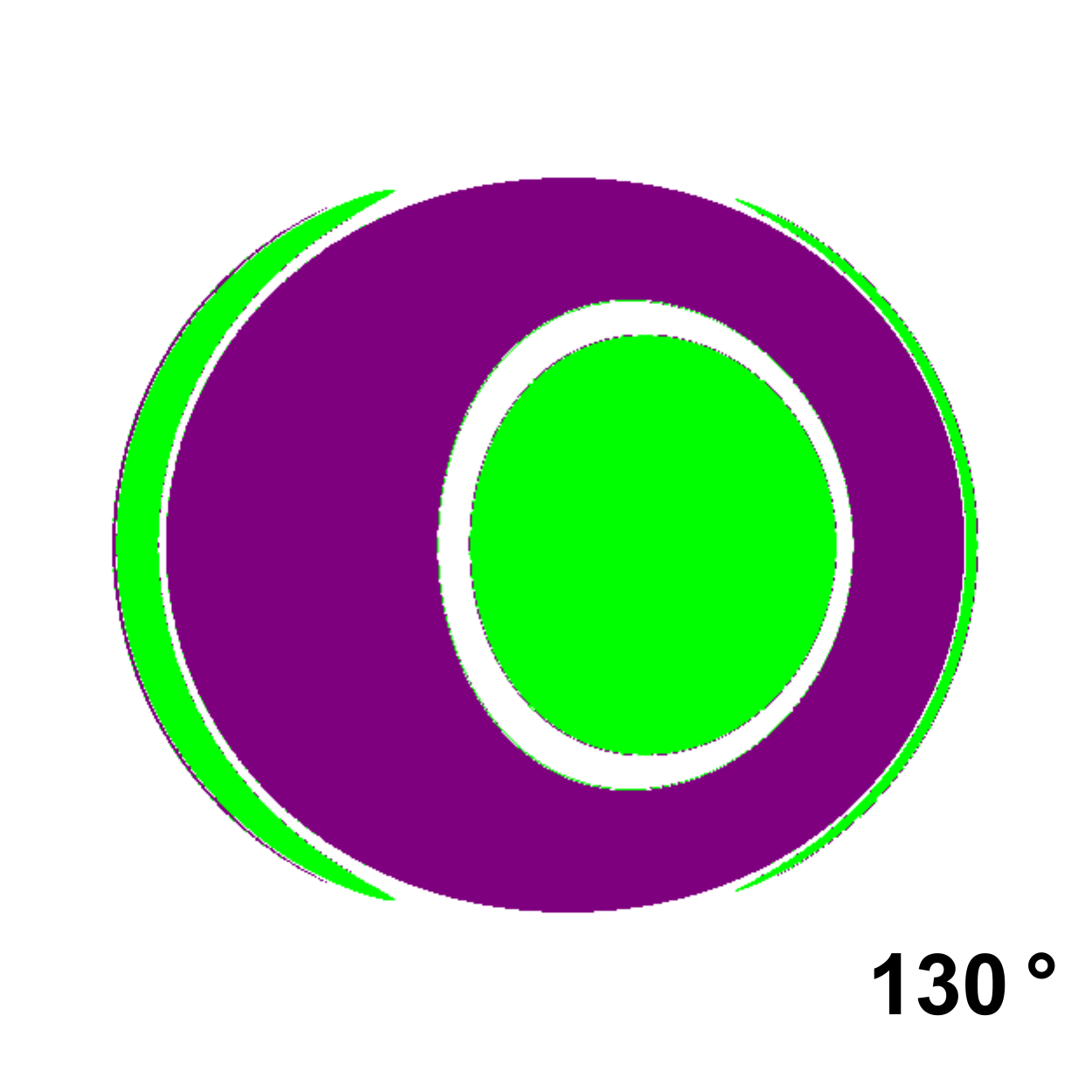}
 \includegraphics[width=3.8cm]{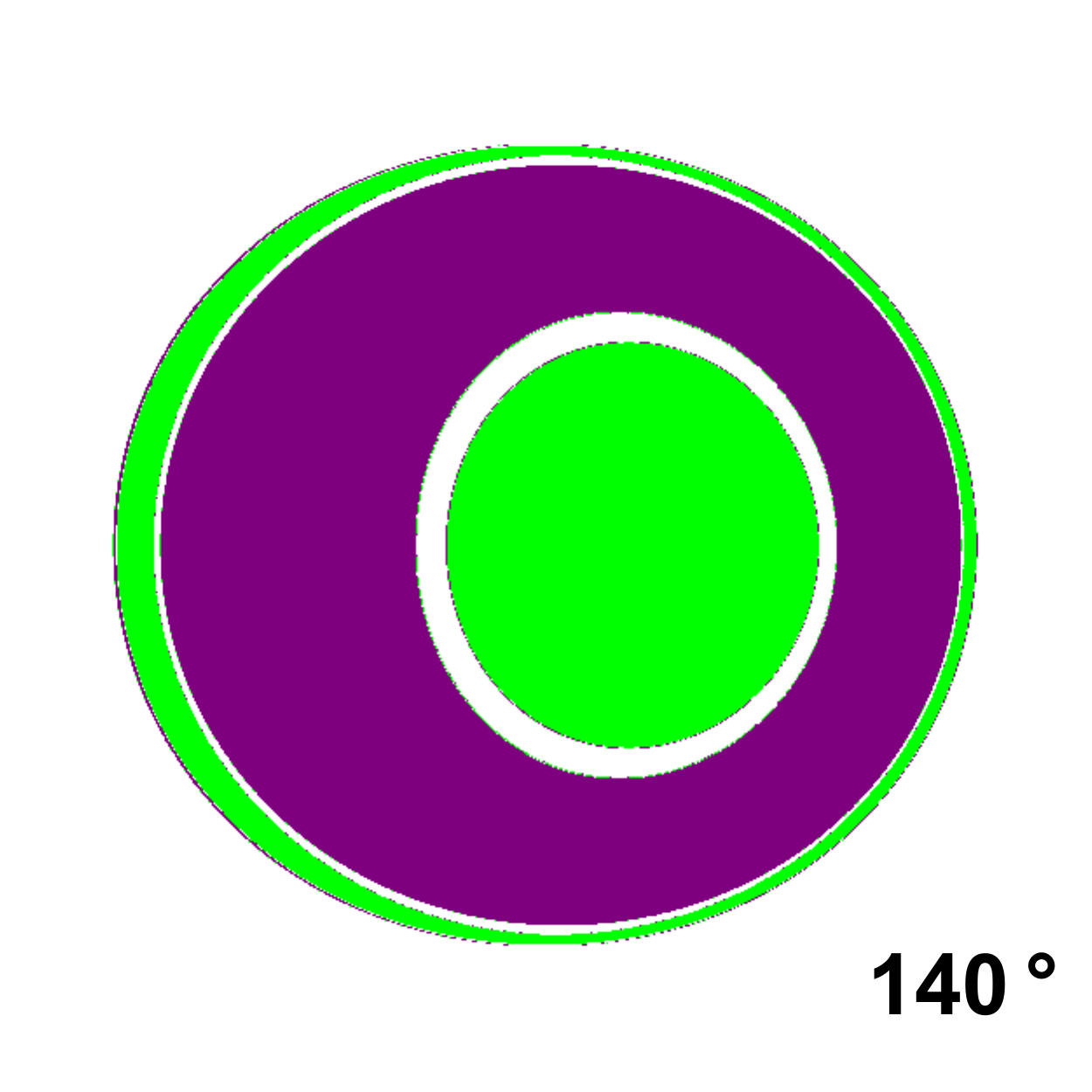}
 \includegraphics[width=3.8cm]{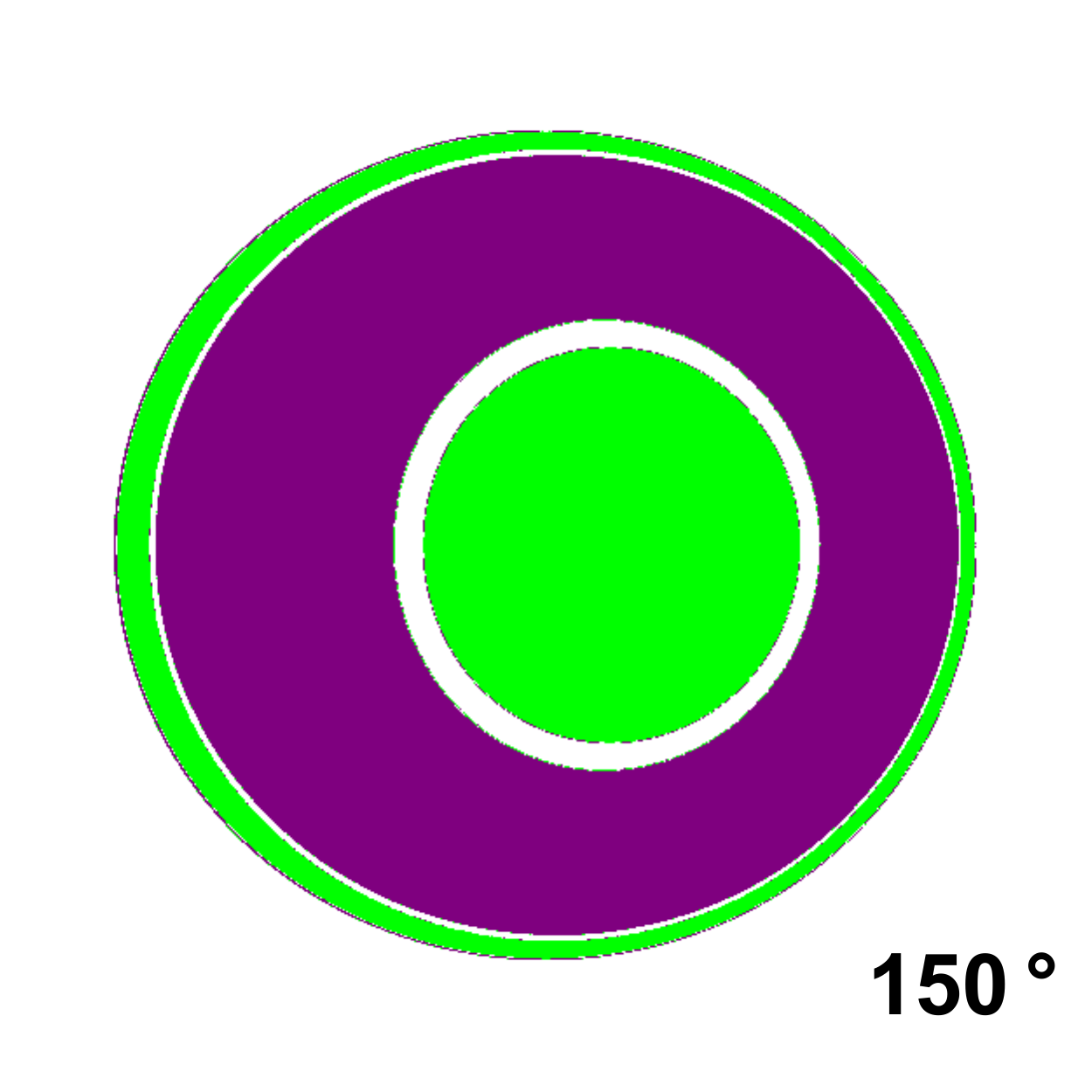}
  \vspace{-0.2cm}
 \includegraphics[width=3.8cm]{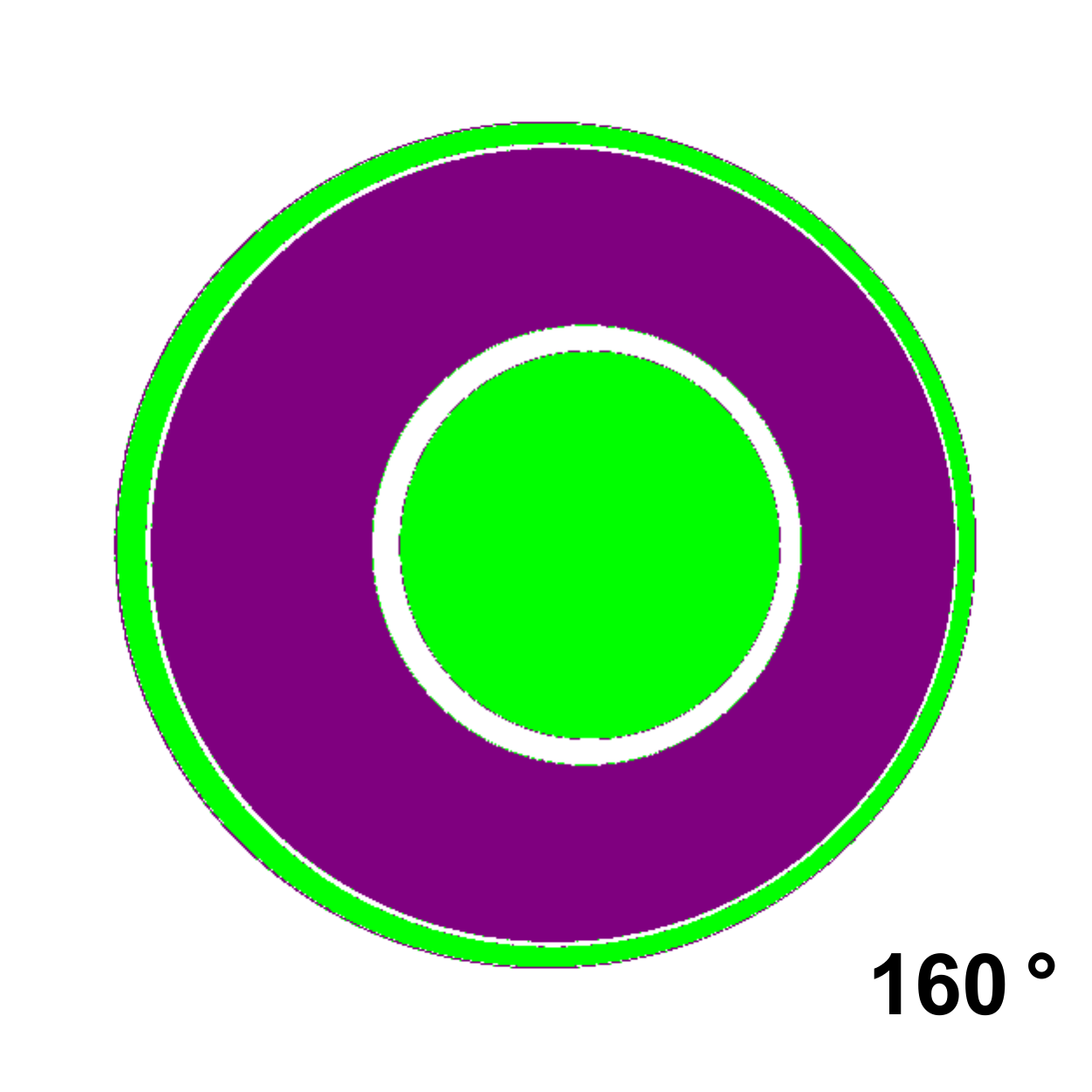}
 \includegraphics[width=3.8cm]{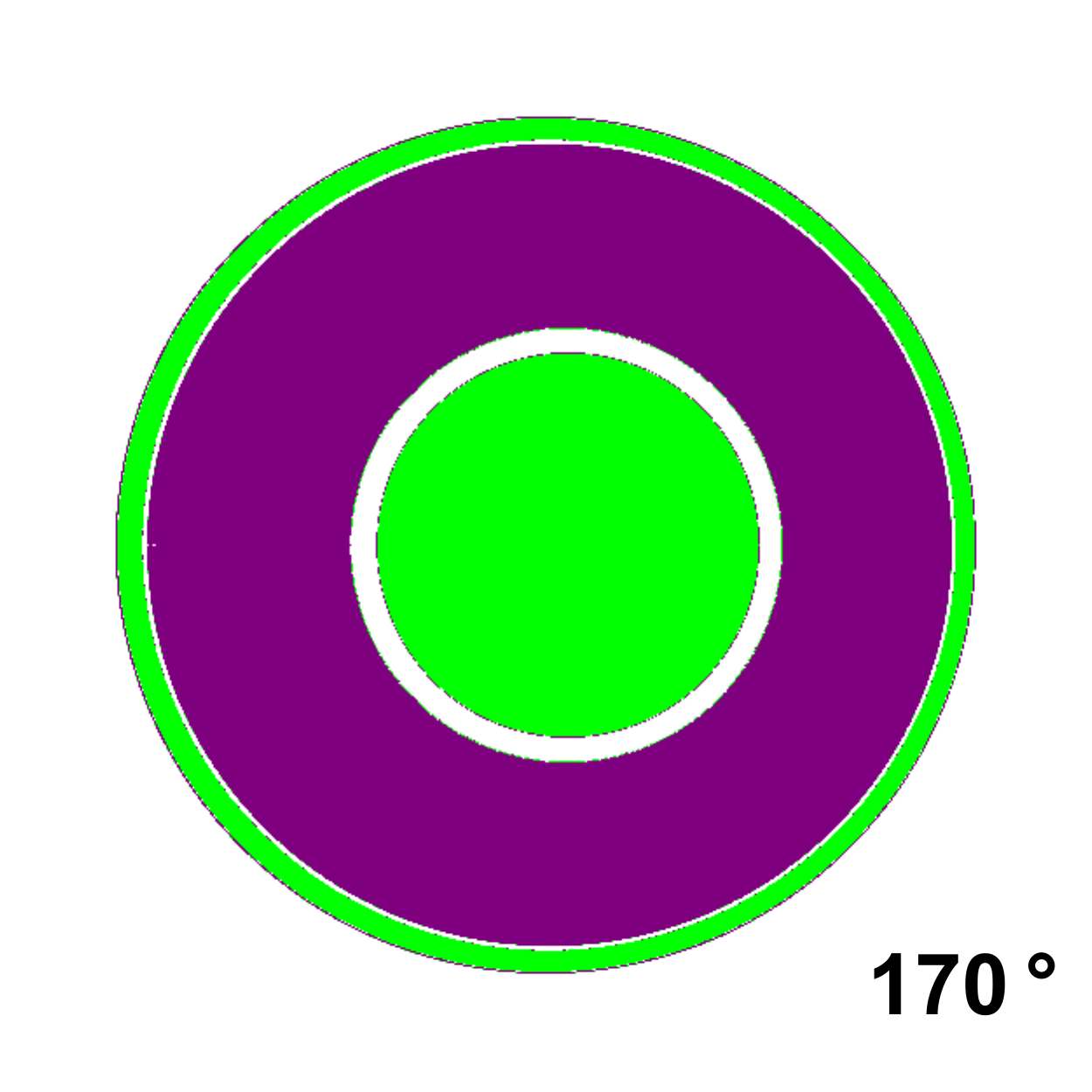}
 \includegraphics[width=3.8cm]{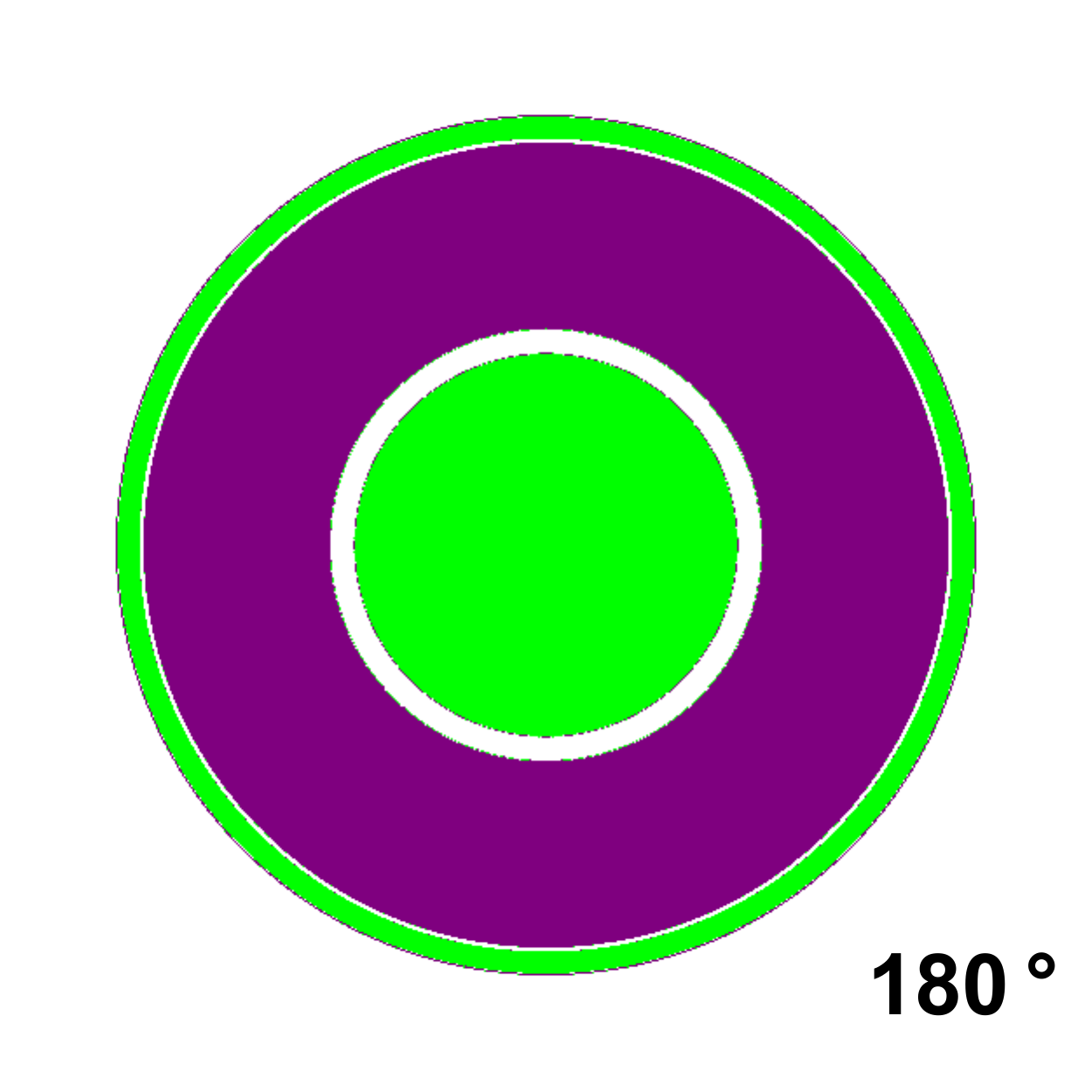}
 \caption{{\footnotesize Shadows cast by a pair of equal-mass extremal black holes of mass $M=1$ separated by $a=2$. 
  Rays falling into the lower (upper) black hole are shown in purple (green).}}
\label{fig:shadows}
\end{figure}

Figure~\ref{fig:shadows} shows the shadows cast by a pair of black holes separated by $a=2$ for various values of the incidence angle $\theta$ (measured from the negative $z$-axis in the anticlockwise direction). Regions coloured purple (green) correspond to the shadow of the lower (upper) black hole. We see that, as anticipated, the shadow of a binary system is not simply the superposition of two singleton shadows, but rather that each black hole has primary shadow -- either ring-shaped or globular -- as well as a hierarchy of secondary features.

As the angle of incidence in Fig.~\ref{fig:shadows} increases, the shadow is distorted but remains qualitatively similar to the $\theta = 0^\circ$ case until we reach $\theta \sim 50^{\circ}$. Here, we clearly see the eyebrow-like features \cite{Nitta:2011in} for the first time. When the angle of incidence is equal to $70^{\circ}$, the green ring-shaped feature has split into two distinct regions; one of which forms the main shadow cast by the upper black hole, whilst the other turns into an eyebrow as we continue to increase the angle of incidence.

In the case $\theta=90^\circ$, in which we regard the system from ``side on'', we observe two main globular shadows of equal size and shape, as well as a Cantor-like hierarchy of self-similar eyebrows around the main shadow (see Fig.~\ref{fig:binaryshadow} for a close-up). This case was also shown in Fig.~2 of Yumoto {\it et al.} \cite{Yumoto:2012kz}.

\subsubsection{On-axis: the $\theta = 0^\circ$ case}
\begin{figure}[h]
\begin{tabular}{c}
{\includegraphics[width=6cm]{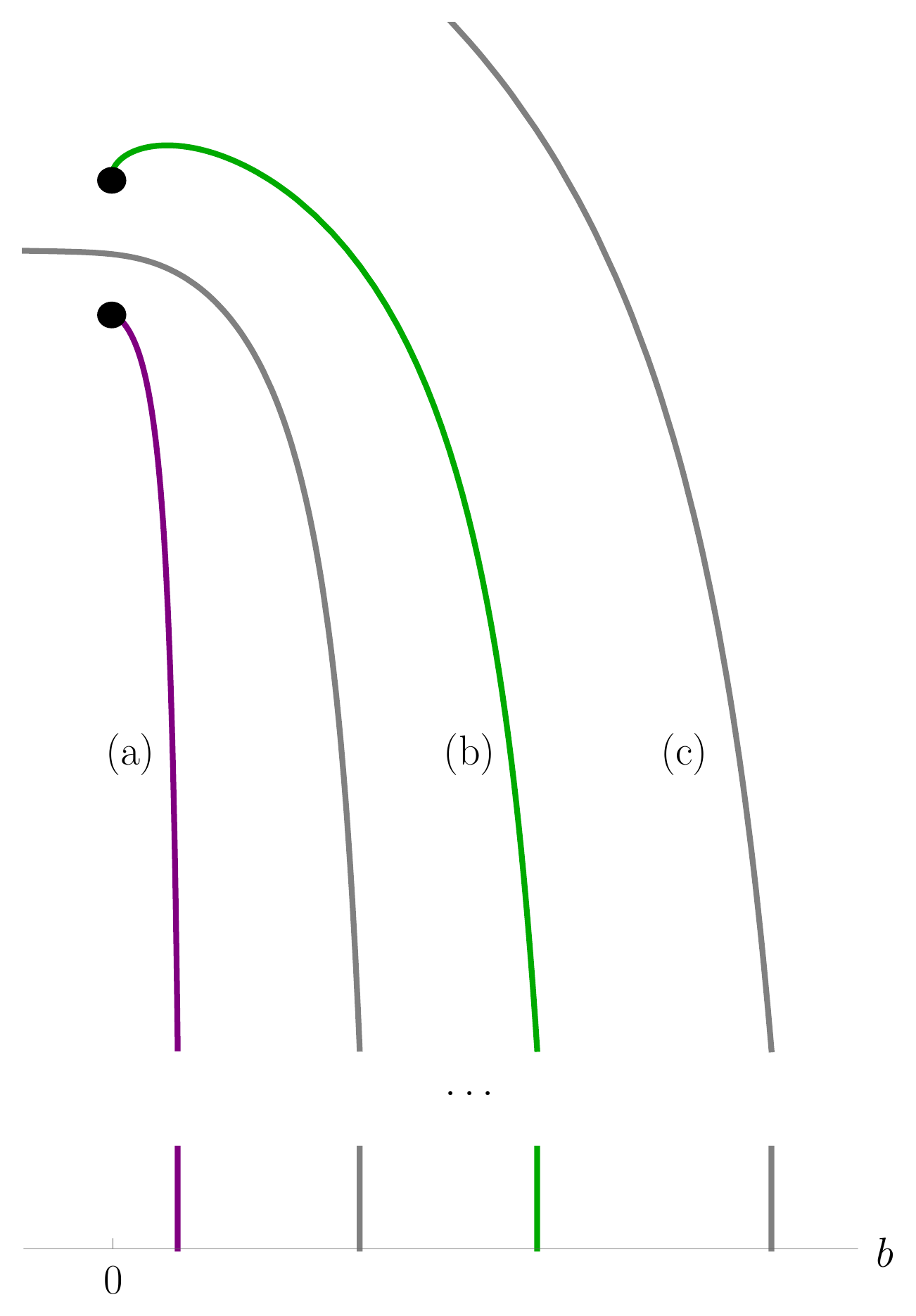}}
\end{tabular}
\begin{tabular}{c}
{\includegraphics[width=6cm]{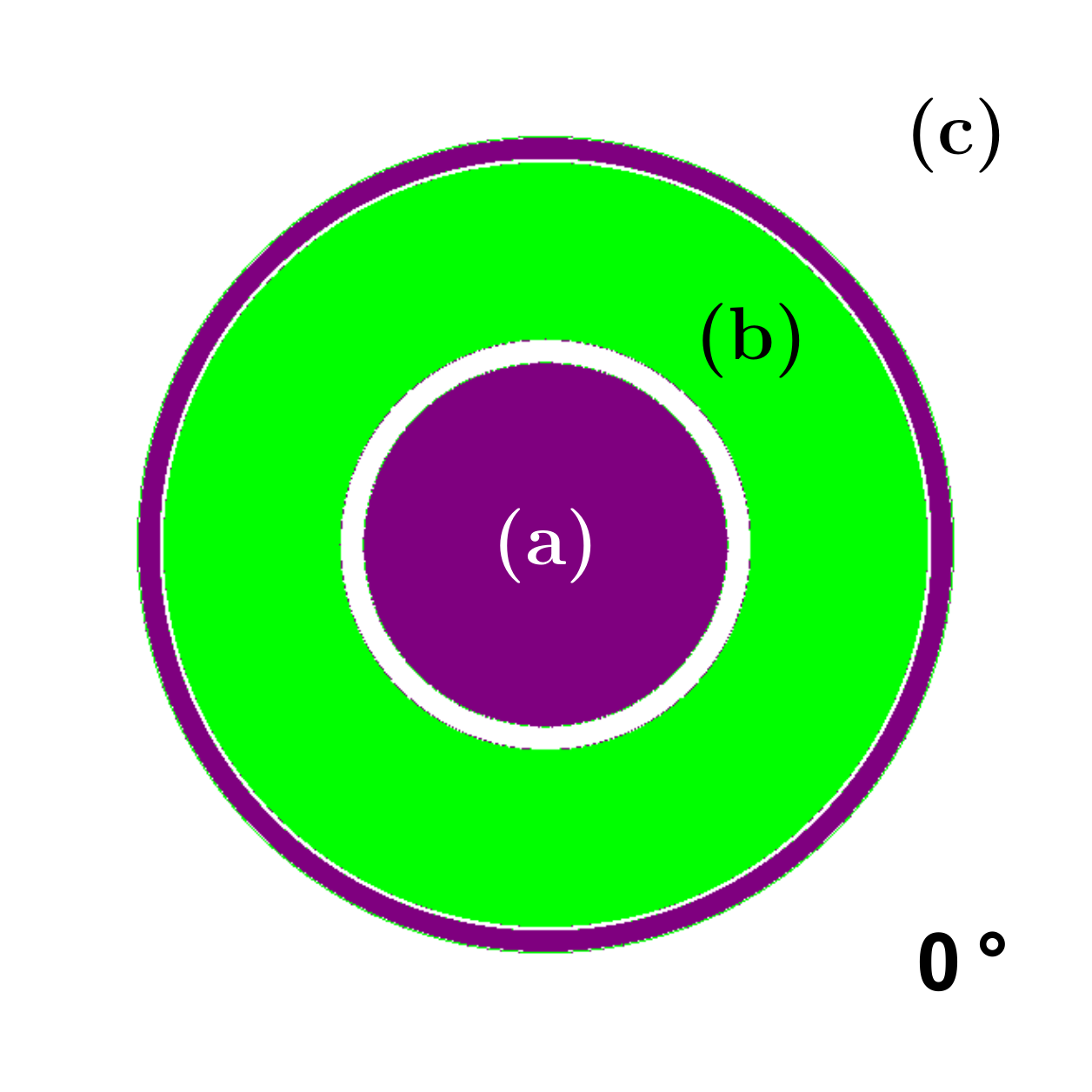}\label{fig:0deg}} \\
{\includegraphics[width=6cm]{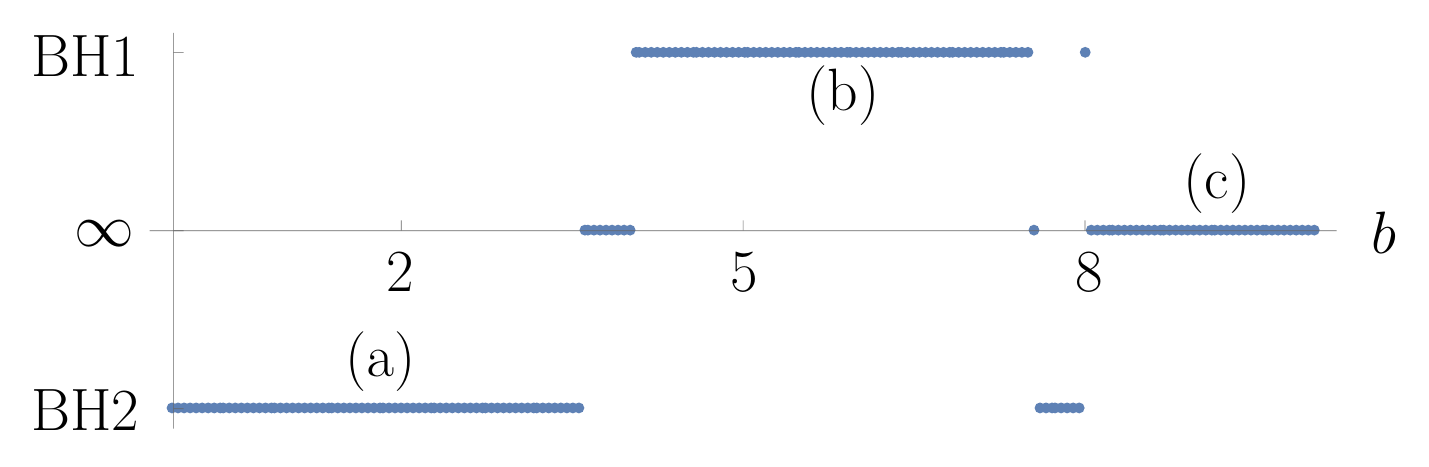}\label{fig:fractalplotb}} \\
\end{tabular}
\caption{The shadow as viewed along the symmetry axis for $a=2$. The 2D shadow is the area of revolution of the 1D shadow, by symmetry. Geodesics on an initial data surface are labelled by their impact parameter. Once again, decision dynamics can be used to order the perpetual orbits in the initial data \label{fig:impactparameter}.}
\end{figure}

The first frame of Fig.~\ref{fig:shadows} depicts the shadow of binary system as seen from the negative $z$-axis ($0^{\circ}$). By symmetry, the $\theta=0^\circ$ shadow may be constructed from the area of revolution of the 1D shadow. Figure \ref{fig:impactparameter} shows a family of null rays, sent towards the two black holes from infinity, initially parallel to the $z$-axis, with impact parameter $b$.

Let us consider the crude features of Fig.~\ref{fig:impactparameter} as $b$ is increased from zero. Near $b\sim 0$, all rays fall into the lower black hole, generating a purple disc in the centre of the 2D shadow. Next comes a `gap', corresponding to rays that pass between the holes and escape; followed by a green ring in $b$ in which geodesics fall into the upper black hole; next, a secondary gap and an outer purple ring. Beyond a certain impact parameter, all rays escape to infinity.

This crude description above overlooks the self-similar properties of the shadow. Let us construct this shadow using decision dynamics (cf.~Fig.~\ref{fig:decision}, Sec.~\ref{subsubsec:ordering} \& \ref{subsec:cantor}). The edge of the inner purple disk corresponds to decision sequence $000\cdots$ and the outer edge of the shadow to sequence $444\cdots$. Embedded in the initial data set are an infinite number of impact parameters $b_X$ which correspond to perpetual orbits. For these values, the scattering process is singular. Here $X$ is any non-terminating decision sequence in base-5 without the digits $1$ or $3$. The ordering of $b_X$ in the initial data set is once again determined by $F(X)$ (Sec.~\ref{subsubsec:ordering}), but here with the ordering reversed (this can be achieved by redefining the parity-reversal~$\widetilde{}$ operation to begin with $P = -1$). The 1D shadow may be constructed iteratively, just as in Sec.~\ref{subsec:cantor}, and thus it has Cantor-like properties, which are inherited by the 2D shadow.

\subsubsection{1D slices of 2D shadows\label{subsec:slices}}

\begin{figure}[h]
\begin{tabular}{cc}
 \subfigure[2D shadow: $a=2$]{
   \includegraphics[width=7cm,trim={0.1cm 0.1cm 0.1cm 0.1cm},clip]{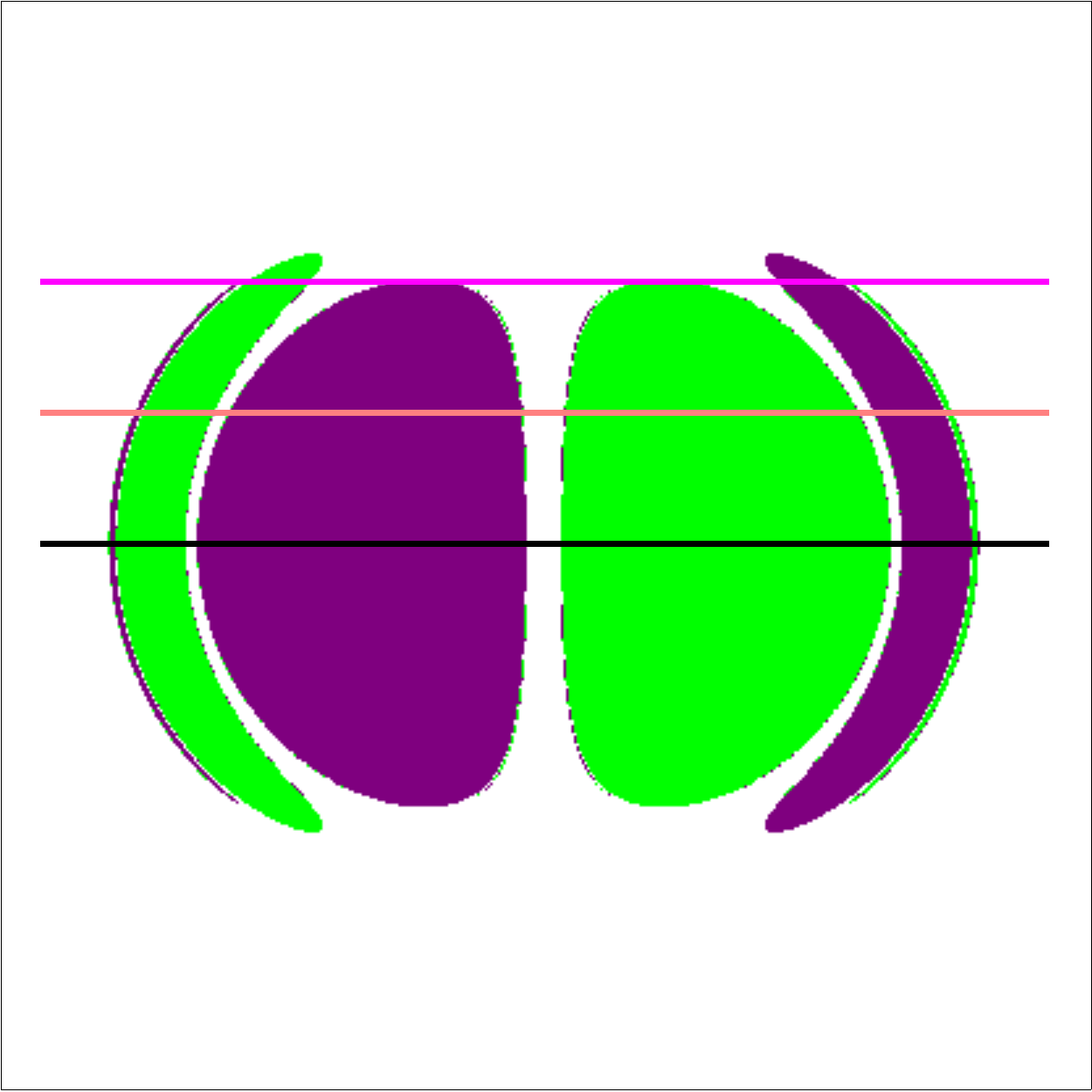} \label{fig:slices_2D_a2}
  }
  &
 \subfigure[1D slices: $a=2$]{
  \begin{tabular}[b]{c}
   \includegraphics[width=7cm]{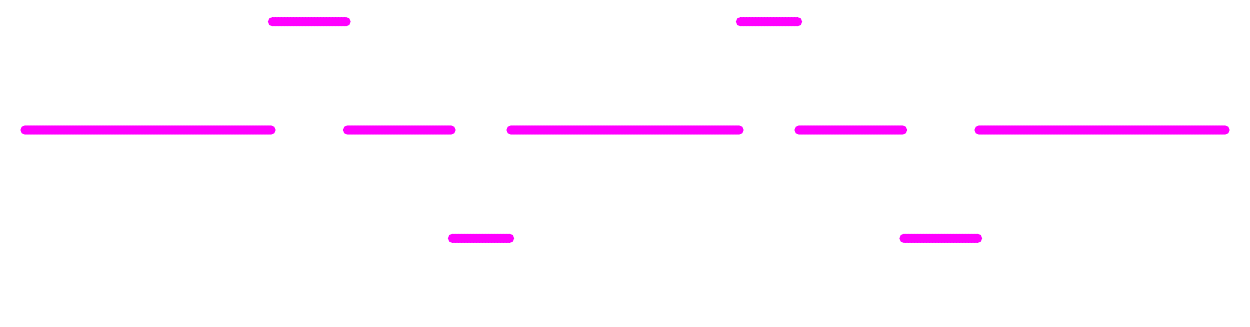} \\
   \includegraphics[width=7cm]{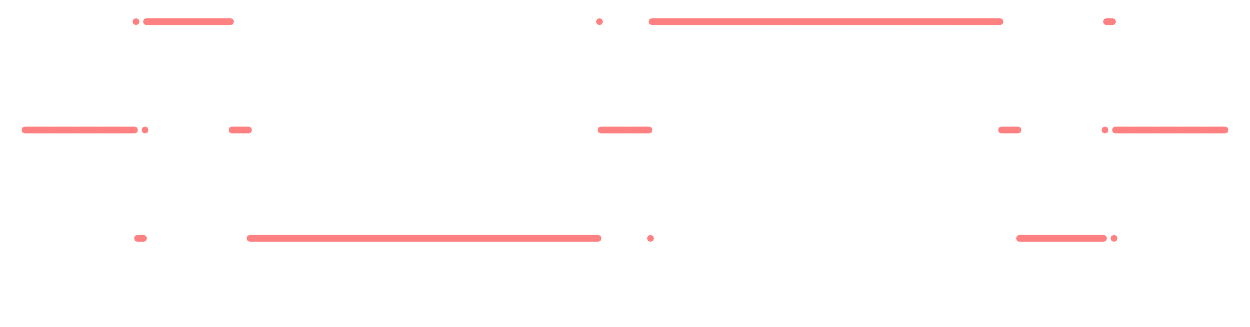} \\
   \includegraphics[width=7cm]{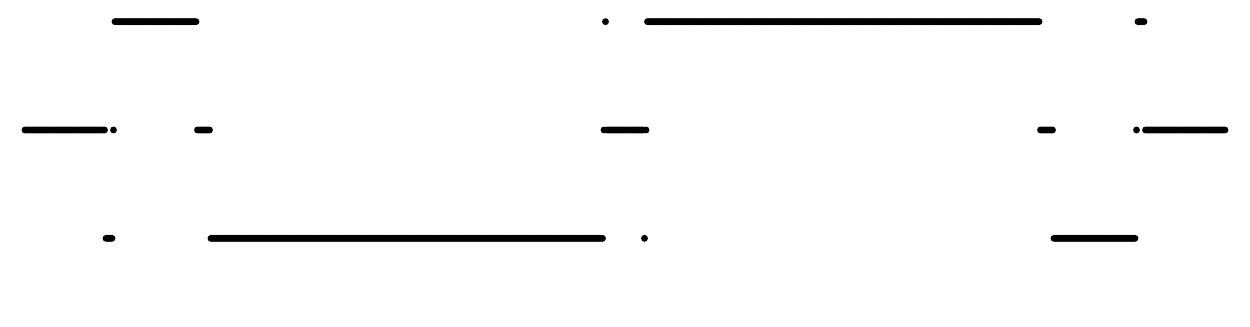}
  \end{tabular} \label{fig:slices_1D_a2}
 } \\
 \subfigure[2D shadow: $a=1$]{
   \includegraphics[width=7cm,trim={0.1cm 0.1cm 0.1cm 0.1cm},clip]{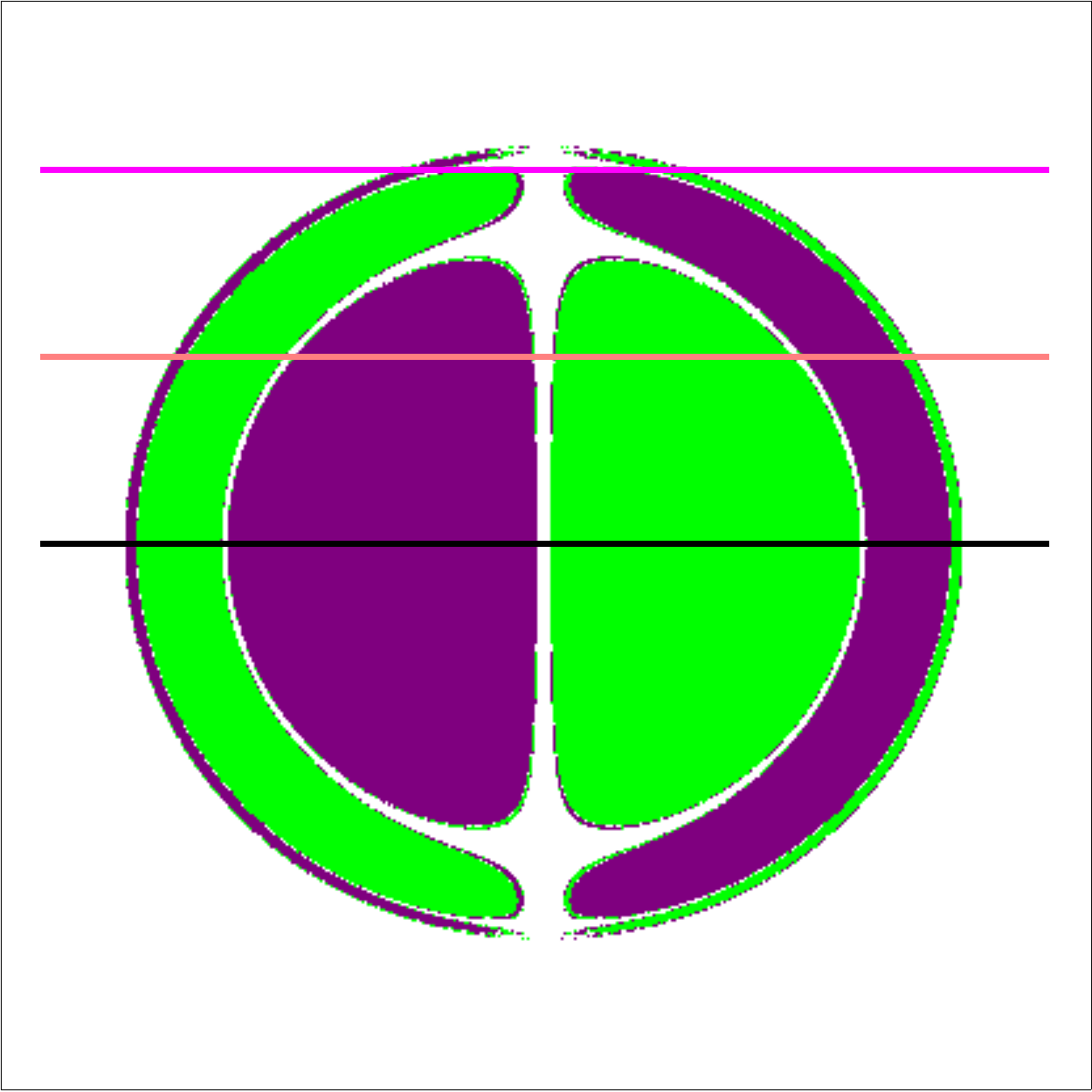} \label{fig:slices_2D_a1}
  }
  &
 \subfigure[1D slices: $a=1$]{
  \begin{tabular}[b]{c}
   \includegraphics[width=7cm]{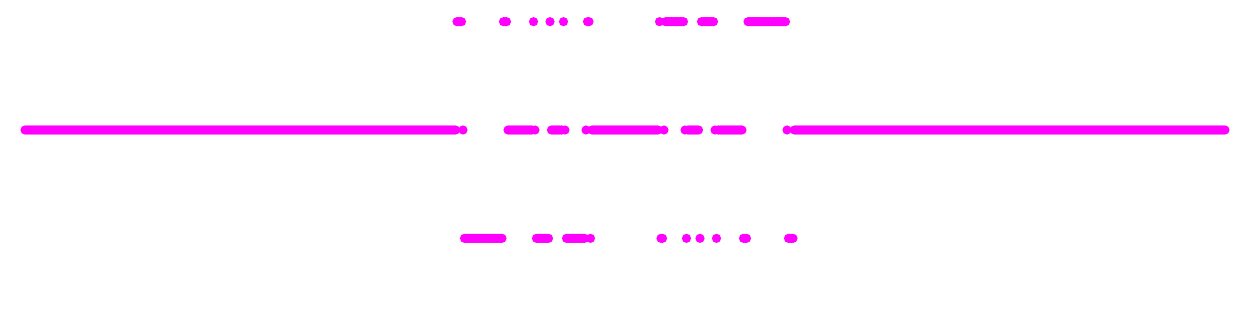} \\
   \includegraphics[width=7cm]{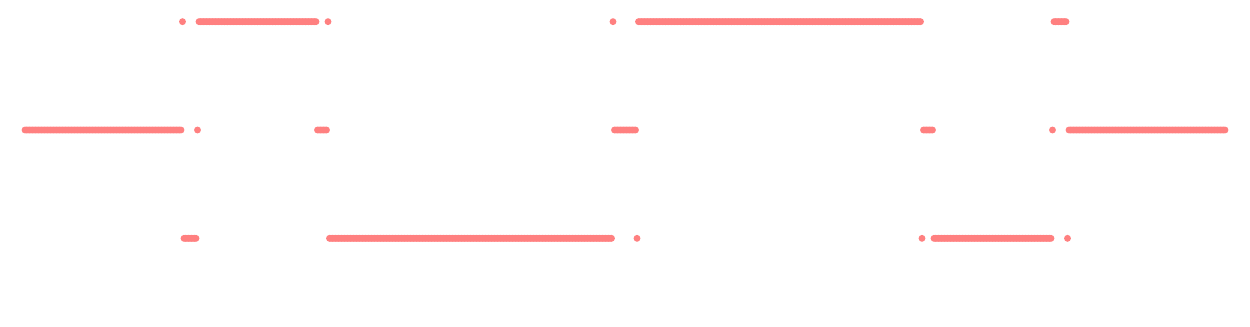} \\
   \includegraphics[width=7cm]{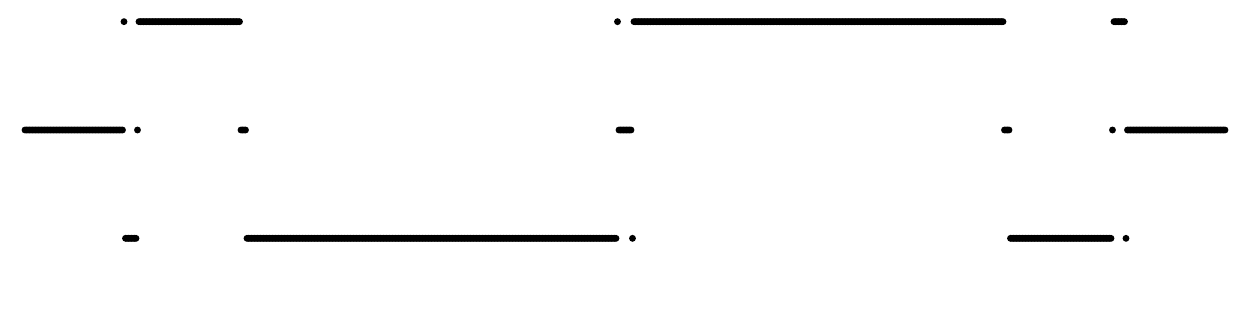}
  \end{tabular} \label{fig:slices_1D_a1}
 }
\end{tabular}
 \caption{Two-dimensional shadows as a union of the one-dimensional shadows of fixed $p_\phi$. Here are shown three 1D `slices' (right) of the $\theta=90^\circ$ 2D shadows (left) for separations $a=2$ (upper) and $a=1$ (lower). Note that the upper slice (magenta) in (b) does not show fractal structure, whereas the upper slice in (d) shows qualitatively different chaotic behaviour associated with the `pocket' of Fig.~\ref{fig:threethroats} (see also Fig.~\ref{fig:frond}).}
 \label{fig:slices}
\end{figure}

We may inspect the $\theta = 90^\circ$ case by decomposing it into a family of 1D shadows of constant $p_\phi$. Each value of $p_\phi$ corresponds to a horizontal slice, running across the image plane. Some examples of 1D slices of 2D shadows are shown in Fig.~\ref{fig:slices}.

In the $a=2$ case, we note that 1D slices taken across the middle of the shadow possess the now-familiar Cantor-like fractal structure. However, certain 1D slices for large $p_\phi$ (across the top of the shadow) do \emph{not} exhibit fractal structure. Instead, these 1D shadows have well-defined sharp edges. This behaviour was anticipated from Fig.~\ref{fig:nonplanar-fundamental}. Figure \ref{fig:nonplanar-fundamental:d}  shows a value of $p_\phi$ for which `exterior' null orbits ($\dot{0}$ and $\dot{2}$) are forbidden. Though two `interior' orbits ($\dot{4}$) remain possible (permitting absorption to occur), these interior orbits no longer dynamically connected to each other. In other words, transitions (`decisions') are not possible, and thus chaotic scattering does not occur.

In the $a=1$ case (Fig.~\ref{fig:slices_1D_a1}), the 1D slices for larger $p_\phi$ show highly chaotic behaviour, which is \emph{not} effectively described by our `decision dynamics' approach. The highly chaotic behaviour is associated with the `pocket' feature shown in Fig.~\ref{fig:threethroats} (and thus, indirectly, with the existence of bounded null orbits, Fig.~\ref{fig:bounded}).

\subsubsection{Varying the separation $a$}

Clearly then, there are important qualitative differences in the shadows for the $a=1$ and $a=2$ cases. In fact, such differences may be anticipated from Figs.~\ref{fig:morphology} and \ref{fig:contour} (and text), where we highlighted a `phase change' in the behaviour of null orbits as the black holes are brought together. This was understood using a classification of stationary points of $h$ in the equatorial plane, which highlighted the critical values $a_1$ and $a_2$ in Eq.~(\ref{eq:a1a2}).

\begin{figure}[h]
\begin{tabular}{cc}
 \subfigure[$a=2$]{
   \includegraphics[width=7cm,trim={1cm 2cm 1cm 2cm},clip]{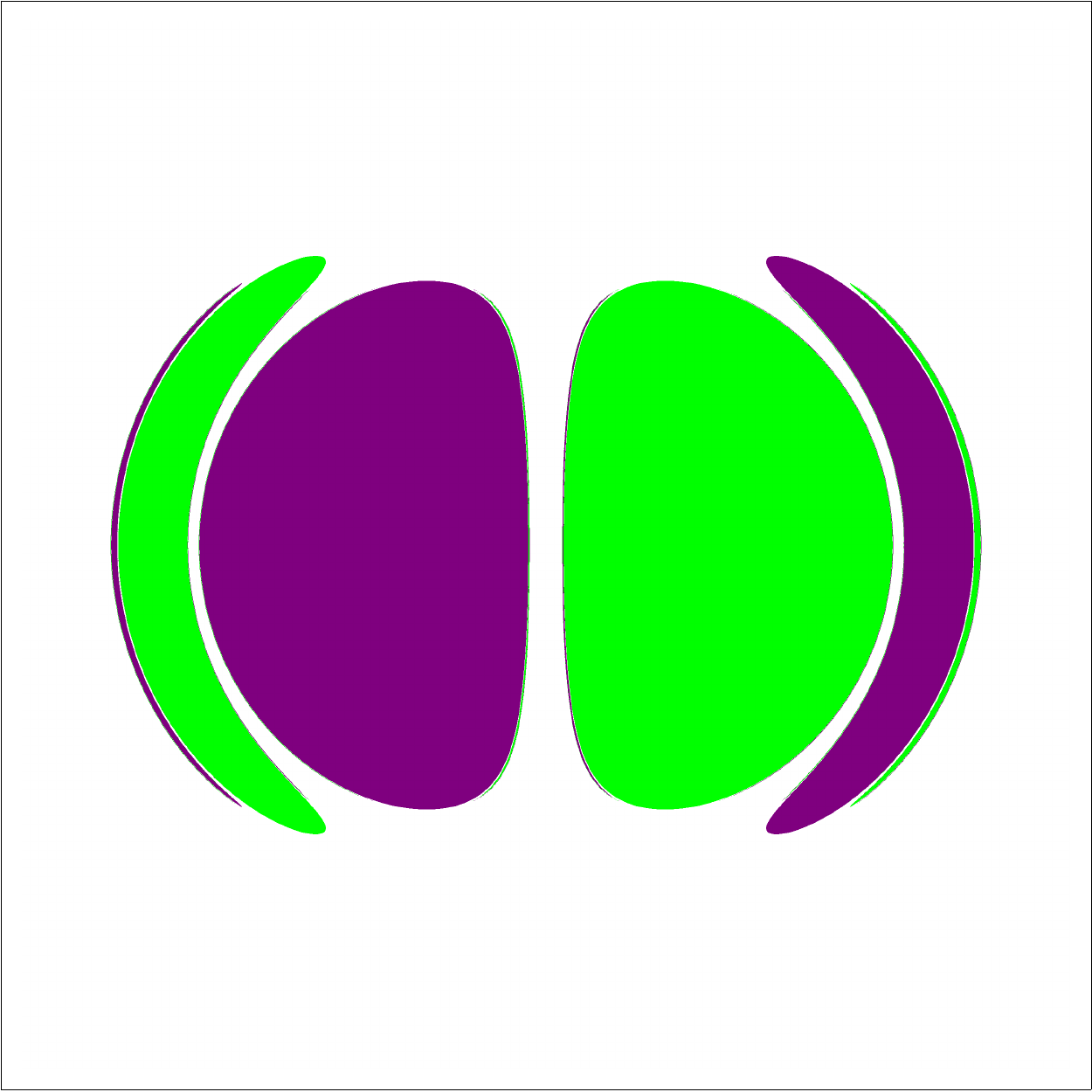}
  } &
 \subfigure[$a=1$]{
   \includegraphics[width=7cm,trim={1cm 1cm 1cm 1cm},clip]{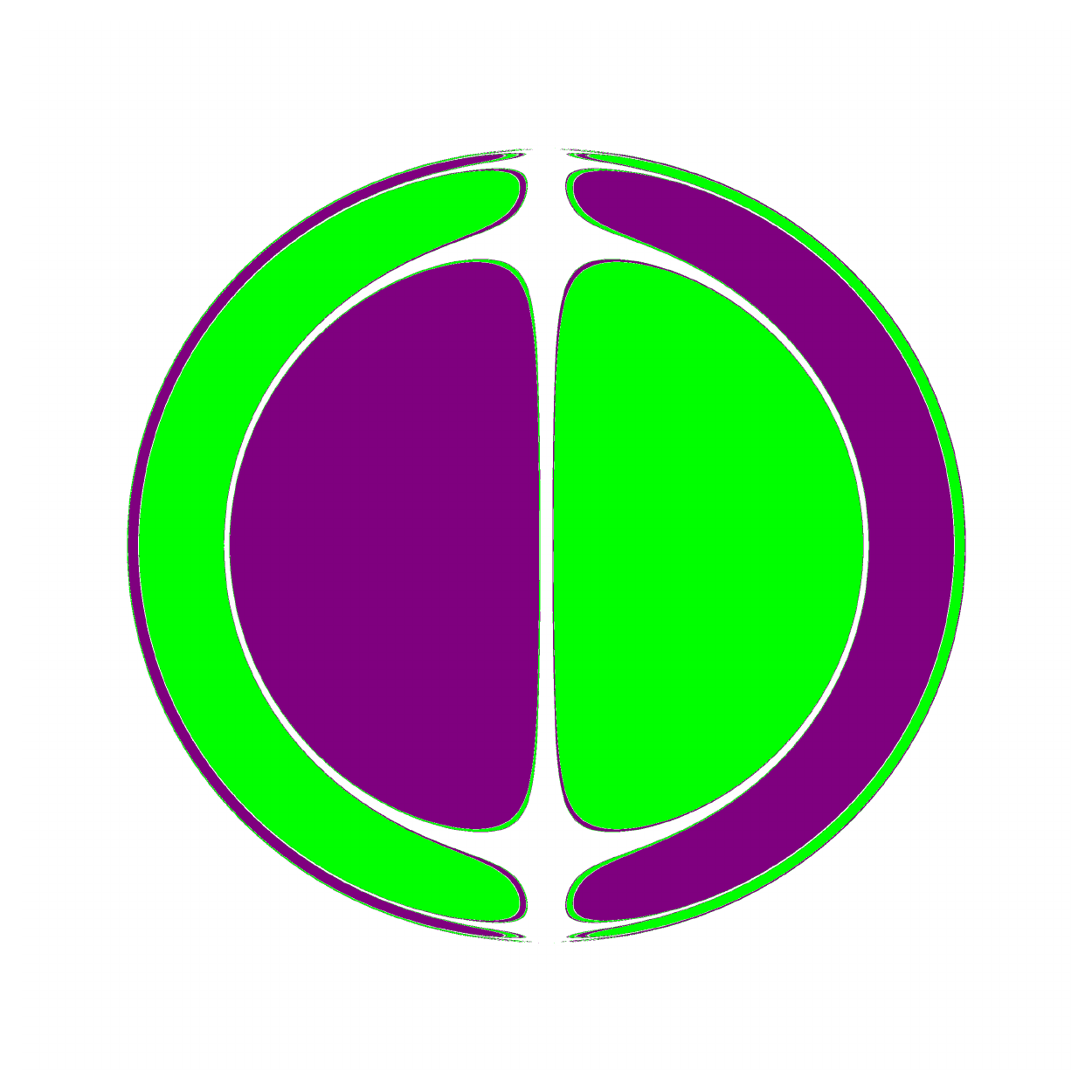}
  } \\ \\
 \subfigure[$a=0.5$]{
   \includegraphics[width=7cm,trim={1cm 1cm 1cm 1cm},clip]{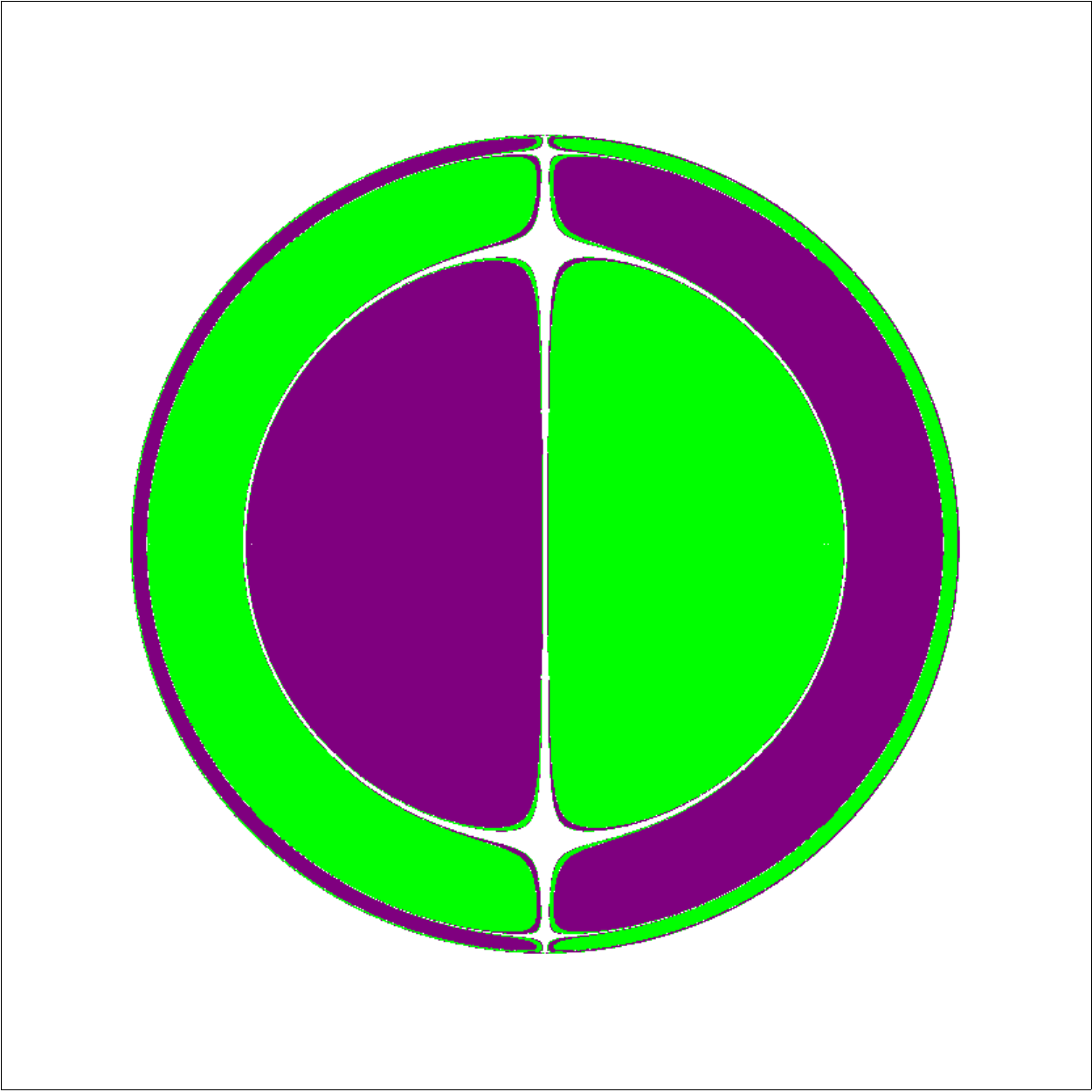}
 } &
 \subfigure[$a=1$, close-up]{
   \includegraphics[width=7cm]{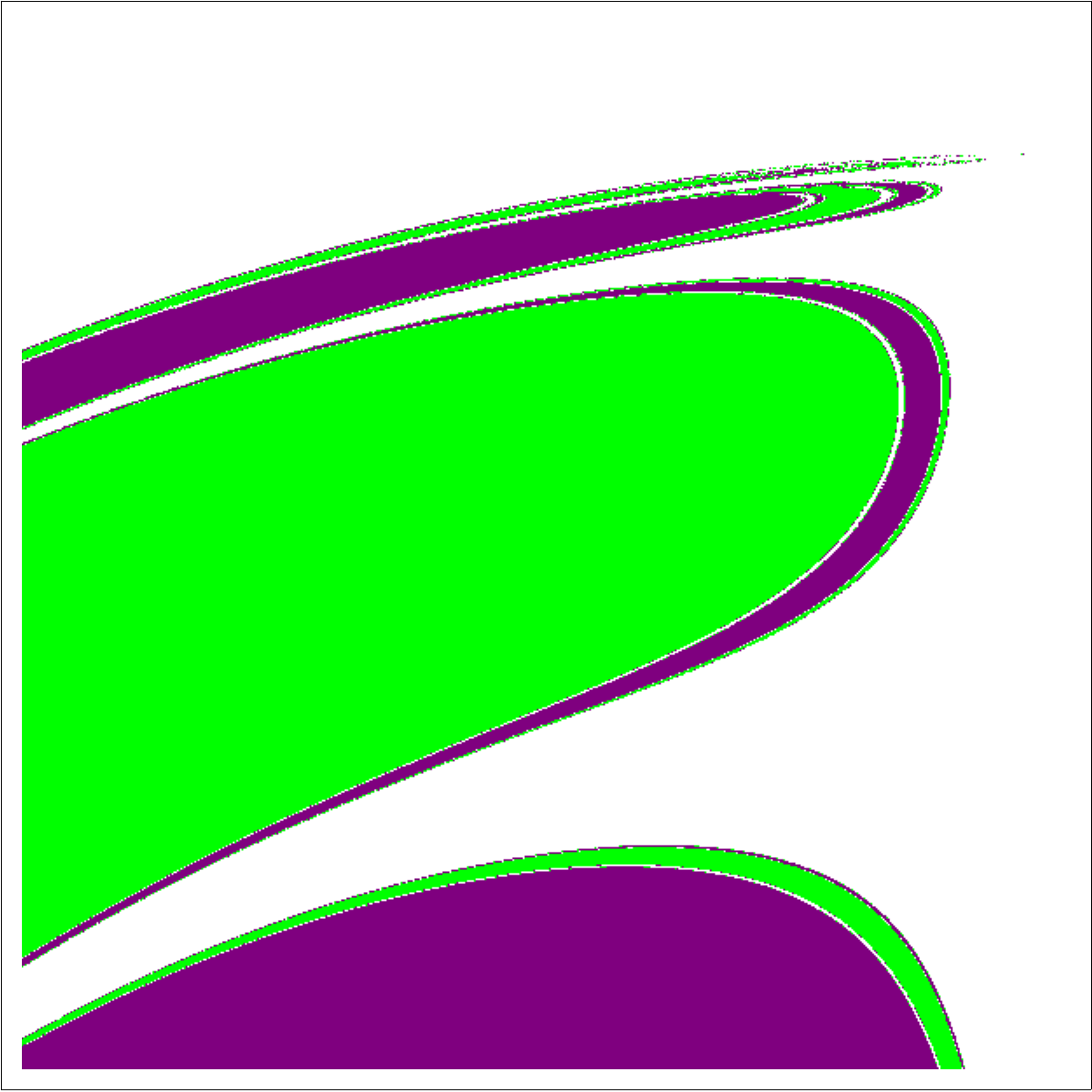}\label{fig:frond}
 }
\end{tabular}
 \caption{The MP binary shadow viewed from $\theta = 90^\circ$ for the cases of equal-mass black holes separated by (a) $a=2$, (b) $a=1$, (c) $a=0.5$. Plot (d) shows a close-up of the upper region of plot (b). The upper `frond' shows highly chaotic behaviour associated with the three-throat system of Fig.~\ref{fig:threethroats}.}
 \label{fig:ashadow}
\end{figure}

Figure \ref{fig:ashadow} shows the shadow cast by MP binaries separated by $a=0.5$, $1$ and $2$ (noting that $0.5 < a_1$, $a_1 < 1 < a_2$ and $a_2 < 2$). Plot \ref{fig:frond} shows a close-up of the upper `fronds' of the shadow in the $a=1$ case. This provides another view of the qualitatively different chaotic behaviour associated with Fig.~\ref{fig:threethroats}. We speculate that the upper frond will exhibit the Wada property \cite{Kennedy:1991, Poon:1996, Sweet:1999, Aguirre:2001}. Further investigation is needed (for example, using the method of Ref.~\cite{Daza:2015}).

\section{Extensions\label{sec:extensions}}
In this section we briefly explore two possible extensions to this work: chaotic scattering of geodesics which pass through the horizons (\ref{subsec:through}); and chaotic scattering on other spacetimes (\ref{subsec:alternative}).

 \subsection{Through the event horizons\label{subsec:through}}
If one is willing to follow null geodesics through the horizons of the binary MP spacetime, then the chaotic scattering phenomenon becomes even richer.

The `points' where $U\rightarrow\infty$ in the MP spacetime (at $x=y=0$, $z=z_\pm$) are \emph{not} spacetime singularities but merely coordinate singularities \cite{Hartle:1972ya}. The coordinate time diverges towards these `points' , $t \rightarrow \infty$ as $U \rightarrow \infty$, but the geodesic affine parameter $\lambda$ remains regular; as do curvature invariants such as $R_{\alpha \beta \mu \nu} R^{\alpha \beta \mu \nu}$. In fact, the $U \rightarrow \infty$ `point' is actually a null surface with finite area \cite{Hartle:1972ya}.

\begin{figure}[h]
\begin{tabular}{ccc}
 \subfigure[Conformal diagram]{
 \includegraphics[height=7cm]{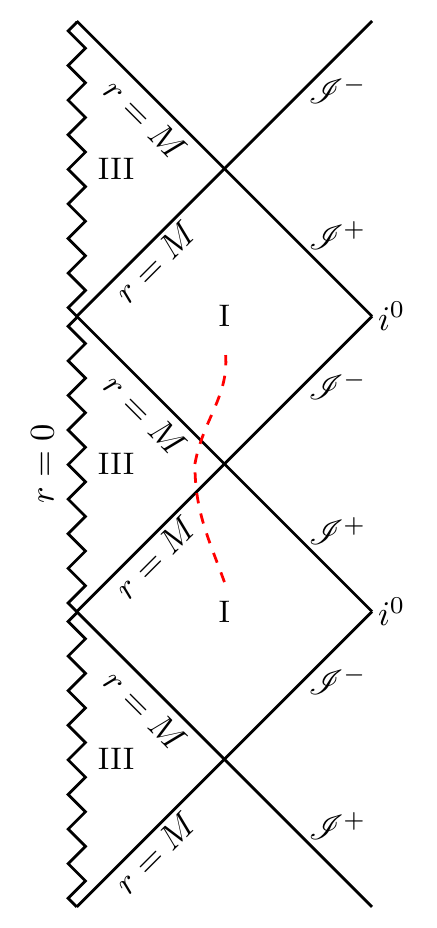} \label{fig:penrosediagram}
  } & \hspace{0.3cm}
  &
 \subfigure[Periodic ray]{
 \includegraphics[height=7cm]{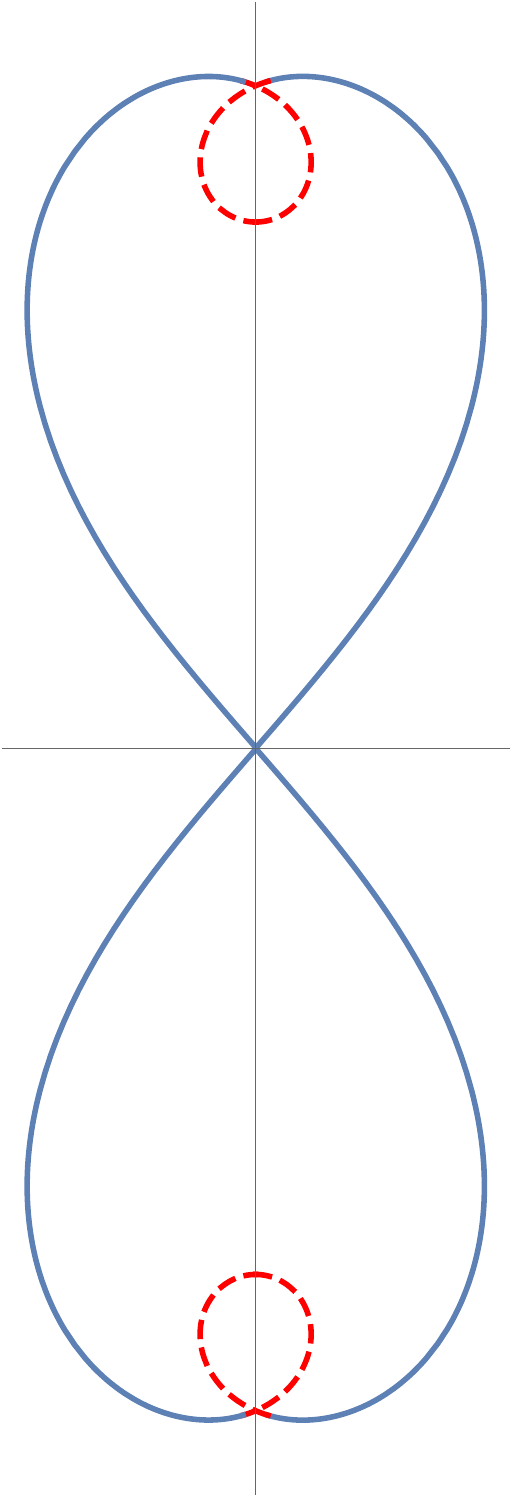}\label{fig:into}
 }
\end{tabular}
 \caption{(a) Conformal diagram for a single extremal Reissner--Nordstr\"{o}m black hole. A light ray with non-zero angular momentum is shown as a red dotted line. Here it passes through a black hole horizon into region III, then emerges from a white hole horizon into a new region I. (b) An example of a periodic null ray that passes through black hole horizons (see text).}
 \label{fig:through}
\end{figure}

In the single black hole case, the MP metric in spherical coordinates, $ds^2 = -U^{-2} dt^2 + U^2 (dr^2 + r^2 d\Omega^2)$ where $U = 1 + M/r$, may be transformed into the standard Reissner--Nordstr\"om metric, $ds^2 = -f(\hat{r}) dt^2 + f^{-1}(\hat{r}) d\hat{r}^2 + \hat{r}^2 d\Omega^2$ where $f(\hat{r}) = (1-M/\hat{r})^2$, by a straightforward change of variables, $\hat{r} \equiv r + M$. Alternatively, one may replace the $t$ coordinate with a retarded ($-$) or advanced ($+$) null coordinate, $w_\pm \equiv t \pm F(r)$ where $dF/dr \equiv U$, giving the line element $ds^2 = \widetilde{g}_{\mu \nu} d\widetilde{x}^\mu d\widetilde{x}^\nu = -U^{-2} dw_\pm^2 \pm dw_\pm dr + U^2 r^2 d\Omega^2$. In the $\{w_\pm,r,\theta,\phi\}$ coordinate system, the components $\widetilde{g}^{\mu \nu}$ remain finite as $U \rightarrow \infty$, and thus the standard Hamiltonian $H = \frac{1}{2} \widetilde{g}^{\mu \nu} \widetilde{p}_{\mu} \widetilde{p}_{\nu}$ may be used. Employing the $w_-$ and $w_+$ coordinates, respectively, we may track a null geodesic passing into ($-$) or out of ($+$) the horizon (see Fig.~\ref{fig:penrosediagram}).

In the double black hole case, we may proceed as follows. Suppose we wish to track a null ray passing through the horizon of the lower black hole. We start by changing to a spherical coordinate system $\{t,r,\theta,\phi\}$ centred at $x=y=0$, $z=z_-$, so that $U(r,\theta) = 1 + M_-/r + M_+/\sqrt{r^2 - 2 a r \cos \theta + a^2}$. Next, we introduce $V(r) \equiv 1 + M_-/r + M_+/a$ and the null coordinate $w_\pm = t \pm F(r)$ such that $dF/dr = V$. The line element becomes
\beq
ds^2 =  - U^{-2} dw_\pm^2 \pm 2V^2 U^{-2} dw_\pm dr + U^{-2} (U^4 - V^4) dr^2 + U^2 r^2 d\omega^2,
\eeq
and the Hamiltonian is
\beq
H = \frac{1}{2} \widetilde{g}^{\mu \nu} \widetilde{p}_\mu \widetilde{p}_\nu = \frac{1}{2 U^2} \left( -(U^4 - V^4) \widetilde{p}_w^2 \pm 2 V^2 \widetilde{p}_{w} \widetilde{p}_r + \widetilde{p}_r^2 + \frac{1}{r^2} \widetilde{p}_\theta^2 + \frac{1}{r^2 \sin^2\theta} \widetilde{p}_\phi^2 \right).
\eeq
The new momenta are related to the old momenta in a straightforward fashion: $\widetilde{p}_r = p_r \mp V^2(r) p_t$, $\widetilde{p}_w = \widetilde{p}_t$,  $\widetilde{p}_\theta = \widetilde{p}_\theta$,  $\widetilde{p}_\phi = \widetilde{p}_\phi$. We note that $\lim_{r\rightarrow 0} U^2 - V^2 = 2 M_+ M_- \cos \theta / a^2$; hence the Hamiltonian formulation is not singular as $r \rightarrow 0$. Thus, we may evolve Hamilton's equations through the coordinate singularity.

It is worth stressing that, in the above, we are considering a coordinate patch in which the event horizon of the lower black hole is at $r=0$ and the singularity is represented by a locus satisfying $U = 0$ where
\beq
U = 1 - \frac{M_-}{r'} + \frac{M_+}{(r')^2 + 2 a r' \cos \theta + a^2}  , \quad \quad \quad r' = -r .
\eeq
In the single black hole case $M_+ = 0$ this is a circle of radius $M_-$; the other black hole has the effect of distorting the circle.

In the case of a single Reissner--Nordstr\"om black hole, it is well-known that null geodesics may be extended  through the (black hole/future) horizon, passing from region I to III as shown in Fig.~\ref{fig:penrosediagram}. Once inside region III ($r < 0$), all null geodesics with non-zero angular momentum will avoid the timelike singularity, and emerge through a (white hole/past) horizon into a new asymptotically flat spacetime (region I$^{\prime}$, $r > 0$).

Similar behaviour is expected in the double black hole case. We may follow a null geodesic through either black hole horizon, from $r > 0$ to $r < 0$ in a coordinate patch centred on the relevant black hole, by using null coordinate $w_-$, as described above. Nearly all null geodesics will come to a turning point where $\dot{r} = 0$; with the exception of a set of measure zero, which collide with the singularity. At the turning point, one may switch to the advanced null coordinate $w_+$ (switching also the momentum $p_{r}^{+} = p_r^{-} - 2V^2 p_t$), and then follow the geodesic from $r < 0$ to $r > 0$. Once in the exterior, one may switch back to the isotropic (MP) coordinates. The geodesic may then plunge into the other black hole; fall into the same black hole again; or escape to infinity. We may follow the geodesic as far as is desired, by repeating the method above.

Figure \ref{fig:into} shows that there exist a class of periodic null orbits which pass into, and out of, black hole event horizons. The red dashed line shows the part of the geodesic that was evolved using the $w_\pm$ null coordinates; the switch from $w_-$ to $w_+$ occurred at the symmetry point.  We take the liberty of using the magnitude of $r$ (e.g. $z = |r| \cos \theta$, etc) in order to show the $r>0$ and $r < 0$ regions on the same plot. One convenient way of visualizing this trajectory is to imagine the horizons as pin holes in a sheet of paper; on passing through the pinhole we continue the trajectory on the opposite side of the paper, before returning through the pinhole. This picture is misleading, of course, as the ray emerges into a new asymptotically flat spacetime each time it emerges from a horizon (see Fig.~\ref{fig:penrosediagram}).

 \subsection{Other spacetimes\label{subsec:alternative}}

In motivating this work, the MP spacetime was introduced as a surrogate for a spacetime of physical interest: two black holes in the final stages of inspiral and merger. However, the MP fails as a surrogate in several ways. In a `realistic' binary the black holes are not significantly charged, yet they may be rotating at a significant fraction of the Kerr bound; they orbit around the centre of mass (thus $p_\phi \neq \text{const}$); and spiral inwards as they lose energy to gravitational-wave emission (thus $p_t \neq \text{const}$). Let us now briefly consider alternative closed-form models and surrogates, not forgetting that Bohn \emph{et al.} \cite{Bohn:2014xxa} have shown that it is possible to study geodesics on numerically generated `realistic' binary spacetimes.

The MP spacetimes are members of more general classes, including: (i) the Israel--Wilson class \cite{Israel:1972vx}; (ii) higher-dimensional MP spacetimes \cite{Hanan:2007}; (iii) the Kastor--Traschen class \cite{Kastor:1992nn}. In each of these cases, the black holes are extremally charged. In the latter case, the spacetime has a positive cosmological constant, which has the effect of pushing the black holes together, mimicking (to some extent) a head-on collision. Shadows of Kastor--Traschen double black hole solutions were studied in Refs.~\cite{Nitta:2011in, Yumoto:2012kz}.

A static spacetime containing two \emph{uncharged} black holes was first studied by Bach and Weyl in 1922 \cite{Bach:2012}. In the absence of charge, a ``Weyl strut'' (a conical deficit angle on the symmetry axis) is required to keep the black holes apart.

Chaotic behaviour is not limited to binary systems. Of course, multi-black-hole spacetimes should exhibit even richer dynamics. But there are also a range of \emph{singleton} systems that exhibit chaotic scattering. Chaotic singletons include perturbed or tidally distorted black holes \cite{Bombelli:1991eg, Abdolrahimi:2015kma}, boson stars \cite{Vincent:2015xta}, and hairy black hole solutions \cite{Cunha:2015yba}.

\section{Conclusions and Discussion\label{sec:conclusions}}

Below we outline the main conclusions of this work. 
\begin{enumerate}
 \item Chaotic scattering \cite{Eckhardt:1988, Ott:Tel:1993} arises generically when a spacetime admits more than one (unstable) fundamental null orbit, provided that the fundamental orbits are distinct but `dynamically connected', such that null rays may transition between the asymptotic neighbourhoods of the fundamental orbits. This leads to the existence of an uncountably infinite number of distinct perpetual orbits, associated with a fractal set of scattering singularities in initial data.
 \item With `decision dynamics', developed here (Fig.~\ref{fig:decision}, Sec.~\ref{subsec:symbolicdyn}), we were able to understand the ordering and organisation of perpetual orbits in initial data for several simple cases. We showed that the 1D shadow could be constructed through an iterative procedure, by successively removing open intervals from initial data (Sec.~\ref{subsec:cantor}). This construction was akin to the usual construction of the Cantor set.
 \item In the case of planar motion in the MP spacetime ($p_\phi=0$), there are three distinct fundamental orbits (Fig.~\ref{fig:fundamental}). These orbits are `dynamically connected', in the sense that a ray can transition between them by making a sequence of `decisions'. As a consequence, there arises a $5$-adic Cantor set of scattering singularities on initial data, and the 1D shadow is manifestly self-similar (Fig.~\ref{fig:cantorzoom}, Sec.~\ref{subsec:selfsimilar}).
 \item In the non-planar case, the number of fundamental orbits varies with angular momentum $p_\phi$ (Fig.~\ref{fig:nonplanar-fundamental}). Where there is just one orbit, or where transitions between orbits are dynamically forbidden, the shadow loses its fractal property (Fig.~\ref{fig:slices}).
 \item The character of the non-planar fundamental orbits changes as the black holes move closer together (Fig.~\ref{fig:morphology}). Null orbits about an individual black hole are forbidden once the black hole separation $a$ is sufficiently small; these orbits are usurped by a equatorially symmetric orbit about the composite system (Fig.~\ref{fig:contour}).
 \item \emph{Stable bounded} null orbits exist for separations $a_1 < a < a_2$, where $a_1 = 4M/\sqrt{27}$ and $a_2 = \sqrt{2}a_1$ (Sec.~\ref{subsec:nonplanar}, Fig.~\ref{fig:bounded}). In the nearly-bounded case, a `pocket' develops, connecting the black holes and spatial infinity via three throats (Fig.~\ref{fig:threethroats}). Qualitatively different chaotic behaviour is associated with this pocket; its effect is visible in parts of the black hole shadow (Figs.~\ref{fig:slices_1D_a1} and \ref{fig:frond}). We speculated that these parts will exhibit the Wada property \cite{Sweet:1999}.
 \item In the MP spacetime, null rays can be followed through horizons, and out again, and thus there arises  more radical possibilities for perpetual orbits and chaotic scattering (Sec.~\ref{subsec:through}, Fig.~\ref{fig:through}).
\end{enumerate}

Let us now discuss some possible implications of this work for more realistic binary black hole systems.

High symmetry of the MP geometry means that one may naturally decompose a 2D shadow into 1D `slices' of fixed $p_\phi$. In 1D shadows, we observed three different qualitative behaviours: (1) well-ordered (Cantor-like) fractal shadows; (2) hard-edged shadows without any fractal features; and (3) highly chaotic fractal regions (Fig.~\ref{fig:slices}). These behaviours arise when: (1) there are several distinct-but-connected fundamental null orbits; (2) there is just one fundamental null orbit, or, multiple null orbits are all isolated; (3) in phase space there arises a `pocket with three throats' harbouring a `randomizing' region (Fig.~\ref{fig:threethroats}).

It is an open question whether all three types of behaviour may occur in simulations of `real' black hole binary shadows. A first investigation of this question might start with inspection of a representative sample of 1D slices from the shadows recently presented by Bohn {\it et al.} \cite{Bohn:2014xxa}.

The highly chaotic behaviour (Fig.~\ref{fig:slices}, Fig.~\ref{fig:frond}) was not anticipated. It only occurs for a limited range of black holes separations, $a_1<a<a_2$. Within this regime, there exists a family of stable bounded null geodesics (Fig.~\ref{fig:bounded}). Stable photon orbits will be explored more fully in Ref.~\cite{Dolan:2016bxj}. In light of this observation, it may prove fruitful to conduct a more systematic search for stable bounded null geodesics in black hole binaries and their surrogates \cite{Bach:2012, Kastor:1992nn}; as well as in singleton systems such as perturbed black holes \cite{Bombelli:1991eg, Abdolrahimi:2015kma}, boson stars \cite{Vincent:2015xta}, and hairy black hole solutions \cite{Cunha:2015yba}.

\acknowledgments
J.S.~acknowledges financial support from the University of Sheffield Harry Worthington Scholarship. S.D.~acknowledges financial support under EPSRC Grant No.~EP/M025802/1, and from the Lancaster-Manchester-Sheffield Consortium for Fundamental Physics under STFC Grant No.~ST/L000520/1. With thanks to Andy Bohn, Jason Cole and Jack Morrice.

\appendix

\section{Translation between symbolic codes\label{sec:translation}}

Given a sequence in the collision dynamics presented by Cornish and Gibbons \cite{Cornish:1996de}, we would like to be able to translate to the decision dynamics language. This can be achieved through the use of the following algorithm. Consider an infinite sequence $a_{1}, a_{2}, a_{3}, \cdots$, where $a_{i} \in \{1, 0, -1\}$, as described by Cornish and Gibbons's symbolic alphabet. Notice that, in the decision dynamics, any decision point must be preceded by $\pm 1$ in the collision dynamics. We refer to these as pivot points, as they will be central to the translation between the two schemes.

\begin{figure}[h]
 \includegraphics[width=7cm]{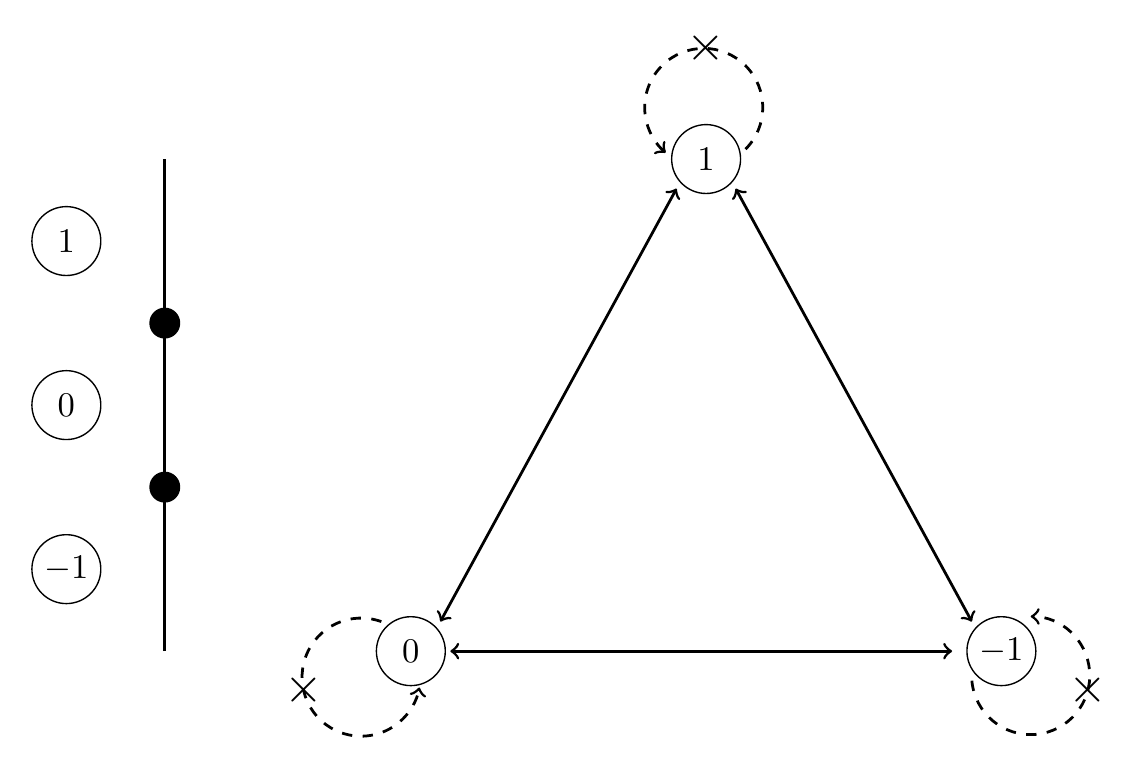}
 \caption{Schematic diagram of the allowed trajectories in the double black hole spacetime, showing the explicit correspondence with Eckhardt's three-disc model.}
 \label{fig:threedisc}
\end{figure}

Consider the first decision point, which follows the pivot point $a_{1} = \pm 1$ in the sequence. If $\left|a_{1} - a_{2} \right| = 2$, then $a_{2} = \mp 1$; that is, $a_{1}$ and $a_{2}$ have the opposite sign. In this case we replace `$a_{1}$' with `$0$' in the decision scheme and use $a_{2}$ as the new pivot point. However, if $\left|a_{1} - a_{2} \right| = 1$, then $a_{2} = 0$, so we must look at the third element of Cornish and Gibbons's sequence. If $\left|a_{1} - a_{3} \right| = 2$ (i.e. $a_{1}$ and $a_{3}$ have the same sign), then replace `$a_{1}, a_{2}$' with `$2$' in the decision dynamics and use $a_{3}$ as the new pivot point. Alternatively, if $\left|a_{1} - a_{3} \right| = 0$ (i.e. $a_{1}$ and $a_{3}$ have the same sign), then we replace `$a_{1}, a_{2}$' with `$4$' in the decision scheme and use $a_{3}$ as the new pivot point. We carry on in the same fashion all the way along the sequence.

The algorithm can be summarized using the following table. In the table, the leftmost column represents the pivot point, whilst the central columns correspond to the subsequent elements of the finite string we wish to translate. The final column gives the corresponding symbol from the decision dynamics alphabet. In each case, the final element of the subsequence under consideration (which is always $\pm 1$) becomes the new pivot point.
\beq
\begin{array}{c|cc||c}
  \pm 1 &  &  & \\ \hline
   & \mp 1 & & 0 \\
   & 0 & \mp 1 & 2 \\
   & 0 & \pm 1 & 4
\end{array}
\eeq


\section{Stationary points and bounded null geodesics\label{appendix:derivation}}
Here we derive the key results of Sec.~\ref{subsec:stationarypoints} and \ref{subsec:bounded}. Note that we use the convention $M=1$ throughout (i.e.~the mass of each black hole is set to unity).

First, consider motion confined to the equatorial plane $z=0$, which is governed by a one-dimensional potential $\hat{h}(\rho) \equiv h(\rho,0) =  \rho \hat{U}^2$, where $\hat{U}(\rho) = 1 + 2/R$ with $R(\rho) = \sqrt{\rho^2+a^2/4}$. It follows that
\beq
\hat{h}_{,\rho} = \frac{\hat{U}}{R^3} p_3(R) , \quad \quad p_3(R) \equiv R^3 - 2R^2 + a^2 .
\eeq
Null circular orbits exist where $\hat{h}_{,\rho} = 0$. By considering the discriminant $\Delta_R ( p_3 )= -a^2(27a^2 - 32)$, and also noting that $\hat{h}_{,\rho}(R=0) > 0$, we see that, for $a > a_2 \equiv \sqrt{32/27}$, there are no roots for $R>0$, and thus no equatorial circular orbits; whereas for $a < a_2$, there are two roots with $R>0$, and thus an inner and outer photon orbit in the equatorial plane. Such orbits are stable (unstable) under radial perturbation if $\hat{h}_{,\rho \rho} < 0$ ($\hat{h}_{,\rho \rho} > 0$). Thus, the inner orbit is stable under radial perturbation, and the outer orbit is unstable. For $a = a_2$, there is but one orbit which is marginally stable ($\hat{h}_{,\rho \rho}=0$) under radial perturbation.

Now let us consider the stability of equatorial circular orbits under perturbation in the $z$-direction. By symmetry, $U_{,z}(\rho,0) = 0$ and thus $h_{,z}(\rho,0) = 0$. Taking a second derivative, $h_{,zz}(\rho,0) = 2 \rho \hat{U} U_{,zz}(\rho,0)$, where
\beq
U_{,zz}(\rho, 0) = \frac{3 a^2 - 4 R^2}{2 R^5} .
\eeq
Thus, the equatorial circular orbits are stable under out-of-plane perturbations if $R > \sqrt{3} a / 2$. Inserting $R = \sqrt{3}a/2$ into $p_3(R) = 0$ gives $a^2 (a - 4/\sqrt{27}) = 0$. Thus, the inner orbit is stable under out-of-plane perturbations only if $a > a_1 \equiv 4/\sqrt{27}$ (and it exists only if $a < a_2$). In summary, we have established that stable circular equatorial orbits exist for MP di-holes with coordinate separations in the range $a_1 < a < a_2$. 

We note that the upper bound ($a < a_{2}$) of this inequality is consistent with Coelho and Herdeiro's result for the existence of equatorial circular orbits (Eq.~(46) in Ref.~\cite{CoelhoHerdeiro2009}), and with the study of W\"unsch \emph{et al.} \cite{Wunsch:2013st}.

Now let us consider equatorial orbits for the special case $a = 1$. For such orbits, the polynomial $p_3(R)$ factorizes into $p_3(R) = (R-1)(R^2 - R - 1)$, and thus the roots are $R = 1$ and $R = \varphi$, where $\varphi$ is the Golden Ratio. The inner orbit at $R=1$ ($\rho = \sqrt{3} / 2$) is stable. The outer orbit at $R = \varphi$ ($\rho = \sqrt{\varphi^2 - 1/4} = \frac{1}{2} 5^{1/4} \varphi^{3/2}$) is unstable under radial perturbation.

It is more challenging to locate the stationary points of $h$ out of the equatorial plane, but progress can be made by introducing elliptic coordinates $\rho = \frac{a}{2} \sinh \xi \sin \eta$, $z = \frac{a}{2} \cosh \xi \cos \eta$, so that $U = 1 + 4 \cosh \xi / (a (\cosh^2 \xi - \cos^2 \eta))$. For simplicity, we make the replacement $X = \cosh \xi$, $Y = \cos \eta$, and we note that $X \in \left(1,\infty \right)$ and $Y \in \left(-1,1\right)$, since we do not want $\rho$ to be zero. The cylindrical coordinates may then be written as $\rho = \frac{a}{2}\sqrt{X^{2} - 1}\sqrt{1-Y^{2}}$, $z = \frac{a}{2} X Y$, such that $U = 1 + 4 X / (a (X^{2}- Y^{2}))$. In these coordinates, the height function reads
\beq
h = \frac{a}{2} \sqrt{X^{2} - 1}\sqrt{1-Y^{2}} \left( 1 + \frac{4X}{a(X^{2}- Y^{2})} \right)^{2}.
\eeq

The stationary point conditions $h_{,X} = 0 = h_{,Y}$ lead to the pair of equations
\begin{eqnarray}
a X^{5} &+& 4 X^{4} - 2a X^{3} Y^{2} + 12 X^{2} Y^{2} - 16 X^{2} + a X Y^{4} = 0, \label{eqn:sp1} \\
a X^{5} &-& 4 X^{4} - 2a X^{3} Y^{2} - 12 X^{2} Y^{2} + 8 X^{2} + a X Y^{4} + 8 Y^{2} = 0. \label{eqn:sp2} 
\end{eqnarray}
Subtracting Eq.~\eqref{eqn:sp2} from Eq.~\eqref{eqn:sp1} and dividing through by a factor of $8$ gives
\beq \label{eqn:sp5}
X^{4} + 3 X^{2} Y^{2} - 3 X^{2} - Y^{2} = 0,
\eeq
which allows us to write $Y^{2}$ in terms of $X$ as
\beq \label{eqn:ysquared}
Y^{2} = \frac{X^{2}(X^{2} - 3)}{1 - 3 X^{2}} = \frac{X^{2} - 3}{X^{-2} - 3}.
\eeq
Remarkably, the relationship between $Y^{2} = \cos^{2} \eta$ and $X = \cosh \xi$ is independent of the value of the coordinate separation $a$. We may now use Eq.~\eqref{eqn:ysquared} to eliminate $Y$ from the sum of Eqs.~\eqref{eqn:sp1} and \eqref{eqn:sp2}. This leads to a quintic in $X$, 
\beq \label{eqn:polyx}
a X^{5} - a X^{3} - 3 X^{2} + 1 = 0.
\eeq

For a general value of $a$, it is not possible to factorize the quintic, or find its roots in closed form in terms of $a$. However, for the special case $a = 1$, the left-hand side of Eq.~\eqref{eqn:polyx} factorizes to give $X^{5} - X^{3} - 3 X^{2} + 1 = (X^{3} + X^{2} + X - 1)(X^{2} - X - 1)$. The first factor has no roots with $X >1$. The second factor has one such root, $X = \varphi$. Using the relation \eqref{eqn:ysquared}, we see that for $X = \varphi$, we have $Y = \pm \varphi^{-2}$. Thus, for $a = 1$, the stationary point conditions are met for $\cosh \xi = \varphi$ and $\cos \eta = \pm \varphi^{-2}$, corresponding to points at $\rho = \frac{1}{2} 5^{1/4} \varphi^{-1/2}$ and $z = \pm 1/(2 \varphi)$. By inserting these results for $a=1$ into $h = \rho U^2$, one can verify that these two saddle points lie on the same contour as the equatorial saddle point, $h = \frac{1}{2} 5^{5/4} \varphi^{3/2}$. Thus $a=1$ is indeed a special case.

Since Eq.~\eqref{eqn:polyx} is linear in $a$, it is possible to substitute in a closed-form value of $X$ and solve to find the corresponding value of the separation $a$. We may then use Eq.~\eqref{eqn:ysquared} as a consistency check, noting that $0 < Y^{2} < 1$. For example, $X = \sqrt{2}$ is a solution to Eq.~\eqref{eqn:polyx} when $a = 5/(2\sqrt{2}) \approx 1.7677695$. It is then straightforward to check that $Y^{2} = 2/5$, which is in the required range.

Considering Eq.~\eqref{eqn:ysquared} for a general value of $a$, and noting that $X > 1$, it is clear that we require $X^{2} < 3$ for a solution. Setting $X^{2} = 3$, Eq.~\eqref{eqn:polyx} implies that $a = 4/\sqrt{27} = a_{1}$. Thus, for $a < a_{1}$ there are no circular photon orbits out of the equatorial plane.

We may perform a first-order perturbative expansion, looking at the limit $X \rightarrow 1$. If we let $X = 1 + \varepsilon$, where $\varepsilon \ll 1$, then $X^{2} \sim 1 + 2\varepsilon$. Thus, from Eq.~\eqref{eqn:ysquared},
\beq
Y^{2} = \frac{(1 + 2 \varepsilon)(2 - 2 \varepsilon)}{2 + 6 \varepsilon} \sim 1 - 2 \varepsilon.
\eeq
So $Y^{2} \sim 1/X^{2}$ in the limit $X \rightarrow 1$. It follows that, for widely separated black holes $a \gg 1$, we have $X \sim 1 + 1/a$ and $Y \sim 1 - 1/a$. In this case, the MP di-hole system will resemble two isolated black holes, each with an unstable circular photon orbit at $\rho \sim 1$ and $z \sim \pm a/2$. 


\bibliographystyle{apsrev4-1}

\bibliography{refs}

\end{document}